\documentclass[twocolumn]{aastex63}
\pdfoutput=1
\usepackage{amsmath}
\usepackage{amsfonts}
\usepackage{amssymb}
\usepackage{url}
\usepackage{xspace}
\usepackage{xcolor}
\usepackage{tikz}
\usepackage{enumitem}

\definecolor{mygrn}{rgb}{0.133, 0.545, 0.133}

\newcommand{\A}{\ensuremath{\mathrm{\AA}}\xspace}

\newcommand{\Hb}{\ensuremath{\rm{H}\beta}\xspace}
\newcommand{\Ha}{\ensuremath{\rm{H}\alpha}\xspace}

\newcommand{\OIII}{[\ion{O}{iii}]\ensuremath{\,\lambda5007}\xspace}

\newcommand{\Lop}{\ensuremath{L_{5100}}}

\defcitealias{Zeltyn24}{Z24}
\newcommand{\Z}{\citetalias{Zeltyn24}}

\defcitealias{ZH19}{ZH19}
\newcommand{\ZH}{\citetalias{ZH19}}

\defcitealias{KH13}{KH13}
\newcommand{\KH}{\citetalias{KH13}}

\defcitealias{RV15}{RV15}
\newcommand{\RV}{\citetalias{RV15}}

\newcommand{\OIIIfull}{[\ion{O}{iii}]\ensuremath{\,\lambda\lambda4959,5007}} 
\newcommand{\OIIfull}
{[\ion{O}{ii}]\ensuremath{\,\lambda3728}} 
\newcommand{\NIIfull}{[\ion{N}{ii}]\ensuremath{\,\lambda\lambda6549,6584}} 
\newcommand{\SIIfull}{[\ion{S}{ii}]\ensuremath{\,\lambda\lambda6718,6732}}

\newcommand{\todo}{\ifmmode \text{\color{red}\Huge{\(\bullet\)}} \else {\color{red}{\Huge$\bullet$}}\fi}
\newcommand{\tido}{\ifmmode {{\color{red}\bullet}} \else {\color{red}$\bullet$}\fi}

\newcommand{\E        }[1]{\ifmmode 10^{#1} \else $10^{#1}$\fi}
\newcommand{\tE        }[1]{\ifmmode \times10^{#1} \else $\times10^{#1}$\fi}
\newcommand{\til}{\ifmmode \sim \else $\sim$\fi}
\renewcommand{\~} {\ifmmode \sim \else $\sim$\fi}

\newcommand{\pc}	{\ifmmode {\rm pc} \else pc\fi}
\newcommand{\kpc}	{\ifmmode {\rm kpc} \else kpc\fi}
\newcommand{\ld}	{\ifmmode {\rm l.d.} \else l.d.\fi}
\newcommand{\kms}	{\ifmmode {\rm km\,s}^{-1} \else km\,s$^{-1}$\fi}
\newcommand{\cc}	{\ifmmode {\rm cm}^{-3}    \else cm$^{-3}$\fi}
\newcommand{\cmii}	{\ifmmode {\rm cm}^{-2}    \else cm$^{-2}$\fi}
\newcommand{\ergs}	{\ifmmode {\rm erg\,s}^{-1} \else erg s$^{-1}$\fi}
\newcommand{\ergcms}	{\ifmmode {\rm erg\,cm}^{-2}\,{\rm s}^{-1} \else erg\,cm$^{-2}$\,s$^{-1}$\fi}
\newcommand{\ergcmsA}	{\ifmmode {\rm erg\,cm}^{-2}\,{\rm s}^{-1}\,{\rm\AA}^{-1}
\else erg\,cm$^{-2}$\,s$^{-1}$\,\AA$^{-1}$\fi}
\newcommand{  \ergcmsHz  }{\ifmmode{\rm erg\,cm}^{-2}\,{\rm s}^{-1}\,{\rm Hz}^{-1}
                       \else ergs\,cm$^{-2}$\,s$^{-1}$\,Hz$^{-1}$\fi}
\newcommand{\kev}	{\ifmmode {\rm keV} \else keV\fi}

\newcommand{\mic}	{\ifmmode {\rm \mu m} \else $\mu$m\fi}
\newcommand{\vFWHM}	{\ifmmode v_{\mbox{\tiny FWHM}} \else $v_{\mbox{\tiny FWHM}}$\fi}
\newcommand{\vBLR}	{\ifmmode v_{\mbox{\tiny BLR}} \else $v_{\mbox{\tiny BLR}}$\fi}
\newcommand{\sigBLR}	{\ifmmode \sigma_{\mbox{\tiny BLR}} \else $\sigma_{\mbox{\tiny BLR}}$\fi}
\newcommand{\vNLR}	{\ifmmode v_{\mbox{\tiny NLR}} \else $v_{\mbox{\tiny NLR}}$\fi}
\newcommand{\tauBLR}	{\ifmmode \tau_{\mbox{\tiny BLR}} \else $\tau_{\mbox{\tiny BLR}}$\fi}

\newcommand{\Hubble}	{\ifmmode {\rm km\,s}^{-1}\,{\rm Mpc}^{-1} \else km\,s$^{-1}$\,Mpc$^{-1}$\fi}
\newcommand{\NDunit}	{\ifmmode {\rm Mpc}^{-3} \else Mpc$^{-3}$\fi}
\newcommand{\LFunit}	{\ifmmode {\rm Mpc}^{-3}\,{\rm mag}^{-1} \else Mpc$^{-3}$\,mag$^{-1}$\fi}
\newcommand{\MFunit}	{\ifmmode {\rm Mpc}^{-3}\,{\rm dex}^{-1} \else Mpc$^{-3}$\,dex$^{-1}$\fi}

\newcommand{\Msun}{\ifmmode M_{\odot} \else $M_{\odot}$\fi}
\newcommand{\Lsun}{\ifmmode L_{\odot} \else $L_{\odot}$\fi}
\newcommand{\Zsun}{\ifmmode Z_{\odot} \else $Z_{\odot}$\fi}
\newcommand{\mpyr}{\ifmmode \Msun\,{\rm yr}^{-1} \else $\Msun\,{\rm yr}^{-1}$\fi}

\newcommand{\qnote}{\ifmmode q_{0} \else $q_{0}$\fi}
\newcommand{\Hnote}{\ifmmode H_{0} \else $H_{0}$\fi}
\newcommand{\hnote}{\ifmmode h_{0} \else $h_{0}$\fi}
\newcommand{\anote}{\ifmmode a_{0} \else $a_{0}$\fi}
\newcommand{\tnote}{\ifmmode t_{0} \else $t_{0}$\fi}

\def\gsim{\;\rlap{\lower 2.5pt \hbox{$\sim$}}\raise 1.5pt\hbox{$>$}\;}
\def\lsim{\;\rlap{\lower 2.5pt \hbox{$\sim$}}\raise 1.5pt\hbox{$<$}\;}

\newcommand{  \Halpha   }{\ifmmode {\rm H}\alpha \else H$\alpha$\fi}

\newcommand{  \ha       }{\Halpha}
\newcommand{  \Hbeta    }{\ifmmode {\rm H}\beta \else H$\beta$\fi}

\newcommand{  \hb       }{\Hbeta}
\newcommand{  \Hgamma   }{\ifmmode {\rm H}\gamma \else H$\gamma$\fi}
\newcommand{  \Hdelta   }{\ifmmode {\rm H}\delta \else H$\delta$\fi}
\newcommand{  \Lya      }{\ifmmode {\rm Ly}\alpha \else Ly$\alpha$\fi}
\newcommand{  \Lyb      }{\ifmmode {\rm Ly}\beta \else Ly$\beta$\fi}
\newcommand{  \Pa       }{\ifmmode {\rm P}\alpha \else P$\alpha$\fi}
\newcommand{  \Pb       }{\ifmmode {\rm P}\beta \else P$\beta$\fi}
\newcommand{  \Bra      }{\ifmmode {\rm Br}\alpha \else Br$\alpha$\fi}
\newcommand{  \Brg      }{\ifmmode {\rm Br}\gamma \else Br$\gamma$\fi}
\newcommand{  \hii      }{\ifmmode {\rm H}\,\textsc{ii} \else H\,\textsc{ii}\fi}
\newcommand{  \hei      }{\ifmmode {\rm He}\,\textsc{i} \else He\,\textsc{i}\fi}
\newcommand{  \heii     }{\ifmmode {\rm He}\,\textsc{ii} \else He\,\textsc{ii}\fi}
\newcommand{  \HeIIuv   }{\ifmmode {\rm He}\,\textsc{ii}\,\lambda1640 \else He\,\textsc{ii}\,$\lambda1640$\fi}
\newcommand{  \HeIIop   }{\ifmmode {\rm He}\,\textsc{ii}\,\lambda4686 \else He\,\textsc{ii}\,$\lambda4686$\fi}
\newcommand{  \CII	}{\ifmmode \left[{\rm C}\,\textsc{ii}\right]\,\lambda157.74\,\mu{\rm m} \else [C\,{\sc ii}]\ $\lambda157.74\,\mu{\rm m}$\fi}
\newcommand{  \cii	}{\ifmmode \left[{\rm C}\,\textsc{ii}\right] \else [C\,{\sc ii}]\fi}

\newcommand{  \ciii     }{\ifmmode \left.{\rm C}\,\textsc{iii}\right] \else C\,\textsc{iii}]\fi}
\newcommand{  \CIII     }{\ifmmode \left.{\rm C}\,\textsc{iii}\right]\,\lambda1909 \else C\,\textsc{iii}]\,$\lambda1909$\fi}
\newcommand{  \civ      }{\ifmmode {\rm C}\,\textsc{iv}  \else C\,\textsc{iv}\fi}
\newcommand{  \CIV      }{\ifmmode {\rm C}\,\textsc{iv}\,\lambda1549 \else C\,\textsc{iv}\,$\lambda1549$\fi}
\newcommand{  \NIIopt   }{\ifmmode \left[{\rm N}\,\textsc{ii}\right]\,\lambda6584 \else [N\,\textsc{ii}]\,$\lambda6584$\fi}
\newcommand{  \nii      }{\ifmmode \left[{\rm N}\,\textsc{ii}\right]  \else [N\,\textsc{ii}]\fi}
\newcommand{  \niii     }{\ifmmode {\rm N}\,\textsc{iii} \else N\,\textsc{iii}\fi}
\newcommand{  \NIII     }{\ifmmode {\rm N}\,\textsc{iii}\,\lambda4640 \else N\,\textsc{iii}\,$\lambda4640$\fi}
\newcommand{  \niv      }{\ifmmode {\rm N}\,\textsc{iv}  \else N\,\textsc{iv}\fi}
\newcommand{  \NIVuv    }{\ifmmode {\rm N}\,\textsc{iv}\,\lambda1486 \else N\,\textsc{iv}\,$\lambda1486$\fi}
\newcommand{  \nv       }{\ifmmode {\rm N}\,\textsc{v}   \else N\,\textsc{v}\fi}
\newcommand{\oi}{\ifmmode \left[{\rm O}\,\textsc{i}\right] \else [O\,{\sc i}]\fi}
\newcommand{\OI}{\ifmmode \left[{\rm O}\,\textsc{i}\right]\,\lambda6300 \else [O\,{\sc i}]$\,\lambda6300$\fi}
\newcommand{\oii}{\ifmmode \left[{\rm O}\,\textsc{ii}\right] \else [O\,{\sc ii}]\fi}
\newcommand{\OII}{\ifmmode \left[{\rm O}\,\textsc{ii}\right]\,\lambda3727 \else [O\,{\sc ii}]\,$\lambda3727$\fi}
\newcommand{\oiii}{\ifmmode \left[{\rm O}\,\textsc{iii}\right] \else [O\,{\sc iii}]\fi}
\newcommand{  \OIIIbf   }{\ifmmode {\rm O}\,\textsc{iii}\,\lambda3133 \else O\,\textsc{iii}\,$\lambda3133$\fi}
\newcommand{  \OIIIuv   }{\ifmmode {\rm O}\,\textsc{iii}\,\lambda1663 \else O\,\textsc{iii}\,$\lambda1663$\fi}
\newcommand{  \oiv      }{\ifmmode {\rm O}\,\textsc{iv}  \else O\,\textsc{iv}\fi}
\newcommand{  \OIVuv    }{\ifmmode {\rm O}\,\textsc{iv}\,\lambda1402  \else O\,\textsc{iv}\,$\lambda1402$\fi}
\newcommand{  \OIVIR    }{\ifmmode {\rm O}\,\textsc{iv}\,25.9\,\mu {\rm m} \else O\,\textsc{iv}\,$25.9\,\mu$m\fi}
\newcommand{  \ovi      }{\ifmmode {\rm O}\,\textsc{vi}   \else O\,\textsc{vi}\fi}
\newcommand{  \Ovi      }{\ifmmode {\rm O}\,\textsc{vi}\,\lambda1035 \else O\,\textsc{vi}\,$\lambda1035$\fi}
\newcommand{  \nei      }{\ifmmode {\rm Ne}\,\textsc{i}   \else Ne\,\textsc{i}\fi}
\newcommand{  \neii     }{\ifmmode {\rm Ne}\,\textsc{ii}  \else Ne\,\textsc{ii}\fi}
\newcommand{  \NeiiIR   }{\ifmmode {\rm Ne}\,\textsc{ii}\,12.8\,\mu {\rm m} \else Ne\,\textsc{ii}\,$12.8\,\mu$m\fi}
\newcommand{  \neiii    }{\ifmmode {\rm Ne}\,\textsc{iii} \else Ne\,\textsc{iii}\fi}
\newcommand{  \neiv     }{\ifmmode {\rm Ne}\,\textsc{iv}  \else Ne\,\textsc{iv}\fi}
\newcommand{  \nev      }{\ifmmode {\rm Ne}\,\textsc{v}   \else Ne\,\textsc{v}\fi}
\newcommand{  \NevIR    }{\ifmmode {\rm Ne}\,\textsc{v}\,24.3\,\mu {\rm m} \else Ne\,\textsc{v}\,$24.3\,\mu$m\fi}
\newcommand{  \nevi     }{\ifmmode {\rm Ne}\,\textsc{vi}  \else Ne\,\textsc{vi}\fi}
\newcommand{  \mgi      }{\ifmmode {\rm Mg}\,\textsc{i} \else Mg\,\textsc{i}\fi}
\newcommand{  \mgii     }{\ifmmode {\rm Mg}\,\textsc{ii} \else Mg\,\textsc{ii}\fi}
\newcommand{  \MgII     }{\ifmmode {\rm Mg}\,\textsc{ii}\,\lambda2798 \else Mg\,\textsc{ii}\,$\lambda2798$\fi}
\newcommand{\sii}{\ifmmode \left[{\rm S}\,\textsc{ii}\right] \else [S\,{\sc ii}]\fi}
\newcommand{  \siii     }{\ifmmode {\rm S}\,\textsc{iii} \else S\,\textsc{iii}\fi}
\newcommand{  \siv      }{\ifmmode {\rm S}\,\textsc{iv} \else S\,\textsc{iv}\fi}
\newcommand{  \sili     }{\ifmmode {\rm Si}\,\textsc{i}   \else Si\,\textsc{i}\fi}
\newcommand{  \silii    }{\ifmmode {\rm Si}\,\textsc{ii}  \else Si\,\textsc{ii}\fi}
\newcommand{  \Siliv    }{\ifmmode {\rm Si}\,\textsc{iv}  \else Si\,\textsc{iv}\fi}
\newcommand{  \SilIVuv  }{\ifmmode {\rm Si}\,\textsc{iv}\,\lambda1400  \else Si\,\textsc{iv}\,$\lambda1400$\fi}
\newcommand{  \AlIII   }{\ifmmode {\rm Al}\,\textsc{iii}\,\lambda1857 \else Al\,\textsc{iii}\,$\lambda1857$\fi}
\newcommand{  \Aliii   }{\ifmmode {\rm Al}\,\textsc{iii} \else Al\,\textsc{iii}\fi}
\newcommand{  \caii     }{\ifmmode {\rm Ca}\,\textsc{ii} \else Ca\,\textsc{ii}\fi}
\newcommand{  \feii     }{\ifmmode {\rm Fe}\,\textsc{ii} \else Fe\,\textsc{ii}\fi}
\newcommand{  \feiii    }{\ifmmode {\rm Fe}\,\textsc{iii} \else Fe\,\textsc{iii}\fi}
\newcommand{  \Kalpha   }{\ifmmode {\rm K}\alpha \else K$\alpha$\fi}

\newcommand{ \Lhb   }{\ifmmode L_{\hb} \else $L_{\hb}$\fi}
\newcommand{ \Lha   }{\ifmmode L_{\ha} \else $L_{\ha}$\fi}
\newcommand{ \fwhb  }{\ifmmode {\rm FWHM}\left(\hb\right) \else FWHM(\hb)\fi}
\newcommand{\sighb  }{\ifmmode \sigma\left(\hb\right) \else $\sigma\left(\hb\right)$\fi}
\newcommand{ \ewhb  }{\ifmmode {\rm EW}\left(\hb\right) \else EW(\hb)\fi}
\newcommand{ \fwha  }{\ifmmode {\rm FWHM}\left(\ha\right) \else FWHM(\ha)\fi}
\newcommand{ \ewha  }{\ifmmode {\rm EW}\left(\ha\right) \else EW(\ha)\fi}
\newcommand{ \Lmg   }{\ifmmode L\left(\mgii\right) \else $L\left(\mgii\right)$\fi}
\newcommand{ \fwmg  }{\ifmmode {\rm FWHM}\left(\mgii\right) \else FWHM(\mgii)\fi}
\newcommand{ \Lciv  }{\ifmmode L\left(\civ\right) \else $L\left(\civ\right)$\fi}
\newcommand{ \fwciv }{\ifmmode {\rm FWHM}\left(\civ\right) \else FWHM(\civ)\fi}
\newcommand{ \fwhm  }{\ifmmode {\rm FWHM} \else FWHM\fi} 
\newcommand{ \voff  }{\ifmmode v_{\rm off} \else $v_{\rm off}$\fi} 
\newcommand{ \vmax  }{\ifmmode v_{\rm max} \else $v_{\rm max}$\fi} 

\newcommand{ \mumg  }{\ifmmode \mu\left(\mgii\right) \else $\mu\left(\mgii\right)$\fi}
\newcommand{ \fmg   }{\ifmmode f\left(\mgii\right) \else $f\left(\mgii\right)$\fi}
\newcommand{ \muciv }{\ifmmode \mu\left(\civ\right) \else $\mu\left(\civ\right)$\fi}
\newcommand{ \fciv  }{\ifmmode f\left(\civ\right) \else $f\left(\civ\right)$\fi}

\newcommand{  \auvo     }{\ifmmode \alpha_{\nu,{\rm UVO}} \else $\alpha_{\nu,{\rm UVO}}$\fi}
\newcommand{  \Ledd     }{\ifmmode L_{\rm Edd} \else $L_{\rm Edd}$\fi}
\newcommand{  \lamLlam  }{\ifmmode \lambda L_{\lambda} \else $\lambda L_{\lambda}$\fi}
\newcommand{  \lLl      }{\ifmmode \lambda L_{\lambda} \else $\lambda L_{\lambda}$\fi}
\newcommand{  \nuLnu    }{\ifmmode \nu L_{\nu} \else $\nu L_{\nu}$\fi}
\newcommand{  \nLn      }{\ifmmode \nu L_{\nu} \else $\nu L_{\nu}$\fi}
\newcommand{  \Luv      }{\ifmmode L_{1350} \else $L_{1350}$\fi}
\newcommand{  \lLop     }{\ifmmode \log\left(\Lop/\ergs\right) \else $\log\left(\Lop/\ergs\right)$\fi}
\newcommand{  \Lthree   }{\ifmmode L_{3000} \else $L_{3000}$\fi}
\newcommand{  \lLthree  }{\ifmmode \log\left(\Lthree/\ergs\right) \else $\log\left(\Lthree/\ergs\right)$\fi}
\newcommand{  \Lsix      }{\ifmmode L_{6200} \else $L_{6200}$\fi}
\newcommand{  \lLisx     }{\ifmmode \log\left(\Lop/\ergs\right) \else $\log\left(\Lop/\ergs\right)$\fi}
\newcommand{  \Lxray    }{\ifmmode L_{\rm X} \else $L_{\rm X}$\fi}

\newcommand{  \Lhard    }{\ifmmode L_{\rm 2-10} \else $L_{\rm 2-10}$\fi}
\newcommand{  \Lsoft    }{\ifmmode L_{\rm 0.5-2} \else $L_{\rm 0.5-2}$\fi}

\newcommand{\Fthree}{\ifmmode F_{3000} \else $F_{3000}$\fi}
\newcommand{\fuv}{\ifmmode f_{\lambda}\left(1450{\rm \AA}\right) \else $f_{\lambda}\left(1450 {\rm \AA}\right)$\fi}
\newcommand{\fthree}{\ifmmode f_{\lambda}\left(3000{\rm \AA}\right) \else $f_{\lambda}\left(3000{\rm \AA}\right)$\fi}
\newcommand{\fH}{\ifmmode f_{\lambda}\left(1.65\micron\right) \else
$f_{\lambda}\left(1.65\micron\right)$\fi}

\newcommand{\fbol}{\ifmmode f_{\rm bol} \else $f_{\rm bol}$\fi}
\newcommand{\fbolwv}{\ifmmode f_{\rm bol}\left(\lambda\right) \else $f_{\rm bol}\left(\lambda\right)$\fi}
\newcommand{\fbolopt}{\ifmmode f_{\rm bol}\left(5100{\rm \AA}\right) \else $f_{\rm bol}\left(5100{\rm \AA}\right)$\fi}
\newcommand{\fbolthree}{\ifmmode f_{\rm bol}\left(3000{\rm \AA}\right) \else $f_{\rm bol}\left(3000{\rm \AA}\right)$\fi}
\newcommand{\fboluv}{\ifmmode f_{\rm bol}\left(1450{\rm \AA}\right) \else $f_{\rm bol}\left(1450{\rm \AA}\right)$\fi}

\newcommand{\fbolbat}{\ifmmode f_{\rm bol}\left(14-150\,\kev\right) \else $f_{\rm bol}\left(14-150\,\kev\right)$\fi}
\newcommand{\fbolhard}{\ifmmode f_{\rm bol}\left(2-10\,\kev\right) \else $f_{\rm bol}\left(2-10\,\kev\right)$\fi}

\newcommand{\fobs}{\ifmmode f_{\rm obs} \else $f_{\rm obs}$\fi}

\newcommand{  \mbh      }{\ifmmode M_{\rm BH} \else $M_{\rm BH}$\fi}
\newcommand{  \lmbh     }{\ifmmode \log\left(\mbh/\Msun\right) \else $\log\left(\mbh/\Msun\right)$\fi} 
\newcommand{  \lledd    }{\ifmmode L/L_{\rm Edd} \else $L/L_{\rm Edd}$\fi}
\newcommand{  \mmedd    }{\ifmmode \dot{m}/\dot{m}_{\rm \,Edd} \else $\dot{m}/\dot{m}_{\rm \,Edd}$\fi}
\newcommand{  \Lbol     }{\ifmmode L_{\rm bol} \else $L_{\rm bol}$\fi}
\newcommand{  \lbol     }{\ifmmode L_{\rm bol} \else $L_{\rm bol}$\fi}
\newcommand{  \lLbol    }{\ifmmode \log\left(\Lbol/\ergs\right) \else $\log\left(\Lbol/\ergs\right)$\fi} 
\newcommand{  \Lagn     }{\ifmmode L_{\rm AGN} \else $L_{\rm AGN}$\fi}
\newcommand{  \lagn     }{\ifmmode L_{\rm AGN} \else $L_{\rm AGN}$\fi}

\newcommand{  \tgrow     }{\ifmmode t_{\rm growth} \else $t_{\rm growth}$\fi}
\newcommand{  \tAD     }{\ifmmode t_{\rm acc} \else $t_{\rm acc}$\fi}
\newcommand{  \tacc    }{\ifmmode t_{\rm acc} \else $t_{\rm acc}$\fi}
\newcommand{  \tUni      }{\ifmmode t_{\rm Universe} \else $t_{\rm Universe}$\fi}

\newcommand{  \Mdotin	}{\ifmmode \dot{M}_{\rm infall} \else $\dot{M}_{\rm infall}$\fi}
\newcommand{  \Mdotbh	}{\ifmmode \dot{M}_{\rm BH} \else $\dot{M}_{\rm BH}$\fi}
\newcommand{  \Mdotad	}{\ifmmode \dot{M}_{\rm AD} \else $\dot{M}_{\rm AD}$\fi}
\newcommand{  \Mdotacc	}{\ifmmode \dot{M}_{\rm acc} \else $\dot{M}_{\rm acc}$\fi}
\newcommand{  \Mdotthin	}{\ifmmode \dot{M}_{\rm thin} \else $\dot{M}_{\rm thin}$\fi}
\newcommand{  \Mdotdisk	}{\ifmmode \dot{M}_{\rm disk} \else $\dot{M}_{\rm disk}$\fi}

\newcommand{  \Mindot	}{\ifmmode \dot{M}_{\rm infall} \else $\dot{M}_{\rm infall}$\fi}
\newcommand{  \Mbhdot	}{\ifmmode \dot{M}_{\rm BH} \else $\dot{M}_{\rm BH}$\fi}
\newcommand{  \Maddot	}{\ifmmode \dot{M}_{\rm AD} \else $\dot{M}_{\rm AD}$\fi}
\newcommand{  \Maccdot	}{\ifmmode \dot{M}_{\rm acc} \else $\dot{M}_{\rm acc}$\fi}
\newcommand{  \Mthdot	}{\ifmmode \dot{M}_{\rm thin} \else $\dot{M}_{\rm thin}$\fi}
\newcommand{  \Mdsdot	}{\ifmmode \dot{M}_{\rm disk} \else $\dot{M}_{\rm disk}$\fi}

\newcommand{  \as	}{\ifmmode a_{\rm *} \else $a_{\rm *}$\fi}
\newcommand{  \avec	}{\ifmmode \vec{a}_{\rm *} \else $\vec{a}_{\rm *}$\fi}
\newcommand{  \re	}{\ifmmode \eta      	 \else $\eta$\fi}
\newcommand{  \RISCO	}{\ifmmode R_{\rm ISCO}  \else $R_{\rm ISCO}$\fi}

\newcommand{  \mseed    }{\ifmmode M_{\rm seed} \else $M_{\rm seed}$\fi}
\newcommand{  \mbul     }{\ifmmode M_{\rm bulge} \else $M_{\rm bulge}$\fi} 
\newcommand{  \mstar    }{\ifmmode M_{*} \else $M_{*}$\fi} 
\newcommand{  \mgal     }{\ifmmode M_{*} \else $M_{*}$\fi} 
\newcommand{  \mhost    }{\ifmmode M_{\rm host} \else $M_{\rm host}$\fi}
\newcommand{  \mmsmall  }{\ifmmode M_{\rm BH}/M_{*} \else $M_{\rm BH}/M_{*}$\fi}
\newcommand{  \mmlarge  }{\ifmmode M_{*}/M_{\rm BH} \else $M_{*}/M_{\rm BH}$\fi}

\newcommand{  \mmdotlarge}{\ifmmode \dot{M}_*/\Mbhdot \else $\dot{M}_*/\Mbhdot$\fi}
\newcommand{  \mmdotsmall}{\ifmmode \Mbhdot/\dot{M}_* \else $\Mbhdot/\dot{M}_*$\fi}

\newcommand{  \mmwp     }{\ifmmode \left(M_{*}/M_{\rm BH}\right) \else $\left(M_{*}/M_{\rm BH}\right)$\fi}
\newcommand{  \ml       }{\ifmmode M_{*}/L_{*} \else $M_{*}/L_{*}$\fi}
\newcommand{  \mlwp     }{\ifmmode \left(M_{*}/L\right) \else $\left(M_{*}/L\right)$\fi}
\newcommand{  \mlk      }{\ifmmode \left(M_{*}/L_{K}\right) \else $\left(M_{*}/L_{K}\right)$\fi}
\newcommand{  \sigs     }{\ifmmode \sigma_{*} \else $\sigma_{*}$\fi}
\newcommand{\Msig}{\ensuremath{M_{\rm BH}\text{--}\sigma_{*}}\xspace}
\newcommand{\MM}{\ensuremath{M_{\rm BH}\text{--}M_{*}}\xspace}
\newcommand{  \Reff     }{\ifmmode R_{\rm e} \else $R_{\rm e}$\fi}
\newcommand{  \Rvir     }{\ifmmode R_{\rm vir} \else $R_{\rm vir}$\fi}
\newcommand{  \Rtwo     }{\ifmmode R_{200} \else $R_{200}$\fi}
\newcommand{  \Rfive    }{\ifmmode R_{500} \else $R_{500}$\fi}
\newcommand{  \Rgrp     }{\ifmmode R_{\rm grp} \else $R_{\rm grp}$\fi}
\newcommand{  \nser     }{\ifmmode n_{\rm s} \else $n_{\rm s}$\fi}
\newcommand{  \LSF      }{\ifmmode L_{\rm SF}  \else $L_{\rm SF}$\fi}
\newcommand{  \LFIR     }{\ifmmode L_{\rm FIR} \else $L_{\rm FIR}$\fi}
\newcommand{  \Lfir     }{\ifmmode L_{\rm FIR} \else $L_{\rm FIR}$\fi}
\newcommand{  \LTIR     }{\ifmmode L_{\rm TIR} \else $L_{\rm TIR}$\fi}
\newcommand{  \Ltir     }{\ifmmode L_{\rm TIR} \else $L_{\rm TIR}$\fi}

\newcommand{  \mdyn     }{\ifmmode M_{\rm dyn} \else $M_{\rm dyn}$\fi} 
\newcommand{  \mgas     }{\ifmmode M_{\rm gas} \else $M_{\rm gas}$\fi} 
\newcommand{  \mh       }{\ifmmode M_{\rm h} \else $M_{\rm h}$\fi}
\newcommand{  \mhalo    }{\ifmmode M_{\rm halo} \else $M_{\rm halo}$\fi}
\newcommand{  \sfr      }{\ifmmode {\rm SFR} \else SFR\fi}

\newcommand{ \Lcii     }{\ifmmode L_{\cii} \else $L_{\cii}$\fi}
\newcommand{ \fwcii  }{\ifmmode {\rm FWHM}\cii \else FWHM\cii\fi}

\newcommand  {\RBLR}        {\hbox{$ {R_{\rm BLR}} $}}

\newcommand{\bj}{\ifmmode b_{\rm J} \else $b_{\rm J}$\fi}

\newcommand{\iab}{\ifmmode i_{\rm AB} \else $i_{\rm AB}$\fi}

\newcommand{\jab}{\ifmmode J_{\rm AB} \else $J_{\rm AB}$\fi}
\newcommand{\hab}{\ifmmode H_{\rm AB} \else $H_{\rm AB}$\fi}
\newcommand{\kab}{\ifmmode K_{\rm AB} \else $K_{\rm AB}$\fi}

\newcommand{\jveg}{\ifmmode J_{\rm Vega} \else $J_{\rm Vega}$\fi}
\newcommand{\hveg}{\ifmmode H_{\rm Vega} \else $H_{\rm Vega}$\fi}
\newcommand{\kveg}{\ifmmode K_{\rm Vega} \else $K_{\rm Vega}$\fi}

\def\arcsec{\hbox{$^{\prime\prime}$}}

\newcommand{  \Chisq    }{\ifmmode \chi^{2} \else $\chi^{2}$}
\newcommand{  \nelec    }{\ifmmode n_{e} \else $n_{e}$\fi}     
\newcommand{  \nh       }{\ifmmode n_{\rm H} \else $n_{\rm H}$\fi}     
\newcommand{  \Ncol     }{\ifmmode N_{\rm col} \else $N_{\rm col}$\fi} 
\newcommand{  \NH       }{\ifmmode N_{\rm H} \else $N_{\rm H}$\fi}     

\def\sun{\hbox{$\odot$}}

\def\arcsec{\hbox{$^{\prime\prime}$}}
\DeclareRobustCommand{\ion}[2]{%
\relax\ifmmode
\ifx\testbx\f@series
{\mathbf{#1\,\mathsc{#2}}}\else
{\mathrm{#1\,\mathsc{#2}}}\fi
\else\textup{#1\,{\mdseries\textsc{#2}}}%
\fi} 

\usepackage[latin1]{inputenc}
\usepackage[T1]{fontenc}
\usepackage{soul}

\published{as ApJ, 1002, 61}

\shorttitle{Hosts of CL-AGNs}
\shortauthors{Zeltyn et al.}
\graphicspath{{./}{figures/}}

\begin{document}

\title{Changing-Look Active Galactic Nuclei in SDSS-V:\\
Host-Galaxy Properties and Black Hole Scaling Relations}


\author[0000-0002-7817-0099]{Grisha Zeltyn}
\affiliation{School of Physics and Astronomy, Tel Aviv University, Tel Aviv 69978, Israel}

\author[0000-0002-3683-7297]{Benny Trakhtenbrot}
\affiliation{School of Physics and Astronomy, Tel Aviv University, Tel Aviv 69978, Israel}
\affiliation{Max-Planck-Institut f{\"u}r extraterrestrische Physik, Gie\ss{}enbachstra\ss{}e 1, 85748 Garching, Germany}
\affiliation{Excellence Cluster ORIGINS, Boltzmannsstra\ss{}e 2, 85748, Garching, Germany}


\author[0000-0002-3719-940X]{Michael Eracleous}
\affiliation{Department of Astronomy and Astrophysics, 525 Davey Lab, The Pennsylvania State University, University Park, PA 16802, USA}
\affiliation{Institute for Gravitation and the Cosmos, The Pennsylvania State University, University Park, PA 16802, USA}

\author[0000-0002-6404-9562]{Scott F. Anderson}
\affiliation{Astronomy Department, University of Washington, Box 351580, Seattle, WA 98195, USA}

\author[0000-0001-5231-2645]{Claudio Ricci}
\affiliation{Department of Astronomy, University of Geneva, ch. d'Ecogia 16, 1290, Versoix, Switzerland}
\affiliation{Instituto de Estudios Astrof\'isicos, Facultad de Ingenier\'ia y Ciencias, Universidad Diego Portales, Av. Ej\'ercito Libertador 441, Santiago, Chile} 
\affiliation{Kavli Institute for Astronomy and Astrophysics, Peking University, Beijing 100871, China}

\author[0000-0002-0761-0130]{Andrea Merloni}
\affiliation{Max-Planck-Institut f{\"u}r extraterrestrische Physik, Gie\ss{}enbachstra\ss{}e 1, 85748 Garching, Germany}

\author[0000-0001-8557-2822]{Jessie Runnoe}
\affiliation{Department of Physics and Astronomy, Vanderbilt University, VU Station 1807, Nashville, TN 37235, USA}

\author{Mirko Krumpe}
\affiliation{Leibniz-Institut f\"ur Astrophysik Potsdam (AIP), An der Sternwarte 16, 14482 Potsdam, Germany}


\author[0000-0003-1908-8463]{James Aird}
\affiliation{Institute for Astronomy, University of Edinburgh, Royal Observatory, Edinburgh EH9 3HJ, UK}

\author[0000-0002-9508-3667]{Roberto J. Assef}
\affiliation{Instituto de Estudios Astrof\'isicos, Facultad de Ingenier\'ia y Ciencias, Universidad Diego Portales, Av. Ej\'ercito Libertador 441, Santiago, Chile}

\author[0000-0001-5609-2774]{Catarina Aydar}
\affiliation{Max-Planck-Institut f{\"u}r extraterrestrische Physik, Gie\ss{}enbachstra\ss{}e 1, 85748 Garching, Germany}
\affiliation{Excellence Cluster ORIGINS, Boltzmannsstra\ss{}e 2, 85748, Garching, Germany}

\author[0000-0002-8686-8737]{Franz E. Bauer}
\affiliation{Instituto de Alta Investigaci{\'{o}}n, Universidad de Tarapac{\'{a}}, Casilla 7D, Arica, 1010000, Chile}

\author[0000-0002-0167-2453]{W.N. Brandt}
\affiliation{Department of Astronomy and Astrophysics, 525 Davey Lab, The Pennsylvania State University, University Park, PA 16802, USA}
\affiliation{Institute for Gravitation and the Cosmos, The Pennsylvania State University, University Park, PA 16802, USA}
\affiliation{Department of Physics, 104 Davey Laboratory, The Pennsylvania State University, University Park, PA 16802, USA}

\author[0000-0002-8725-1069]{Joel R. Brownstein}
\affiliation{Department of Physics and Astronomy, University of Utah, 115 S. 1400 E., Salt Lake City, UT 84112, USA}

\author[0000-0003-0426-6634]{Johannes Buchner}
\affiliation{Max-Planck-Institut f{\"u}r extraterrestrische Physik, Gie\ss{}enbachstra\ss{}e 1, 85748 Garching, Germany}

\author[0000-0002-6252-3750]{Kaushik Chatterjee}
\affiliation{South-Western Institute for Astronomy Research (SWIFAR), Yunnan University, Kunming, Yunnan 650500, People's Republic of China}
\affiliation{Key Laboratory of Survey Science of Yunnan Province, Yunnan University, Kunming, Yunnan 650500, People's Republic of China}

\author[0000-0003-1752-679X]{Laura Duffy}
\affiliation{Institute for Gravitation and the Cosmos, The Pennsylvania State University, University Park, PA 16802, USA}

\author[0000-0002-8606-6961]{Lorena Hern\'andez-Garc\'ia}
\affiliation{Instituto de Estudios Astrof\'isicos, Facultad de Ingenier\'ia y Ciencias, Universidad Diego Portales, Av. Ej\'ercito Libertador 441, Santiago, Chile}
\affiliation{Centro Interdisciplinario de Data Science, Facultad de Ingenier\'ia y Ciencias, Universidad Diego Portales, Av. Ej\'ercito Libertador 441, Santiago, Chile}

\author{H\'ector Hern\'andez-Toledo}
\affiliation{Universidad Nacional Aut\'{o}noma de M\'{e}xico. Instituto de Astronom\'ia. A.P. 70-264, 04510. Ciudad de M\'{e}xico, M\'{e}xico}

\author[0000-0002-6610-2048]{Anton M. Koekemoer}
\affiliation{Space Telescope Science Institute, 3700 San Martin Drive, Baltimore, MD 21218, USA}

\author[0000-0002-6770-2627]{Sean Morrison}
\affiliation{Department of Astronomy, University of Illinois at Urbana-Champaign, Urbana, IL 61801, USA}

\author[0000-0002-1656-827X]{Castalia Alenka Negrete Pe\char"F1 aloza}
\affiliation{Universidad Nacional Aut\'{o}noma de M\'{e}xico. Instituto de Astronom\'ia. A.P. 70-264, 04510. Ciudad de M\'{e}xico, M\'{e}xico}

\author[0000-0001-7116-9303]{Mara Salvato}
\affiliation{Max-Planck-Institut f{\"u}r extraterrestrische Physik, Gie\ss{}enbachstra\ss{}e 1, 85748 Garching, Germany}

\author[0000-0001-7240-7449]{Donald P. Schneider}
\affiliation{Department of Astronomy and Astrophysics, 525 Davey Lab, The Pennsylvania State University, University Park, PA 16802, USA}
\affiliation{Institute for Gravitation and the Cosmos, The Pennsylvania State University, University Park, PA 16802, USA}

\author[0000-0003-1659-7035]{Yue Shen}
\affiliation{Department of Astronomy, University of Illinois at Urbana-Champaign, Urbana, IL 61801, USA}
\affiliation{National Center for Supercomputing Applications, University of Illinois at Urbana-Champaign, Urbana, IL 61801, USA}

\author[0000-0003-2656-6726]{Marzena \'Sniegowska}
\affiliation{Astronomical Institute, Czech Academy of Sciences, Bo\v{c}n\'{i} II 1401, Prague, 14100, Czech Republic}
\affiliation{School of Physics and Astronomy, Tel Aviv University, Tel Aviv 69978, Israel}

\correspondingauthor{Grisha Zeltyn}
\email{grishazeltyn@tauex.tau.ac.il, bennyt@tauex.tau.ac.il}

\vspace{1cm}
\begin{abstract}

Changing-look active galactic nuclei (CL-AGNs) exhibit dramatic spectral variability on unexpectedly short timescales, challenging standard accretion flow models. Despite growing samples, the physical drivers of this extreme variability, and the potential link to host-galaxy properties, remain unknown.
Regardless of the underlying mechanism, the transition between AGN-dominated and host-dominated spectra offers a unique opportunity to study relations between AGNs and their hosts within the same objects.
We present intermediate-resolution spectroscopy of 23 CL-AGNs identified by the Sloan Digital Sky Survey V (SDSS-V), obtained with the Very Large Telescope/X-shooter and Gemini-N/GMOS.
An analysis of the \MgII\ emission line observed in the spectra demonstrates that the majority of these sources cannot be driven by variable obscuration.
Our CL-AGNs roughly follow the \Msig\ and \MM\ relations of inactive galaxies, with a median black hole-to-stellar mass ratio of 0.38\,\%. 
We find no evidence that the stellar population properties of our CL-AGNs, including stellar mass, age, young stellar fraction, and star-formation rate, differ from those of type 2 AGNs in SDSS.
These results suggest that CL-AGNs reside in typical AGN host galaxies and that their extreme variability is likely unrelated to host-galaxy environment, supporting the idea that CL-AGNs are not a distinct population, but rather represent a phase of normal AGN activity. This result, in turn, implies that CL-AGNs can serve as useful probes of the AGN-host connection, providing access to both AGN-dominated and host-dominated spectra of the same systems.

\end{abstract}


\section{Introduction}
\label{sec:intro}


Optical/UV emission from active galactic nuclei (AGNs) is well known to exhibit stochastic variability, typically ranging from a few to tens of percent, over timescales spanning days to years \citep[e.g.,][]{Ulrich97,VB04,MacLeod12,Li18,Stone22}. However, recent advances in time-domain astronomy have revealed more dramatic variations in the behavior of AGNs. In particular, optical/UV changing-look AGNs (CL-AGNs) are defined as objects that exhibit the (dis)appearance of broad emission lines and the quasar-like continuum.\footnote{The term ``changing-look'' is also used in the X-ray literature to describe transitions between states with drastically different levels of line-of-sight column densities \citep[e.g.,][]{Matt03}, which are a distinct phenomenon from the optical/UV CL-AGNs discussed in this work.} These dramatic changes far exceed the variability amplitudes typically seen in AGNs, and can occur on timescales as short as a few months (see, e.g., \citealt{RT23} for a recent review). Because CL-AGNs often show transitions between AGN-dominated spectra, where the broad-line and continuum emission characteristic of type 1 AGNs are prominent, and host-dominated spectra, in which the galaxy emission dominates, they provide a valuable opportunity to study the AGN-host relationship within the same systems.

We note that there remains some ambiguity in the definition of CL-AGNs, as the apparent (dis)appearance of broad-line emission can depend on the quality of the available spectra as well as on the specific line used for classification \cite[e.g.,][]{Homan20, Yang20_CLAGN}. It has been suggested that it is more appropriate to regard CL-AGNs as a subset of the more general population of extremely variable AGNs (EVAGNs) or quasars (\citealt{Rumbaugh18}), whose large-amplitude variability may or may not be accompanied by changes in broad-line visibility \cite[see, e.g., the discussion in][]{Guo20_mgii}. However, in this work, we adopt the term CL-AGN for consistency with the literature.

Currently, $\sim1000$ CL-AGNs exhibiting variability in the (rest-frame) UV-optical have been reported in the literature. Many studies have performed detailed analyses of individual objects \cite[e.g.,][]{Denney14, LaMassa15, Husemann16,Ruan16, Runnoe16, Gezari17, Sheng17, Ross18, Stern18, Hutsemekers19, Hutsemekers20, Trakhtenbrot19, Wang19, Ricci20, Yu20, Guolo21, Nagoshi21,Zeltyn22, Yang23,Duffy25b, Duffy25}, while others have focused on more sizable samples and their statistical properties \cite[e.g.,][]{MacLeod16, Yang18, Potts21, Green22, LN22, LN23, Temple23, Guo24_desiI, Guo25, Ps24, Wang24, Zeltyn24, Yang25}. 
Many of these studies favor substantial changes in the accretion flow as the primary driver of the spectral changes, as supported, for example, by concurrent infrared response of the torus and the absence of obscuration signatures in the optical/UV and X-ray regimes.  Although some work has suggested that variable obscuration, perhaps from dusty gas clumps, might explain some spectral changes \cite[as is the case for some X-ray CL-AGNs; see, e.g.,][]{Risaliti05, Maiolino10, Markowitz14, Hernandez17, Liu22}, the short timescales involved pose a significant challenge for such models \citep[e.g.,][]{Potts21,Zeltyn22}.


One open question regarding CL-AGNs is whether their host-scale properties have any bearing on their occurrence. 
Studying CL-AGN hosts can provide important constraints on the mechanisms responsible for their extreme variability, particularly on whether large-scale galactic properties affect the small-scale accretion processes that drive the observed transitions.
In parallel, CL-AGNs present an opportunity to study the AGN-host relation more broadly: if their variability is linked to the host, it may reveal how galaxy-scale processes influence AGN behavior; if not, CL-AGNs may serve as unique probes of the general AGN population, allowing access to AGN-dominated and host-dominated spectra of the same systems.

Several studies have addressed this issue with mixed results. \citet{Charlton19} analyzed broadband optical imaging of four dimming CL-AGNs and reported that their stellar populations and morphologies are similar to those of typical AGN hosts. Similarly, \citet{Yu20} analyzed the host spectra of five CL-AGNs and found them to be consistent with other AGNs in terms of their location on the star-forming (SF) main sequence, while \citet{Jin22} reported that CL-AGNs have stellar populations similar to those of other AGNs. Furthermore, some studies have found that CL-AGNs obey the same scaling relation between BH mass (\mbh) and stellar velocity dispersion (\sigs) as other AGNs \cite[e.g.,][see below]{Yu20, Jin22, Yang25}. 
Recently, \citet{Verrico25} modeled the star formation histories of 39 CL-AGNs and found no evidence that their extreme variability is linked to the large-scale formation history of their host galaxies, suggesting that CL-AGN variability is more likely driven by processes on subnuclear scales.

Some studies, however, suggest that CL-AGNs may inhabit distinct host environments. \citet{Dodd21} reported that CL-AGNs preferentially reside in the ``green valley,'' and proposed that these systems are in the late stages of the transition between SF and quenched galaxies. In this scenario, CL-AGNs are driven by episodic accretion bursts in galaxies with relatively low cold gas content.
Similarly, \citet{Yang25} has found CL-AGNs to reside between typical AGNs and quenched galaxies on the \MM\ relation, suggesting that CL-AGNs might be a transitional phase between active and inactive galaxies. \citet{Liu21}, who differentiated between local CL Seyferts and CL quasars, found that the former are typically located in gas-poor galaxies, whereas the latter tend to reside in star-forming galaxies, implying that different mechanisms might be at play for these two populations. Possibly related to this behavior, \citet{Wang23} observed that within a sample of ``partially obscured'' AGNs (i.e., type 1.8/1.9), CL-AGNs tend to show a relative absence of young stellar populations. 

Such studies face several challenges. For example, they may rely on stellar properties from existing galaxy catalogs, which may be unreliable for sources with significant AGN emission \cite[see, e.g.,][for discussions of how AGN contamination can bias derived stellar properties]{Ciesla15,Buchner24}. In addition, the use of relatively low-resolution spectra can lead to large uncertainties and unreliable measurements of \sigs\ \cite[e.g.,][]{Toloba11, Scott18}. Moreover, unaccounted-for biases in CL-AGN samples, stemming from different detection methods and differences in the criteria for classifying an EVAGN as a CL-AGN---a challenge common to many CL-AGN studies---further complicate these investigations.

If, as some studies suggest, the host properties of CL-AGNs are similar to those of other AGNs, then CL-AGNs can bridge the observational gap between AGN-dominated spectra (i.e., luminous type 1 quasars) and host-dominated spectra (type 2 or low-luminosity AGNs; see \citealt{CS22,Negrete25} for a quantitative approach to distinguishing between AGN- and host-dominated spectra).
In the former, black hole (BH) properties are relatively well constrained, whereas in the latter, host galaxy properties are accessible.
Spectral transitions observed in CL-AGNs thus offer a rare opportunity to connect supermassive BH (SMBH) and host galaxy properties within the same systems, offering a direct observational approach that is independent of the model assumptions and degeneracies that may be affecting some other techniques (e.g.,  image- and/or spectral energy distribution-decomposition methods).
This investigation is particularly interesting given the tight correlations observed between \mbh\ and various host galaxy properties \cite[e.g.,][]{FM00,Gebhardt00,Tremaine02,Gultekin09,KH13,MM13}. In particular, the tight correlation between \mbh\ and \sigs\ may arise from feedback mechanisms during SMBH growth that influence the host galaxy evolution \cite[see, e.g., the review by][and references therein]{KP15}. Moreover, the slope of the \Msig\ relation can offer insights into the nature of this feedback \cite[e.g.,][]{SR98,King03}.

To explore the SMBH-host relation for AGNs, one has to estimate \mbh\ for AGNs.
For reverberation-mapped (RM) AGNs, \mbh\ can be estimated by multiplying the virial product by the virial scale factor ($f$):
\begin{equation}
\label{eq:mbh_rm}
    \mbh=f\frac{R_{\rm BLR}\Delta V^2}{G},
\end{equation}
where $\Delta V$ is the width of an emission line, a proxy for the broad-line region (BLR) velocity, and $R_{\rm BLR}$ is the BLR \textit{characteristic} radius \cite[inferred from time lags between variability in continuum and BLR emission; see, e.g.,][]{Peterson93,Kaspi00}.
The dimensionless factor $f$ accounts for BLR geometry and kinematics, and is calibrated under the assumption that RM AGNs, on average, follow the same \Msig\ relation as inactive galaxies, and is typically of order unity \cite[e.g.,][]{Onken04, Park12, Grier13, Woo13, Woo15, Batiste17, Grier17}.
With such \mbh\ estimates, one can investigate the relation of \mbh\ to other host properties, such as stellar mass \cite[\mstar; e.g.,][]{Li23}. 
Studies have also explored the slope of the virial product vs. \sigs\ relation for RM AGNs, finding RM AGNs to exhibit a shallower slope compared to quiescent galaxies \cite[e.g.,][]{Woo13, Woo15, Batiste17}. However, it remains unclear whether this difference reflects a physical distinction, or arises from factors such as unreliable \sigs\ measurements due to AGN contamination \cite[e.g.,][]{GH06}, selection biases \cite[e.g.,][]{Lauer07, Shankar16}, or extinction in the BLR \cite[e.g.,][]{Caglar20,Caglar23, MejiaRestrepo22}. Another complication arises from extreme AGN variability, as the virial product may change between accretion states, leading to different inferred BH masses depending on the epoch of observation \cite[e.g.,][]{Guo25}.
To claim that \mbh\ follows the same relation, 
one must further assume that $f$ does not vary systematically with \mbh\ and luminosity (Equation \ref{eq:mbh_rm}).

For non-RM AGNs, \mbh\ is estimated using locally calibrated single-epoch mass prescriptions, which use Equation \ref{eq:mbh_rm} with $R_{\rm BLR}$ inferred using empirical correlations between AGN luminosity and BLR size \cite[e.g.,][]{Kaspi00, Peterson04, Bentz13, Cho23}. 
These prescriptions, despite having large systematic uncertainties, allow the study of the SMBH-host scaling relations of non-RM AGNs and of larger AGN samples \cite[e.g.,][]{Xiao11, Bennert15, Bennert21, RV15, Caglar20,Caglar23, Winkel25}.


In this work, we present spectroscopy and analysis of the host galaxies of a subsample of dimming CL-AGNs, obtained with the Very Large Telescope (VLT)/X-shooter and Gemini-N/GMOS. The subsample studied here is based on CL-AGNs identified through repeat spectroscopy in the Sloan Digital Sky Survey V (SDSS-V; \citealt{Kollmeier26}) of legacy SDSS AGNs, and is mostly drawn from the larger sample assembled in \citet[][hereafter Z24]{Zeltyn24}, along with a few newly identified sources. We compare the CL-AGN hosts to SF galaxies and other AGNs in terms of their stellar populations, their location relative to the SF main sequence, and their position in the \Msig\ and \MM\ planes.

A companion SDSS-V study by C. Aydar et al.\ (2026, in preparation) investigates the reliability of simultaneously recovering galaxy and AGN properties (e.g., \sigs, \mstar, single-epoch \mbh) from SDSS-V spectra. Their analysis uses several AGN samples, including a larger CL-AGN sample that partially overlaps with the one presented here.

This paper is organized as follows. Section \ref{sec:data_obs} describes our sample selection and presents the spectroscopic observations used in this paper.
Section \ref{sec:analysis} describes the spectral decomposition tools employed to derive the properties of AGNs and their host galaxies.
Section \ref{sec:results_discussion} presents the host properties of our dimming CL-AGN sample and compares them with those of the general galaxy and AGN populations.
Finally, Section \ref{sec:conclusions} summarizes our key findings.
Throughout this work, we adopt a flat $\Lambda$ cold dark matter cosmology with $H_0=70\,\kms\,\rm{Mpc}^{-1}$ and $\Omega_{\rm{\Lambda}}=0.7$.

\section{Data and Observations}
\label{sec:data_obs}

We conducted medium-resolution spectroscopy of 23 dimming CL-AGNs using the X-shooter instrument \citep{Vernet11} mounted on the VLT and the GMOS instrument \citep{Hook04} at the Gemini North Telescope (Gemini-N).


\subsection{Dimming CL-AGN Sample Selection}

Our dimming CL-AGN subsample is based on CL-AGNs identified through SDSS-V \cite[][]{Kollmeier26} repeat spectroscopy of legacy SDSS AGNs, with 19 sources taken from the first-year SDSS-V CL-AGN sample presented in detail in \Z, along with four newly identified ones (see Table \ref{tab:obs}).
The SDSS-V spectra were obtained through the Black Hole Mapper program (S.F. Anderson et al. in preparation) using the Baryon Oscillation Spectroscopic Survey \cite[BOSS;][]{Smee13} spectrograph, mounted on the Sloan Foundation 2.5\,m telescope \citep{Gunn06} at Apache Point Observatory. The spectra cover the wavelength range 3570--10400\,\AA\ with a spectral resolution of $\sim1900$.

\begin{figure*}
    \centering
    \includegraphics[width=1.0\textwidth]{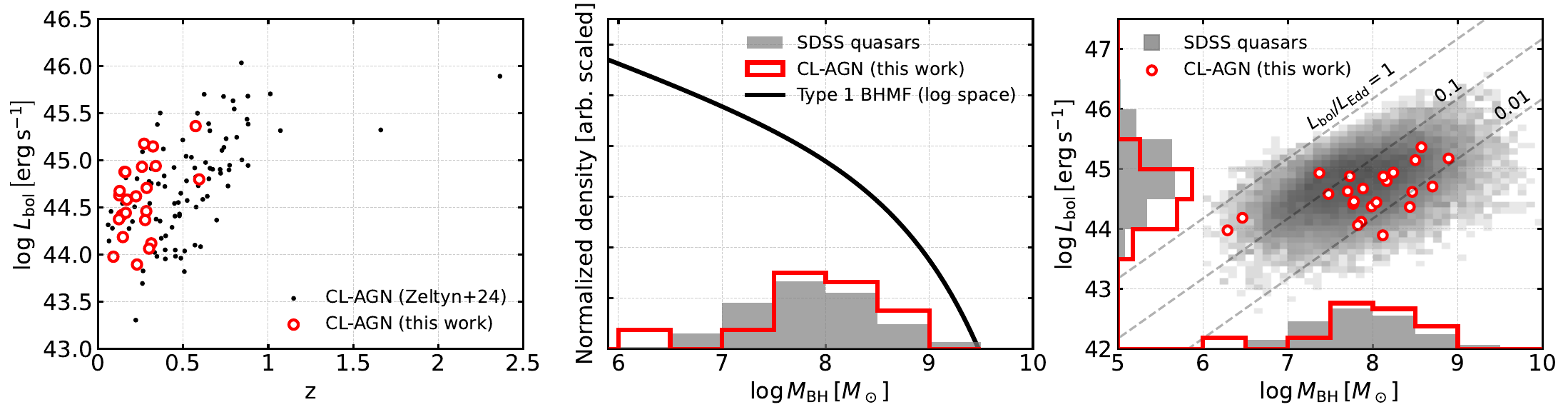}   
    \caption{{\it Left:} the distribution of the \Z\ sample of CL-AGNs on the luminosity--redshift plane, along with the subsample observed with VLT/X-shooter and Gemini-N/GMOS studied here (red circles). {\it Middle:}: the \mbh-normalized density distributions of SDSS DR16 $z<0.6$ quasars \citep[][gray region]{WS22} and of the CL-AGNs analyzed in this work (red). The shape of the logarithmic BH mass function (BHMF) of broad-line AGNs, taken from \citet[][black]{Ananna22}, is shown for reference with arbitrary scaling. {\it Right:} \Lbol\ vs. \mbh\ distributions for the same sets of AGNs. Dashed diagonal lines indicate Eddington ratios of $L_{\rm bol} /L_{\rm Edd}= 0.01, 0.1$, and $1$. All \mbh\ values used in this figure were derived from the legacy SDSS spectra using the prescription of \citet[][Equation \ref{eq:cho23}]{Cho23}, based on the \ha\ line. For the two highest-redshift sources from our CL-AGN sample, \mbh\ was estimated from alternative lines (see Section~\ref{subsub:bh_mass_method} for details). The \Lbol\ values, likewise derived from the legacy SDSS spectra, were adopted from \citet{WS22} and \Z.}
    \label{fig:sample_dr16}
\end{figure*}

For our dimming CL-AGN subsample selection, we required that objects exhibit broad-line emission in an earlier, AGN-dominated ``bright'' spectrum from legacy SDSS data \cite[DR16;][]{Ahumada20_DR16, Lyke20_DR16Q}, and later show a host-dominated ``dim'' spectrum as captured by SDSS-V (BOSS pipeline v6.2.1; \citealt{Bolton12, Dawson13}; S. Morrison et al. in preparation), in which the AGN emission has faded and the object was identified as a CL-AGN (see \Z\ for details of the identification and selection procedure). The dim-state spectrum had to be bright enough for spectroscopy with a sufficiently high signal-to-noise ratio ($S/N \gtrsim10$ per pixel) using long-slit spectroscopy with the VLT/X-shooter or Gemini-N/GMOS (resulting in $V$-band magnitudes in the range of 18.4--21.4).\footnote{This brightness requirement may introduce a mild selection bias toward intrinsically brighter or lower-redshift CL-AGNs, relative to the full SDSS-V CL-AGN sample.}
Taking into account both the selection constraints described above and practical limitations such as sky accessibility and allocated telescope time, we were ultimately able to observe 23 dimming CL-AGNs. Our CL-AGNs cover a redshift range of $z=\,$0.09--0.6, and show extreme flux variability ($\gtrsim1\,\rm{mag}$) in at least one broad emission line on observed timescales of 4--21\,yr. The left panel of Figure \ref{fig:sample_dr16} shows the distribution of the \Z\ sample of CL-AGNs on the luminosity--redshift plane, along with the subsample observed as part of the current work, where the bolometric luminosity (\Lbol) is that of the bright, legacy SDSS state.
Kolmogorov--Smirnov (KS) tests comparing the luminosity and \mbh\ distributions of our observed subsample with the full \Z\ CL-AGN sample find no statistically significant differences, indicating that our targets are representative of the parent population (see Figure~\ref{fig:sample_dr16} for details on the \Lbol\ and \mbh\ estimates).
The properties of our sources are tabulated in Table \ref{tab:obs}.

\begin{deluxetable*}{llccccccccl}
\tablecaption{VLT/X-shooter and Gemini-N/GMOS spectroscopy. \label{tab:obs}}
\tablewidth{\columnwidth}
\tablehead{
\colhead{Name} & \colhead{Instrument} & \colhead{$z$} & \colhead{Legacy} & \colhead{SDSS-V} & \colhead{X-shooter} & \colhead{Coverage} & \colhead{Exposure} & \colhead{$R$} & \colhead{New State} \\ [-2ex]
\colhead{} & \colhead{} & \colhead{} & \colhead{SDSS} & \colhead{} & \colhead{/GMOS} & \colhead{} & \colhead{} & \colhead{} & \colhead{} \\[-1ex]
\colhead{} & \colhead{} & \colhead{} & \colhead{(MJD)} & \colhead{(MJD)} & \colhead{(MJD)} & \colhead{(\AA)} &  \colhead{(s)}  & \colhead{} & \colhead{}  \\ [-1ex]
\colhead{(1)} & \colhead{(2)} & \colhead{(3)} & \colhead{(4)} & \colhead{(5)} & \colhead{(6)} & \colhead{(7)} & \colhead{(8)} &  \colhead{(9)}  & \colhead{(10)}
}
\startdata
J0044$-$0106 & X-Shooter & 0.2282 & 52531 &  59187 & 60569 & 2990--24790 & 5760 & 3200 & Consistent \\
J0124$+$0040 & X-Shooter & 0.594 & 55484 &  59203 & 60575 & 2990--24790 & 5760 & 5000 & Consistent \\
J0158$+$0013 & X-Shooter & 0.145 & 55449 &  59165 & 60287 & 2990--24790 & 8640 & 4710 & Brightened \\
J0159$+$0033 & X-Shooter & 0.312 & 51871 &  59165 & 60598 & 2990--24790 & 5760 & 4400 & Brightened \\
J0206$-$0414 & X-Shooter & 0.1387 & 57742 &  59226 & 60572 & 2990--24790 & 5760 & 3650 & Brightened \\
J0213$-$0253\tablenotemark{\footnotesize a} & GMOS & 0.168 & 57336 & 60228 & 60531 & 3780--5460 & 4170 & 3730 & - \\
J0245$+$0037 & X-Shooter & 0.2994 & 51871 &  59166 & 60296 & 2990--24790 & 5760 & 4140 & Dimmed \\
J0801$+$3417 & GMOS & 0.338 & 52584 & 59165 & 60597 & 4420--6110 & 1880 & 4270 & - \\
J0845$-$0027 & X-Shooter & 0.1544 & 51901 &  59222 & 60687 & 2990--24790 & 2880 & 3430 & Consistent \\
J0846$+$0000 & X-Shooter & 0.2575 & 51901 &  59227 & 60368 & 2990--24790 & 5760 & 3980 & Brightened \\
J0855$+$0329 & X-Shooter & 0.124 & 52225 &  59304 & 60758 & 2990--24790 & 2880 & 4480 & Consistent \\
J0903$+$0106 & X-Shooter & 0.1217 & 51924 &  59257 & 60712 & 2990--24790 & 5760 & 3200 & Brightened \\
J0904$-$0042 & X-Shooter & 0.2824 & 51929 &  59253 & 60385 & 2990--24790 & 5760 & 3270 & Consistent \\
J0916$+$0000 & X-Shooter & 0.2225 & 51955 &  59257 & 60740 & 2990--24790 & 8640 & 3780 & Consistent \\
J0927$+$0433 & X-Shooter & 0.3218 & 52254 &  59306 & 60355 & 2990--24790 & 5760 & 3860 & Consistent \\
J0927$+$0503 & X-Shooter & 0.1261 & 52707 &  59292 & 60687 & 2990--24790 & 2880 & 3450 & Consistent \\
J0932$+$0403 & X-Shooter & 0.2758 & 52264 &  59284 & 60724 & 2990--24790 & 8640 & 3350 & Consistent \\
J0933$+$0101 & X-Shooter & 0.1599 & 51965 &  59310 & 60687 & 2990--24790 & 2880 & 3260 & Consistent \\
J0953$+$0529\tablenotemark{\footnotesize a} & GMOS & 0.285 & 52725 & 60030 & 60674 & 4220--5910 & 2480 & 4100 & - \\
J1042$+$1212\tablenotemark{\footnotesize a} & GMOS & 0.270 & 53090 & 60056 & 60710 & 4170--5860 & 1600 & 4050 & - \\
J1104$+$0118\tablenotemark{\footnotesize a} & GMOS & 0.575 & 52374 & 60050 & 60648 & 5030--7430 & 2480 & 3610 & - \\
J2209$-$0038 & X-Shooter & 0.08895 & 55499 &  59386 & 60559 & 2990--24790 & 8640 & 3200 & Brightened \\
J2336$+$0040 & X-Shooter & 0.1615 & 55449 &  59164 & 60563 & 2990--24790 & 5760 & 3240 & Consistent \\
\enddata
\tablecomments{The columns indicate (1) the object identifier, (2) the instrument used for the spectrum, (3) the redshift, (4)--(6) the MJD of the legacy SDSS bright-state spectrum, the SDSS-V dim-state spectrum, and the X-shooter/GMOS spectrum (for objects observed on multiple nights, the mean MJD is listed), (7) the spectral coverage of the X-shooter/GMOS observation, (8) the total X-shooter/GMOS exposure time, (9) the effective spectral resolution near the Ca~\textsc{ii} H+K feature (rest-frame 3950\,\AA), and (10) the X-shooter state relative to the SDSS-V ``dim-state'' spectrum (not applicable to GMOS spectra due to limited wavelength coverage).}
\tablenotetext{a}{Newly identified CL-AGNs in SDSS-V, not presented in \Z.}
\end{deluxetable*}

\subsection{VLT/X-shooter Observations and Data Reduction}
\label{subsec:vlt}

Spectroscopy for 18 sources was obtained using the X-shooter instrument.\footnote{Program IDs: 112.25S8 and 114.27B8 (PI: Zeltyn).} X-shooter is a wide-wavelength, medium-resolution spectrograph consisting of three arms covering the ultraviolet-blue (UVB; 2990--5560\,\AA), visible (VIS; 5340--10,200\,\AA), and near-infrared (NIR; 9940--24,790\,\AA) wavelength ranges.
To obtain accurate measurements of the stellar velocity dispersions (\sigs), we aimed for sufficiently high spectral resolution ($\Delta\sigs\lesssim25\,\kms$). To that end, each spectrograph arm was configured as follows: the UVB arm used a 1.6\,\arcsec\ slit (expected instrumental spectral resolution $R\simeq3200$), the VIS arm used a 1.5\,\arcsec\ slit ($R\simeq5000$), and the NIR arm used a 0.9\,\arcsec\ slit ($R\simeq5600$). 
For the UVB and VIS arms, we used 2 pixel binning in the dispersion direction.

Targets were observed using observation blocks (OBs). For the UVB/VIS arms, each OB consisted of four 720\,s exposures, dithered along the slit with a 5\,\arcsec\ offset between positions. For the NIR arm, we used four 70\,s long exposures, and each exposure was further split into seven 100\,s long subexposures. To obtain the desired minimum $S/N\approx10$ per pixel for our spectra, brighter targets required a single OB, fainter targets required two OBs, and the faintest objects required three OBs. In some cases, OBs of the same target were executed during separate nights, with no significant difference between the observed spectra.

All X-shooter data were reduced and calibrated using the standard pipelines in \texttt{EsoReflex} v2.11.5 \cite[][]{EsoReflex}. 
Telluric absorption corrections were applied using \texttt{Molecfit} v1.5.9 \cite[][]{Molecfit1,Molecfit2}, except for one source (J0845$-$0027), for which the correction could not be applied successfully.
Spectrophotometric standards, which were also used for absolute flux calibration, were obtained as part of the X-shooter calibration plan, which observes a standard star for each arm every three days. 
Each OB spectrum was visually inspected alongside the most recent SDSS-V spectrum to verify the quality of the reduction, based on features such as the host galaxy continuum shape and the depth of stellar absorption features.
For all targets with multiple OBs, the spectra from individual OBs were coadded (see details in Section \ref{subsubsec:measuring_sigma}). 
For all analyses and modeling, we masked the edges of the spectra that are dominated by noise, specifically, the reddest region of the UVB arm and the bluest part of the VIS arm. In practice, this masking resulted in the exclusion of the 5555--5605\,\AA\ region in the observed frame. The UVB and VIS spectra were stitched together without scaling, owing to their similar slit sizes.

Reliable measurement of \sigs\ (Section \ref{subsubsec:measuring_sigma}) requires estimating the effective spectral resolution for each spectrum, which depends on whether the source is slit-filling. For the X-shooter data, the relatively wide 1.5\,\arcsec/1.6\,\arcsec\ slits may result in some observations not being fully slit-filling, leading to a higher effective resolution than the nominal value.
To determine the effective seeing and the corresponding $R$ of our observations, we use the 2D spectra, as described in Appendix~\ref{app:resolution}.
The resulting resolution at 3950\,\AA\ (near the Ca~\textsc{ii} H+K absorption feature) is listed for each object in Table \ref{tab:obs}, and ranges from 3200 to 5000 for the X-shooter data.


\subsection{Gemini-N/GMOS Observations and Data Reduction}
\label{subsec:gemini}

Spectroscopy was obtained for five sources using the normal long-slit mode of the GMOS instrument.\footnote{Program ID: 2024B-309515 (PI: Eracleous).} Observations were conducted in single slit mode with a 0.5\,\arcsec\ wide slit.
Our spectra were centered on the Ca~\textsc{ii} H+K ($\lambda=3934$, 3968\,\AA\ in the rest-frame) absorption feature, which is commonly used for measuring $\sigma_\ast$ (Section \ref{subsubsec:measuring_sigma}).
Four of the targets were observed using the B1200 grating ($R\simeq3740$ at the blaze wavelength), which provides continuous spectral coverage of $\simeq 1700$\,\AA.
The highest-redshift target, J1104+0118 at $z=0.575$, was observed instead with the R831 grating ($R\simeq4400$), with a wavelength coverage of $\simeq2400$\,\AA.
For targets observed using the B1200 grating, we used 2-pixel binning in both the spatial and dispersion directions, resulting in a pixel scale of 0.46\,\AA$\,\rm{pix}^{-1}$. For the target observed using the B831 grating, 2-pixel binning was applied only in the spatial direction, producing a pixel scale of 0.34\,\AA$\,\rm{pix}^{-1}$.
Taken together, the GMOS spectra cover observed wavelengths from 3780 to 7430\,\AA\ across the full sample.

The total exposure times for each target, ranging from 1600 to 4170\,s, were determined to achieve an expected $S/N\gtrsim10$ per pixel at the Ca~\textsc{ii} H+K region, based on the online GMOS-N exposure-time calculator. 
All observations were divided into four subexposures, dithered along the slit, except for the faintest target (J0213-0253), for which six subexposures were used.

All GMOS spectra were reduced using the automated pipelines in the \texttt{DRAGONS} platform v3.2.2 \cite[][]{DRAGONS}.
Spectrophotometric standard stars, chosen from the CALSPEC database \cite[see][and references therein]{Bohlin20} and the Gaia SPSS V2 catalog \citep{Pancino21},\footnote{\url{https://gaiaextra.ssdc.asi.it}.} were observed immediately before or after each science target.

Given the limited wavelength coverage of the GMOS observations (see Table \ref{tab:obs}), these data were used only for \sigs\ measurements (Section \ref{subsubsec:measuring_sigma}).
Since all GMOS observations were slit-filling (owing to the narrow  0.5\,\arcsec\ slit), we adopt the nominal instrumental resolution for these sources.
The resulting resolution at 3950\,\AA\ (near the Ca~\textsc{ii} H+K absorption feature) ranges from 3610 to 4270, and is listed in Table \ref{tab:obs}.


\subsection{New Spectra and AGN States}


Table \ref{tab:obs} summarizes our observations and presents the most recent states of the sources relative to their SDSS-V state. These classifications are qualitative, based on visual inspection of the broad emission lines and the quasar-like blue continuum. Specifically, 11 of our X-shooter sources remain in a state consistent with that observed in SDSS-V, one source exhibits further dimming, while six sources show brightening to an intermediate state (i.e., still dimmer than the legacy SDSS bright state). For all five GMOS sources, we are unable to determine the most recent state due to the limited wavelength coverage. 

We note that our classifications may be biased toward detecting brightening events, since the SDSS-V state corresponds to a relatively dim phase, making additional dimming more difficult to identify. As a result, the relative occurrence of dimming versus brightening events in our sample should not be interpreted as representative of the underlying variability behavior, and we do not draw any statistical conclusions from the relative occurrence of these states. Accordingly, this qualitative classification does not enter into our quantitative analysis.

All the legacy SDSS, SDSS-V, and newly obtained spectra of our subsample are displayed in Figure \ref{fig:all_spectra} in Appendix \ref{app:spectra_plots}.


\subsection{Aperture Effects}
\label{subsec:aperture_effects}

In this work, some of the quantities derived from the SDSS-V spectra depend on the total light of the host galaxy (i.e., stellar mass and star formation rate; see Sections \ref{subsec:ssp} and \ref{subsubsec:sfr}).
Since the dim-state SDSS-V spectra were obtained with an SDSS 2\,\arcsec\ fiber, and given the typical sizes of galaxies and the relatively low redshifts of most of the sources, we must account for the light missing outside the fiber.
However, directly measuring the fraction of light falling outside the fiber is challenging because dim-state imaging is generally not available.
We therefore estimate this aperture loss indirectly, by constructing for each source a sample of SDSS galaxies matched in redshift and synthetic $r$-band magnitude that were likewise observed with 2\,\arcsec\ fibers, and take the mean offset between their \texttt{cmodel} $r$-band magnitude (tracing the total galaxy light) and the synthetic $r$-band magnitude as our estimate of the light loss, $\Delta_r$. The full procedure is detailed in Appendix \ref{app:aperture_corrections}.
Across our sample, $\Delta_r$ ranges from 0.46 to 1.9\,mag, with a median of 1.2\,mag.
Throughout this work, unless stated otherwise, we use the uncorrected (fiber-based) values in our figures and mark the mean aperture correction with arrows for reference.


\section{Analysis}
\label{sec:analysis}

This section describes the methods used to derive various properties of the CL-AGNs in our sample and their host galaxies.
All spectra were shifted to the rest frame using the SDSS pipeline redshift and corrected for Galactic extinction using the \citet[]{SFD98} dust maps and a Milky Way (MW) extinction law \cite[][$R_V=3.1$]{ODonnell94}.
In what follows, AGN properties are derived from the legacy SDSS ``bright-state'' spectra, while host galaxy properties are measured from the ``dim-state'' spectra obtained with either VLT/X-shooter, Gemini/GMOS, or SDSS-V.
All derived quantities are listed in Table \ref{tab:quantities}.
 
\subsection{AGN Spectral Measurements}
\label{subsec:pyqsofit}

To extract the AGN continuum and line emission properties, we used the spectral fitting code \texttt{PyQSOFit} \citep{QSOFit}. 
For every VLT/X-shooter target we have fitted the legacy-SDSS, SDSS-V, and X-shooter spectra. For the Gemini-N/GMOS targets, we fitted only the legacy SDSS and SDSS-V spectra because of GMOS's limited wavelength coverage.
For each spectrum, when the wavelength range permitted, we fitted the broad \ha, \hb, and \MgII\ emission lines and the narrow \SIIfull, \ha, \NIIfull, \hb, \OIIIfull, and \OIIfull\ emission lines.
Broad Balmer lines were modeled using two Gaussians each, while the broad \mgii\ line was modeled using one Gaussian, except for two objects, where some or all of the spectra required two Gaussians in order to properly fit the line shape.
Narrow emission lines were modeled using one Gaussian each, except for three spectra where the \oiii\ doublet needed additional broad wing components.
Within each spectral complex (\ha\ or \hb) the widths and velocity shifts of all narrow lines were tied, and the intensity ratios for the \oiii\ and \nii\ doublets were fixed at 1:3.
For seven objects, visual inspection of some or all of their spectra revealed that tying the narrow line widths and offsets significantly degraded the fit quality. In these cases, we relaxed these constraints, allowing these parameters to vary independently.
Because we used \texttt{PyQSOFit} primarily for measuring emission-line widths and fluxes, we did not include any host galaxy component in the spectral fitting, but instead fitted each line complex independently, modeling the underlying continuum with a power law. 

Uncertainties in the continuum and line measurements were obtained via a Monte Carlo (MC) refitting approach that relies on the error spectra, using 200 realizations for each spectrum.
\\

\subsection{Measuring BH mass}
\label{subsub:bh_mass_method}

Throughout this work, we rely on the bright-state (legacy SDSS) spectra for studying \mbh, following standard approaches used for objects with AGN-dominated spectra. 
We estimate \mbh\ using a single-epoch prescription based on the luminosity and width of the broad \Ha\ emission line. We adopt an \Ha-based prescription because, unlike the commonly used \Hb-based estimators, it does not rely directly on continuum luminosity. This approach makes it less sensitive to host galaxy contamination, a key advantage for our sample, where the host contribution can be substantial even during the bright state.

Among the available \Ha-based prescriptions, we adopt the relation from \citet{Cho23}, which provides an updated calibration of this estimator.\footnote{We verified that the \citet{Cho23} masses for our sample are in good agreement with other \Ha-based prescriptions \citep[][]{GH05, Bonta25}.} Unlike other \Ha-based prescriptions for \mbh, the one by \cite{Cho23} is based on a direct determination of the \Ha\ size-luminosity relation using \Ha\ RM. The adopted formula is
\begin{align}
\label{eq:cho23}
    \log\left(\frac{\mbh}{10^6\,\Msun}\right) &= \log f+0.456+2\log\left(\frac{\rm FWHM[\Ha]}{ 1000\,\kms}\right)\nonumber\\
    &+0.61\log\left(\frac{L[\Ha]}{10^{42}\,\ergs}\right)\,,
\end{align}
where $f$ is the dimensionless virial scaling factor that encapsulates the kinematics and geometry of the BLR \cite[e.g.,][and references therein]{Shen13,MejiaRestrepo18}.

In this work, we explore two complementary approaches: (1) adopt a fixed value of $f=1.12$, as suggested by \citet{Cho23} drawing  on the earlier work by \citet{Woo15}; or (2)
infer the value of $f$ that is favored by our own measurements, under the assumption that CL-AGNs follow the same \Msig\ relation as inactive galaxies.

For the two highest-redshift sources where the \ha\ line is not covered by the bright-state spectroscopy, we adopted alternative prescriptions. For J0124+0040, \mbh\ was estimated using the \hb\ line width and the continuum luminosity at rest-frame 5100\,\AA. For J1104+0118, we used the \mgii\ line and the 3000\,\AA\ continuum instead, due to an atypical \hb\ profile that \texttt{PyQSOFit} could not reliably decompose. In both cases, we used the corresponding prescriptions from \citet{MejiaRestrepo22}.

Systematic uncertainties in single-epoch mass estimates are typically considered to reach $\sim 0.5$\,dex \citep{Shen13}, with $\sim 0.3$\,dex attributed to uncertainty in the virial scaling factor $f$, and the remaining $\sim 0.4$\,dex to uncertainties in the virial product itself, which are in turn dominated by the uncertainties in the $\RBLR-L$ relation(s). The errors in \mbh\ due to spectral measurement uncertainties, obtained through MC resampling (see Section \ref{subsec:pyqsofit}), were much smaller ($<0.05$\,dex for all objects) and were considered negligible compared to the systematics, and thus ignored in our analysis.
Additional sources of systematic uncertainty may affect our CL-AGN sample, both because the \Ha-based single-epoch calibration was derived for higher-luminosity objects \cite[][]{Cho23} and because CL-AGNs may introduce further unknown systematics associated with their unusual variability.
The \mbh\ estimates for our sample range $\log\mbh/\Msun=6.3$--8.9, as illustrated in the middle and right panels of Figure \ref{fig:sample_dr16}, and are listed in Table \ref{tab:quantities}.

\setlength{\tabcolsep}{4pt}
\begin{deluxetable*}{lcccccccccccc}
\tablecaption{Derived quantities. \label{tab:quantities}}
\tablewidth{\columnwidth}
\tablehead{
\colhead{Name} &  \colhead{\sigs} & \colhead{\hspace{-8pt}$\log\,(\mbh/M_\odot)$} & \colhead{\hspace{-5pt}$\Delta_r$} & \colhead{\hspace{-4pt}SFR} & \colhead{\hspace{-4pt}$\Delta_{\rm{SFR}}$} & \colhead{\hspace{-5pt}$\log\,(\mstar/\Msun)$} & \colhead{$\Delta_{\mstar}$} & \colhead{$t_{\rm \ast, M}$} & \colhead{$f_{\rm 1Gyr}$} & \colhead{\hspace{-4pt}$E(B-V)$} &
\colhead{$\log\,f_{\rm FWHM}$} & \colhead{$\log\,f_{\rm \sigma}$} \\ [-2ex]
\colhead{} &  \colhead{(\kms)} & \colhead{} & \colhead{\hspace{-5pt}(mag)} & \colhead{\hspace{-4pt}($M_\odot\,\rm{yr}^{-1}$)} & \colhead{\hspace{-4pt}($M_\odot\,\rm{yr}^{-1}$)} & \colhead{} & \colhead{(dex)} & \colhead{(Gyr)} & \colhead{} & \colhead{\hspace{-4pt}(mag)} &
\colhead{} & \colhead{} \\ [-2ex]
\colhead{(1)} &  \colhead{(2)} & \colhead{\hspace{-8pt}(3)} & \colhead{\hspace{-5pt}(4)} & \colhead{\hspace{-4pt}(5)} & \colhead{\hspace{-4pt}(6)} & \colhead{\hspace{-5pt}(7)} & \colhead{(8)} & \colhead{(9)} & \colhead{(10)} & \colhead{\hspace{-4pt}(11)} & \colhead{(12)} &
\colhead{(13)}
}
\startdata
    \hline
J0044$-$0106 & $166\pm20$ & $8.12$ & $1.1$ & $1.4$ & $2.8$ & $10.4^{+0.2}_{-0.1}$ & $0.45$ & $5.9^{+2.5}_{-0.7}$ & $0.0002$ & $0.23$ &  0.07$\pm$0.23 &  0.49$\pm$0.23 \\ 
J0124$+$0040 & $40\pm34$\tablenotemark{\footnotesize a} & $8.16$ & $0.52$ & $6.9$ & $4.9$ & $10.4^{+0.2}_{-0.0}$ & $0.21$ & $6.6^{+0.3}_{-2.8}$ & $0.02$ & $0.32$ & -2.80$\pm$1.61 & -2.04$\pm$1.61 \\ 
J0158$+$0013 & $118\pm11$ & $6.46$ & $1.1$ & $0.08$ & $0.21$ & $10.0^{+0.2}_{-0.4}$ & $0.46$ & $9.9^{+1.1}_{-2.8}$ & $0.02$ & $0.23$ &  1.07$\pm$0.18 &  1.25$\pm$0.18 \\ 
J0159$+$0033 & $103\pm15$ & $7.86$ & $0.85$ & $3.1$ & $3.6$ & $10.6^{+0.0}_{-0.2}$ & $0.34$ & $8.0^{+0.0}_{-2.0}$ & $0.001$ & $0.17$ & -0.59$\pm$0.28 & -0.05$\pm$0.28 \\ 
J0206$-$0414 & $136\pm8$ & $7.77$ & $1.5$ & $0.042$ & $0.15$ & $10.2^{+0.2}_{-0.1}$ & $0.6$ & $6.5^{+3.0}_{-0.6}$ & $0.0002$ & $0.098$ &  0.03$\pm$0.12 &  0.58$\pm$0.12 \\ 
J0213$-$0253 & $101\pm14$ & $7.48$ & $1.3$ & $0.37$ & $1.1$ & $10.2^{+0.1}_{-0.8}$ & $0.53$ & $8.3^{+0.6}_{-5.7}$ & $0.1$ & $0.33$ & -0.24$\pm$0.27 &  0.13$\pm$0.27 \\ 
J0245$+$0037 & $131\pm8$ & $7.82$ & $0.91$ & $1.8$ & $3$ & $10.2^{+0.1}_{-0.8}$ & $0.53$ &  &  & $0.34$ & -0.09$\pm$0.12 &  0.48$\pm$0.12 \\ 
J0801$+$3417 & $158\pm29$ & $8.24$ & $0.78$ & $0.98$ & $1.1$ & $10.6^{+0.2}_{-0.2}$ & $0.31$ & $4.9^{+1.6}_{-2.1}$ & $0.04$ & $0.077$ & -0.15$\pm$0.35 &  0.30$\pm$0.35 \\ 
J0845$-$0027 & $214\pm13$ & $7.73$ & $1.2$ & $0.51$ & $1.4$ & $10.1^{+0.3}_{-0.0}$ & $0.49$ & $2.8^{+4.7}_{-0.0}$ & $0.3$ & $0.53$ &  0.94$\pm$0.11 &  1.27$\pm$0.11 \\ 
J0846$+$0000 & $117\pm13$ & $7.37$ & $1.1$ & $1.4$ & $2.7$ & $10.4^{+0.1}_{-0.3}$ & $0.43$ & $6.9^{+3.1}_{-2.1}$ & $0.02$ & $0.18$ &  0.14$\pm$0.21 &  0.33$\pm$0.21 \\ 
J0855$+$0329 & $135\pm6$ & $7.70$ & $1.4$ & $1$ & $3.2$ & $10.3^{+0.3}_{-0.1}$ & $0.56$ & $9.5^{+1.0}_{-1.3}$ & $0.008$ & $0.15$ &  0.09$\pm$0.09 &  0.55$\pm$0.09 \\ 
J0903$+$0106 & $161\pm5$ & $7.98$ & $1.5$ & $0.18$ & $0.56$ & $10.5^{+0.2}_{-0.1}$ & $0.61$ & $9.4^{+1.5}_{-1.5}$ & $0$ & $0$ &  0.14$\pm$0.06 &  0.98$\pm$0.06 \\ 
J0904$-$0042 & $156\pm11$ & $7.78$ & $0.92$ & $1$ & $1.3$ & $10.7^{+0.1}_{-0.2}$ & $0.37$ & $8.2^{+0.7}_{-2.0}$ & $0.02$ & $0.23$ &  0.29$\pm$0.14 &  1.00$\pm$0.14 \\ 
J0916$+$0000 & $136\pm17$ & $8.46$ & $1.3$ & $1.4$ & $4$ & $10.2^{+0.2}_{-0.1}$ & $0.51$ & $7.7^{+2.3}_{-4.7}$ & $0$ & $0.014$ & -0.66$\pm$0.23 & -0.33$\pm$0.23 \\ 
J0927$+$0433 & $218\pm11$ & $8.50$ & $0.78$ &  &  & $10.9^{+0.1}_{-0.1}$ & $0.31$ & $8.1^{+0.4}_{-1.0}$ & $0.003$ & $0.11$ &  0.20$\pm$0.10 &  0.74$\pm$0.10 \\ 
J0927$+$0503 & $138\pm26$ & $7.88$ & $1.3$ & $2.9$ & $8.1$ & $10.4^{+0.1}_{-0.2}$ & $0.54$ & $7.0^{+1.2}_{-1.8}$ & $0.0002$ & $0.14$ & -0.05$\pm$0.36 &  0.40$\pm$0.36 \\ 
J0932$+$0403 & $218\pm11$ & $8.43$ & $1$ & $0.78$ & $0.93$ & $10.9^{+0.0}_{-0.2}$ & $0.41$ & $9.8^{+0.1}_{-1.6}$ & $0.03$ & $0.088$ &  0.27$\pm$0.10 &  0.98$\pm$0.10 \\ 
J0933$+$0101 & $180\pm18$ & $8.12$ & $1.2$ & $0.14$ & $0.31$ & $10.5^{+0.0}_{-0.1}$ & $0.48$ & $3.8^{+0.4}_{-1.8}$ & $0$ & $0$ &  0.21$\pm$0.19 &  0.74$\pm$0.19 \\ 
J0953$+$0529 & $138\pm32$ & $8.70$ & $0.87$ & $5.5$ & $6$ & $10.7^{+0.1}_{-0.9}$ & $0.35$ & $9.3^{+0.2}_{-4.3}$ & $0.09$ & $0.42$ & -0.86$\pm$0.44 & -0.04$\pm$0.44 \\ 
J1042$+$1212 & $153\pm33$ & $8.89$ & $0.88$ & $0.28$ & $0.36$ & $10.5^{+0.3}_{-0.1}$ & $0.35$ & $3.5^{+4.6}_{-0.7}$ & $0.008$ & $0.24$ & -0.86$\pm$0.41 & -0.03$\pm$0.41 \\ 
J1104$+$0118 & $260\pm33$ & $8.57$ & $0.46$ & $1.5$ & $0.82$ & $10.6^{+0.2}_{-0.4}$ & $0.18$ & $1.7^{+1.6}_{-1.0}$ & $0.6$ & $0$ &  0.42$\pm$0.24 &  0.27$\pm$0.24 \\ 
J2209$-$0038 & $49\pm8$ & $6.29$ & $1.9$ & $0.13$ & $1.1$ & $9.4^{+0.3}_{-0.1}$ & $0.77$ & $3.7^{+4.1}_{-0.1}$ & $0.1$ & $0.049$ & -0.43$\pm$0.31 & -0.19$\pm$0.31 \\ 
J2336$+$0040 & $188\pm9$ & $8.04$ & $1.4$ & $4.2$ & $15$ & $10.1^{+0.2}_{-0.1}$ & $0.55$ & $5.2^{+1.0}_{-1.4}$ & $0.1$ & $0.24$ &  0.39$\pm$0.09 &  1.26$\pm$0.09 \\ 
    \hline
\enddata
\tablecomments{The columns indicate (1) the object identifier, (2) the stellar velocity dispersion (Section \ref{subsubsec:measuring_sigma}), (3) the single-epoch BH mass (systematic uncertainty of 0.5\,dex; \citealt{Shen13}; Section \ref{subsub:bh_mass_method}), (4) the $r$-band aperture light-loss correction (Appendix \ref{app:aperture_corrections}), (5) the star-formation rate (systematic uncertainty of 0.3\,dex; \ZH; Section \ref{subsubsec:sfr}), (6) the SFR correction due to aperture losses (Section \ref{subsubsec:sfr}), (7) the stellar mass (Section \ref{subsec:ssp}), (8) the stellar mass correction due to aperture losses (Section \ref{subsec:ssp}), (9) the mass-weighted stellar age (Section \ref{subsec:ssp}), (10) the mass-weighted fraction of young ($<1$\,Gyr) stars (Section \ref{subsec:ssp}), (11) the color excess (Section \ref{subsec:ssp}), and (12)--(13) the inferred virial scaling factor using the FWHM and the line dispersion as the BLR velocity proxy, assuming CL-AGNs follow the \KH\ \Msig\ relation (Section \ref{subsubsec:msig})}.
\tablenotetext{a}{Removed from the analyses involving \sigs\ due to a high relative error.}
\end{deluxetable*}


\subsection{Measuring \sigs}
\label{subsubsec:measuring_sigma}

Throughout this work, we rely on the ``dim-state'' medium-resolution X-shooter/GMOS spectra for studying \sigs.
The measurements are performed using the penalized fitting method code \texttt{pPXF} \citep{Cappellari23}. This method fits the observed galaxy's spectrum using a convolution of stellar population templates and a line-of-sight velocity distribution (LOSVD), finding the best fit using a maximum penalized likelihood approach, and is widely used for \sigs\ measurements in extragalactic surveys \citep[e.g.,][]{Ahn12_sdssDR9, vanderWel21, Koss22}.
\texttt{pPXF} also takes into account the effective spectral resolution of the observed spectra, which were estimated as described in Appendix~\ref{app:resolution}.

Specifically, to construct our galaxy templates, we use stellar population models that are based on the empirical X-shooter Spectral Library \citep[][]{Verro22}. These population models have a relatively high spectral resolution ($R\sim10,000$), which is essential in order to precisely recover \sigs\ in our medium-resolution spectra, as template libraries have difficulties measuring \sigs\ that is lower than the libraries' nominal resolution \citep[see detailed discussions in, e.g.,][]{Boardman16,Boardman17,Gannon20}.

For each spectrum, we fit the spectral region between 3880--4200\,\AA, which covers the Ca~\textsc{ii} H+K stellar absorption features. While these features pose some challenges for measuring \sigs---owing to their strong dependence on spectral type, a steep local continuum, and large intrinsic broadening---they remain the most viable choice in analyses of visual-regime spectra where high AGN continuum contamination may be present \citep[][]{GH06}.
Our fitting model also includes fourth-degree additive Legendre polynomials, which can minimize template mismatch and correct for imperfect sky subtraction or scattered light \cite[e.g.,][]{Cappellari17}. For the LOSVD modeling, the first four Gauss-Hermite moments were allowed to vary.

Before fitting the spectrum, we masked several spectral regions to avoid potential contamination by AGN emission lines. Following \citet{Koss22}, we classify emission lines as either ``strong'' (masked regions cover $2500\,\kms$), or ``weak'' ($1000\,\kms$), where the definition of what constitutes a ``strong'' or ``weak'' emission line reflects its strength relative to \Hb in typical AGNs \citep[e.g.,][]{Tran00}. The full list of masked regions is presented in Table \ref{tab:mask_lines}. In addition, Table \ref{tab:mask_lines} also lists regions around ``very strong'' emission lines and the \Ha\ line, which are relevant to the analysis described in Section \ref{subsec:ssp}.
Due to the overlap with the H$\epsilon$ and [Ne\,\textsc{iii}] emission lines near 3967\,\AA, we are effectively masking the region around the Ca H $\lambda$3968 stellar absorption feature.
We also mask the stitching region between the UVB and VIS arms in X-shooter spectra, ranging from 5555 to 5605\,\AA\ in the observed frame.

\begin{deluxetable}{@{}l@{\extracolsep{\fill}}l@{}}
\tablecaption{Masked AGN emission lines for \texttt{pPXF} and \texttt{FIREFLY} fitting.}
\label{tab:mask_lines}
\tablewidth{\columnwidth}
\tablecolumns{2}
\setlength{\tabcolsep}{70pt}
\tabletypesize{\scriptsize}
\tablehead{
\\[-2ex]
Emission & Wavelength\\[-1ex] 
line & (\AA)
}
\startdata
    {\Ha} & 6562.8 \\
    \hline
    \multicolumn{2}{c}{``Very strong''} \\
    \hline
    {\Hb} & 4861.33 \\
    {[O \textsc{iii}]} & 4958.92 \\
    {[O \textsc{iii}]} & 5006.84 \\
    \hline
    \multicolumn{2}{c}{``Strong''} \\ 
    \hline
    {[Ne \textsc{v}]} & 3462 \\
    {[O \textsc{ii}]} & 3727 \\
    {[Ne \textsc{iii}]} & 3868.71 \\
    {$\rm{H}\zeta \rm{+} \rm{He}\,\textsc{i}$} & 3889.1 \\
    {$\rm{H}\epsilon\rm{+}$[Ne \textsc{iii}]} & 3967.41 \\
    {$\rm{H}\delta$} & 4101.76 \\
    {$\rm{H}\gamma$} & 4340.47 \\
    {[O \textsc{iii}]} & 4363.21 \\
    {He \textsc{i}} & 4471 \\
    {He \textsc{ii}} & 4686.00 \\
    {Fe \textsc{vii}} & 5720 \\
    {He \textsc{i}} & 5876 \\
    {[Fe \textsc{vii}]} & 6087 \\
    {[N \textsc{ii}]} & 6547.96 \\
    {[N \textsc{ii}]} & 6583 \\
    {[S \textsc{ii}]} & 6716 \\
    {[S \textsc{ii}]} & 6731 \\
    \hline
    \multicolumn{2}{c}{``Weak''} \\ 
    \hline
    {He \textsc{i}} & 4026 \\
    {[S \textsc{ii}]} & 4071.24 \\
    {[Fe \textsc{v}]} & 4229 \\
    {[Ar \textsc{iv}]} & 4711.30 \\
    {[Ar \textsc{iv}]} & 4740 \\
    {[Fe \textsc{iv}]} & 5146 \\
    {[Fe \textsc{vii}]} & 5159 \\
    {[Fe \textsc{vi}]} & 5176 \\
    {[N \textsc{i}]} & 5200 \\
    {[Ca \textsc{v}]} & 5309 \\
    {[Fe \textsc{vi}]} & 5485 \\
    {[Cl \textsc{iii}]} & 5518 \\
    {[Cl \textsc{iii}]} & 5538 \\
    {[Fe \textsc{vi}]} & 5677 \\
    {[N \textsc{ii}]} & 5755 \\
    {[Fe \textsc{vii}]} & 6087 \\
    {[O \textsc{i}]} & 6300 \\
    {[Fe \textsc{x}]} & 6374 \\
\enddata
\tablecomments{Masked regions cover $10,000\,\kms$ for \Ha, $5000\,\kms$ for ``very strong'',  $2500\,\kms$ for ``strong'', and $1000\,\kms$ for ``weak'' emission lines.}
\end{deluxetable}

After running the \texttt{pPXF} code to identify the best-fit parameters for each spectrum, we visually inspected the residuals to verify the quality of the fits. An example of a good fit is presented in Figure \ref{fig:ppxf_example}. All \texttt{pPXF} fits for our objects are presented in Figure \ref{fig:ppxf_fits} in Appendix \ref{app:ppxf_plots}.

To estimate the uncertainties associated with the \sigs\ measurements, we complement the formal \texttt{pPXF} errors (derived from the covariance matrix of each fit) with an MC error analysis. Following \citet{CE04} and \citet{Cappellari23}, we estimate uncertainties using a wild bootstrap of the residuals, generating 100 mock spectra that are refit with the \texttt{pPXF} penalization turned off (\texttt{bias}=0). The uncertainty on \sigs\ is taken as the standard deviation of the resulting distribution. For each object, we conservatively adopt the larger of the bootstrap-based uncertainty and the formal \texttt{pPXF} error.
For one object, J0124+0040, this method yielded a large relative uncertainty ($\sim85$\,\%) on \sigs. We therefore exclude this object from analyses involving \sigs\ (Section \ref{subsubsec:msig}).
The \sigs\ values derived for our sample range from 49 to 260\,\kms, with associated uncertainties between 5 and 33\,\kms, and are listed in Table \ref{tab:quantities}.

\begin{figure}
    \centering
    \includegraphics[width=1.0\columnwidth]{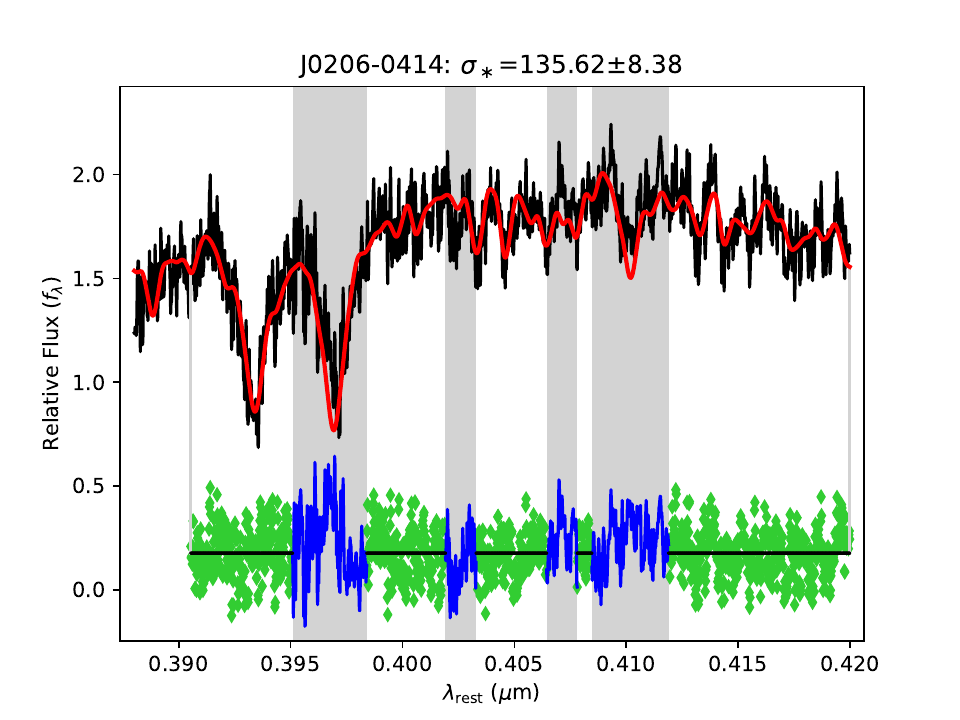}
    \caption{Example of the \texttt{pPXF} fitting and \sigs\ measurement for J0206$-$0414. The X-shooter data are shown in black, the best-fitting model in red, and the residuals are displayed near the bottom, where green points were considered for the fit, while blue ones were masked (masked spectral regions are indicated in gray and listed in Table \ref{tab:mask_lines}).}
    \label{fig:ppxf_example}
\end{figure}


\subsection{Stellar Population Modeling}
\label{subsec:ssp}

To model the stellar populations in the hosts of our CL-AGNs, we applied the \texttt{FIREFLY} spectral fitting code \cite[][]{Firefly} on the ``dim-state'' SDSS-V spectra.\footnote{Here and in Section~\ref{subsubsec:sfr}, we used the SDSS-V data rather than the X-shooter spectra to ensure uniformity across the full sample, which includes objects observed with GMOS, whose more limited wavelength coverage would not allow the derivation of stellar properties.}
\texttt{FIREFLY} derives the stellar population properties by iteratively fitting combinations of single-burst stellar population model templates to spectroscopic data.
Our choice of \texttt{FIREFLY} was motivated by two main considerations. First, the availability of a large comparison sample of SDSS galaxies analyzed with the same code (\citealt{Comparat17}; see Section~\ref{subsubsec:control}). Second, because CL-AGNs are highly variable, methods that rely on broadband spectral energy distribution fitting (e.g., \texttt{CIGALE}) are less suitable, as they combine multiwavelength data obtained at different epochs. 

Before fitting, \texttt{FIREFLY} applies a high-pass filter (HPF) to remove large-scale continuum modes from the spectrum, fits the stellar populations to the HPF-filtered spectrum, and then models the recovered large-scale continuum shape with an MW extinction law \citep[$R_V=3.1$;][]{ODonnell94} to infer the color excess, $E(B-V)$.
This fitting procedure returns the stellar mass, stellar age, and spectrum of each of the single-burst stellar populations comprising each of our galaxies, from which we derive the total stellar mass ($M_\ast$), mass-weighted age ($t_{\rm \ast, M}$), and mass-weighted fraction of young ($<1\,$Gyr) stars for each source ($f_{\rm 1Gyr}$).

Specifically, for our \texttt{FIREFLY} fitting procedure, we used the MILES stellar library \cite[][covering 3500--7430\,\AA]{MILES11} combined with the stellar velocity dispersion obtained with \texttt{pPXF} (Section \ref{subsubsec:measuring_sigma}). All other input parameters, including the initial mass function (IMF), metallicity range, etc, were adopted from our reference sample (\citealt{Comparat17}; see Section \ref{subsubsec:control} below) or left at their default values. A complete list of the \texttt{FIREFLY} input parameters is provided in Table \ref{tab:firefly}.

Similarly to our \texttt{pPXF} analysis (Section \ref{subsubsec:measuring_sigma}), we masked several spectral regions that can potentially be affected by AGN contamination.
Since in \texttt{FIREFLY} we fit a wider wavelength range, we had to consider additional masked regions, which are all listed in Table \ref{tab:mask_lines}. Specifically, in addition to ``weak'' and ``strong'' emission lines, we also include ``very strong'' lines (masked regions cover $5000\,\kms$), and \Ha\ (masked region covers $10,000\,\kms$).

For the uncertainties, we adopted the formal ones provided by the  \texttt{FIREFLY} fits, which are estimated using the parameters' likelihood distributions.
We deemed measurements with uncertainties larger than 1\,dex as unreliable. One source (J0245+0037) showed such a large uncertainty in stellar age and was therefore excluded from analyses involving stellar age and the young stellar fraction, but retained in the overall sample and in other analyses.
The log$(\,\mstar/M_\odot)$ values for our sample range from 9.4 to 10.9 with a median value of 10.4, the $t_{\rm \ast, M}$ values cover the range of 1.7--9.9\,Gyr, with a median value of 6.9\,Gyr, while $f_{\rm 1Gyr}$ ranges from 0 to 0.6, with a median value of 0.02.
All derived quantities are listed in Table \ref{tab:quantities}, and the \texttt{FIREFLY} fits are presented in Figure \ref{fig:firefly_fits} in Appendix \ref{app:firefly_plots}.

Since our \mstar\ measurements are derived from SDSS spectra obtained with a 2\,\arcsec\ fiber, some galaxy light lies outside the aperture.
Using the $r$-band light-loss estimates ($\Delta_r$) from Appendix \ref{app:aperture_corrections}, and assuming a constant mass-to-light ratio across each galaxy (i.e., no difference between regions inside and outside the fiber), we derived aperture corrections of $\Delta_{\mstar} = -0.4\,\Delta_r$ in the range of 0.18--0.77\,dex with a median of 0.46\,dex.

\begin{deluxetable}{@{}l@{\extracolsep{\fill}}l@{}}
\tablecaption{\texttt{FIREFLY} input parameters}
\label{tab:firefly}
\tablewidth{\columnwidth}
\tablecolumns{2}
\setlength{\tabcolsep}{80pt}
\tabletypesize{\scriptsize}
\tablehead{
\\[-2ex]
Input parameter & Value
}
\startdata
    model key & m11 \\ 
    model flavor & MILES \\ 
    IMF & Kroupa \\ 
    age limits & [0.001\,Gyr, `AoU']\tablenotemark{\footnotesize a} \\ 
    Z limits & {[-3, 3]} \\ 
    Milky Way reddening & False \\ 
    hpf mode & on \\ 
    dust law & O'Donnell 94\tablenotemark{\footnotesize b} \\ 
    max ebv & 0.7 \\
    num dust vals & 200 \\             
    dust smoothing length & 200 \\
    max iterations & 10 \\
    pdf sampling & 300 \\    
\enddata
\tablenotetext{a}{Fits are not allowed to exceed the age of the universe at each object`s redshift.}
\tablenotetext{b}{The \citealt{ODonnell94} dust law is not included in the standard \texttt{FIREFLY} implementation and was added to the code in this work for consistency with the rest of the analysis.}

\end{deluxetable}


\subsection{Star Formation Rate}
\label{subsubsec:sfr}

Throughout this work, we rely on the ``dim-state'' SDSS-V spectra to measure star formation rates (SFRs).
We follow the prescription of \citet[][hereafter ZH19]{ZH19}, which estimates SFRs based on the \OII\ narrow emission-line flux, calibrated against the commonly used \Ha-based SFR indicator for SF galaxies, and assuming the \citet{Kroupa01} IMF. The prescription uses the \OIII\ line to account for and remove the contribution of AGNs to the \oii\ emission.
This prescription is metallicity dependent, and we estimate the metallicity for our CL-AGN using the stellar mass--metallicity relation \cite[][]{KE08}, where the stellar mass was derived from the \texttt{FIREFLY} fitting procedure (Section \ref{subsec:ssp}).
As the adopted \oii-based SFR prescription is calibrated against \Ha-based SFRs, it inherits the main systematic uncertainties affecting \Ha-based tracers (e.g., the assumed IMF and uncertainties in dust attenuation corrections). In addition, \oii\ itself is sensitive to metallicity and excitation conditions (see, e.g., \citealt{Kennicutt12} and references therein).

We adopt emission-line-based estimates rather than those derived from \texttt{FIREFLY}, as \texttt{FIREFLY} assumes single-burst stellar populations and hence yields average SFRs over extended timescales rather than tracing the current star-formation activity.
We also do not use D4000-based SFR estimates \cite[e.g.,][]{Brinchmann04}, since the D4000 index is sensitive to AGN contamination, which may still be present even in our dim-state spectra. Emission-line-based estimators are thus more appropriate for tracing ongoing star formation in our sample.

We measured the \oii\ and \oiii\ emission-line fluxes using \texttt{PyQSOFit} (Section \ref{subsec:pyqsofit}), and corrected them for host galaxy dust extinction using the best-fit $E(B-V)$ values derived from the \texttt{FIREFLY} fitting procedure (Section \ref{subsec:ssp}), assuming MW-like extinction (\citealt{ODonnell94}; $R_V = 3.1$).
We adopt an uncertainty of 0.3\,dex for the SFRs, corresponding to the formal scatter associated with this estimator (\ZH).

For one object (J0927+0433), the SFR prescription yielded a negative value because the \oiii-based AGN correction exceeded the observed \oii\ flux. For this source, we therefore used the \oii\ flux alone to derive an upper limit on the SFR.
The SFR values we derive for our sample range from 0.04 to $7\,M_\odot\,\rm{yr}^{-1}$ with a median value of $1\,M_\odot\,\rm{yr}^{-1}$, and are listed in Table \ref{tab:quantities}.

Our SDSS-V-based SFR measurements are underestimated because the SDSS 2\arcsec\ fiber captures only part of the galaxy light.
Following the methodology in Appendix \ref{app:aperture_corrections}, we estimated the $r$-band aperture loss for each target, $\Delta_r$.
To obtain a rough first-order correction, we assumed that the regions inside and outside the fiber have similar average properties, and applied this offset to the \oii\ and \oiii\ line luminosities, i.e., $\Delta_{L\oii} = \Delta_{L\oiii} = -0.4\,\Delta_r$.
We then used the corrected line luminosities, together with the aperture-corrected \mstar\ values (Section \ref{subsec:ssp}), to derive the corrected SFRs using the \ZH\ prescription. The resulting SFR corrections, $\Delta_{\rm{SFR}}$, span 0.19 to 0.96\,dex, with a median of 0.45\,dex.


\subsection{Comparison samples}
\label{subsubsec:control}

To contextualize the host galaxy properties derived for our CL-AGNs, we compare them to large and homogeneously selected samples of low-redshift (1) star-forming (SF) galaxies and (2) type 2 AGNs.
To derive the properties of these two populations, we use two catalogs: (1) classifications based on line ratio diagnostics and emission-line measurements from the SDSS DR7 \cite[][]{Abazajian09_DR7} MPA/JHU catalogs \cite[][]{Kauffmann03, Brinchmann04},\footnote{\url{https://wwwmpa.mpa-garching.mpg.de/SDSS/DR7/}} which modeled the spectra of all narrow emission-line galaxies in DR7, regardless of nuclear activity. 
(2) stellar masses, stellar ages, and young stellar population fractions from the SDSS DR14 \texttt{FIREFLY} catalog \citep{Comparat17}, which were derived using the same fitting code and parameter configuration adopted for our CL-AGN analysis.\footnote{To ensure consistency with our CL-AGN analysis, we use \texttt{FIREFLY}-based stellar masses for the comparison samples, rather than the photometry-based values directly available from the MPA/JHU catalog.}
Because the \texttt{FIREFLY} catalog allowed fits that include stellar populations with unphysical ages (i.e., exceeding the age of the universe at the system's redshift), we chose to remove all galaxies in which more than 10\,\% of the stellar mass originated from such unphysical stellar populations.

To ensure consistency with our CL-AGN analysis, we derive SFRs for our comparison samples using the same \oii\ estimator as we did for our CL-AGN sample (see Section \ref{subsubsec:sfr}).\footnote{The photometry-based SFR values for SF galaxies available from the MPA/JHU catalog show good agreement with those derived using the \oii\ estimator, with a scatter of $\sim0.2\,$dex and a systematic offset of $\sim0.1\,$dex at high stellar masses (see discussion in \ZH, and references therein).} We first applied basic quality cuts to the catalog as detailed in \ZH. These include S/N cuts on key emission lines, a minimum fiber light coverage to ensure representative metallicity measurements, and removal of objects with unphysical Balmer decrements.
Following \ZH, we correct line fluxes for dust extinction using the observed Balmer decrement and the MW extinction curve of \citet{ODonnell94}, assuming $R_V = 3.1$, and intrinsic \Ha/\Hb ratios of 2.86 and 3.1 for SF galaxies and AGNs, respectively.
We then applied the \oii\ SFR metallicity-dependent prescription to the DR7 catalog objects.
For metallicity estimation, we followed the recommendations by \ZH.
For SF galaxies, we estimate the metallicity using the \nii\ to \oii\ ratio (available in the MPA/JHU catalog).
For AGNs, we estimated the metallicity using the relation to \mstar\ (\citealt{KE08}; \mstar\ values are available in the SDSS \texttt{FIREFLY} catalog). The SFR AGN estimator also relies on the \oiii\ flux, which is available in the MPA/JHU catalog.
After applying all cuts described above, the final comparison samples comprise 62,259 SF galaxies and 10,615 type 2 AGNs.

\section{Results and Discussion}
\label{sec:results_discussion}

We now examine and discuss the properties of our CL-AGN sample, beginning with the nature of their variability, followed by an analysis of host galaxy properties and implications for BH mass estimates.

\subsection{Revisiting the nature of CL-AGNs}
\label{subsec:nature_clagn}

\begin{deluxetable*}{lcccc|ccc}
\tablecaption{Broad-line \Ha\ and \mgii\ fluxes and predicted ratios based on different extinction laws.}
\label{tab:lines}
\tablewidth{\columnwidth}
\tablehead{
\colhead{Name} & \multicolumn{2}{c}{\ha\ Flux} & \colhead{\ha\ Flux} & \colhead{\mgii\ Flux} & \multicolumn{3}{c}{Predicted ``Bright-state'' \mgii/\ha\ Flux Ratio}\\ [-4ex]
\colhead{} & \colhead{} & \colhead{} & \colhead{Ratio} & \colhead{} & \colhead{} & \colhead{} & \colhead{}\\ [-3ex]
\colhead{} & \colhead{Legacy SDSS} & \colhead{X-shooter} & \colhead{} & \colhead{X-shooter} & \colhead{Maiolino+01} & \colhead{MW $R_V=4.4$} & \colhead{MW $R_V=3.1$}\\ [-2ex]
\colhead{(1)} & \colhead{(2)} & \colhead{(3)} & \colhead{(4)} & \colhead{(5)} & \colhead{(6)} & \colhead{(7)} & \colhead{(8)}
}
\startdata
J0158$+$0013 &  350$\pm$10  &  310$\pm$4  & 1.13$\pm$0.05 & 120$\pm$14 & 0.44 & 0.45 & 0.48 \\
J0159$+$0033 &  950$\pm$40  &  310$\pm$6  & 3.10$\pm$0.13 & 270$\pm$23 & 1.86 & 2.21 & 4.23 \\
J0206$-$0414 & 1420$\pm$30  &  840$\pm$5  & 1.68$\pm$0.03 & 600$\pm$36 & 1.02 & 1.10 & 1.49 \\
J0245$+$0037 & 1150$\pm$40  &  230$\pm$3  & 4.92$\pm$0.17 & 190$\pm$8  & 2.35 & 2.99 & 7.44 \\
J0801$+$3417 & 1590$\pm$50  &  540$\pm$25 & 2.95$\pm$0.17 & 520$\pm$28 & 2.00 & 2.36 & 4.37 \\
J0846$+$0000 & 1360$\pm$110 &  540$\pm$4  & 2.52$\pm$0.20 & 250$\pm$13 & 0.86 & 0.99 & 1.68 \\
J0855$+$0329 & 2470$\pm$30  &  670$\pm$7  & 3.71$\pm$0.07 & 320$\pm$49 & 1.16 & 1.41 & 2.98 \\
J0904$-$0042 &  530$\pm$40  &  130$\pm$17 & 4.09$\pm$0.61 &  80$\pm$8  & 1.68 & 2.08 & 4.65 \\
J0916$+$0000 & 2170$\pm$100 &  670$\pm$4  & 3.26$\pm$0.14 & 300$\pm$14 & 1.02 & 1.22 & 2.39 \\
J0927$+$0433 & 2940$\pm$70  & 1050$\pm$4  & 2.79$\pm$0.07 & 870$\pm$12 & 1.65 & 1.93 & 3.46 \\
J0927$+$0503 & 5120$\pm$60  & 1670$\pm$10 & 3.06$\pm$0.04 & 620$\pm$48 & 0.79 & 0.94 & 1.78 \\
J0932$+$0403 & 1230$\pm$30  &  200$\pm$15 & 6.20$\pm$0.49 & 110$\pm$10 & 1.92 & 2.53 & 7.18 \\
J2336$+$0040 & 1020$\pm$30  &  420$\pm$9  & 2.43$\pm$0.09 & 160$\pm$41 & 0.70 & 0.80 & 1.34 \\
\enddata
\tablecomments{
The columns indicate (1) the object identifier, (2)--(3) the broad \Ha\ fluxes from the bright legacy SDSS and dim X-shooter states in units of $10^{-17}\,\rm{erg}\,\rm{s}^{-1}\,\rm{cm}^{-2}$, (4) the \Ha\ flux ratio between the bright legacy SDSS and the dim X-shooter spectra, (5) the broad \mgii\ flux from the X-shooter spectrum in units of $10^{-17}\rm{erg}\,\rm{s}^{-1}\,\rm{cm}^{-2}$, and (6)--(8) the predicted \mgii/\Ha\ flux ratio in the bright state assuming the dim-state results from dust obscuration, using the \citet{Maiolino01} extinction law and MW-like extinction laws \citep[][with $R_V=4.4$ and $R_V=3.1$]{ODonnell94}.
}
\end{deluxetable*}

Out of our 23 CL-AGNs, 16 have reliable and simultaneous spectral coverage of the \mgii, \hb, and \ha\ regions in their dim-state spectra. Of these, only two already had such coverage in their SDSS-V spectra, while the remaining 14 benefit from the broad wavelength range of X-shooter. The other sources either lack \mgii\ coverage, have data of insufficient quality (e.g., due to noise or strong sky/telluric residuals), or were observed with GMOS.
Visual inspection shows that 13 of these 16 objects display strong broad \mgii\ emission even in the dim state, while the broad Balmer lines are weak or absent \citep[see also][]{Guo20_mgii,Yang20_CLAGN}. At face value, this disfavors an obscuration-driven explanation, since \mgii, being at shorter (UV) wavelengths, should be more strongly affected by dust extinction than the optical Balmer lines.

To examine this more quantitatively, we note that in principle the relative changes in \mgii\ and \ha\ fluxes between the bright and dim states could directly test the role of dust obscuration versus intrinsic accretion variability. However, due to the limited wavelength coverage of the legacy SDSS ``bright-state'' spectra, all 16 objects with both lines observed in the dim state have bright-state spectra that include \ha\ only, with no \mgii\ coverage.
Consequently, we cannot directly compare the \mgii\ flux between states. Instead, we measure the observed change in the broad \ha\ flux between the bright (legacy SDSS) and dim (X-shooter/SDSS-V) spectra for each object. Assuming this change is entirely due to dust attenuation, we calculate the amount of extinction required to produce such change, adopt a specific extinction law, and then deredden the observed dim-state \mgii\ flux to determine the expected bright-state \mgii\ flux under the assumption of variable obscuration.
The resulting predicted bright-state \mgii-to-\ha\ flux ratios can then be compared with those typically observed in unobscured, broad-line AGNs to evaluate the plausibility of the extinction-driven scenario.
We base this calculation on the \ha\ line rather than the \hb\ line because estimating the dim-state \hb\ flux is significantly more challenging: in most cases, the broad \hb\ line is absent or extremely weak; placing upper limits depends strongly on the host-AGN spectral decomposition, particularly the treatment of stellar \hb\ absorption.
In contrast, \ha\ is brighter and thus can be more reliably measured in the dim state.

\begin{figure}
    \centering
    \includegraphics[width=1.0\columnwidth]{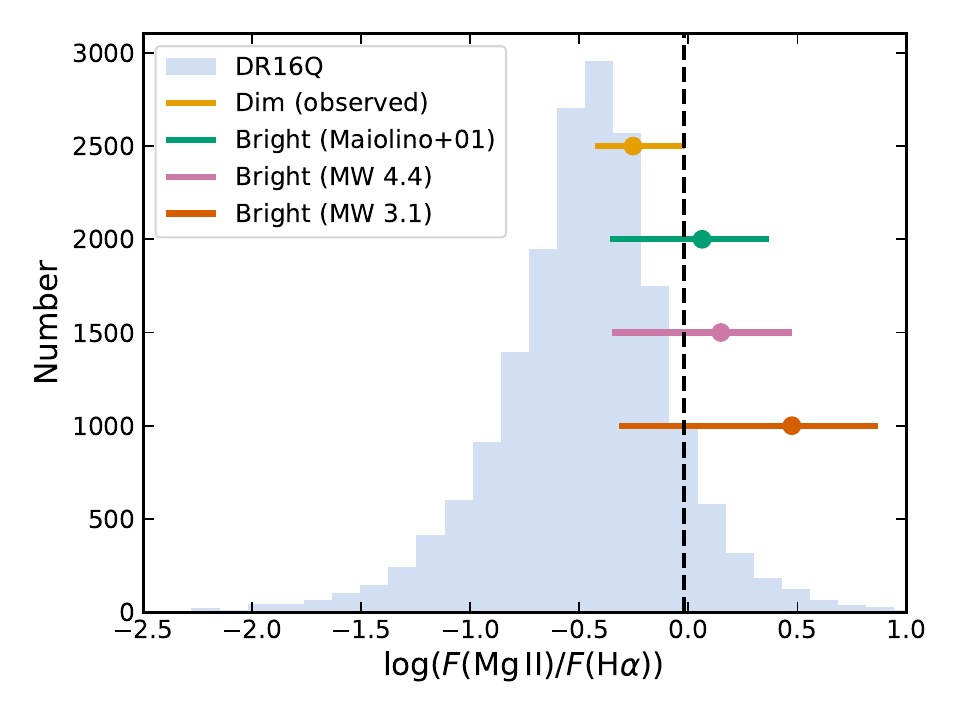}
    \caption{Broad \mgii\ to broad \ha\ line flux ratios for our CL-AGNs, compared with SDSS DR16 quasars. The $F(\mgii)/F(\ha)$ distribution of DR16 quasars, taken from \cite{WS22}, is shown in light blue, and the $90^{\rm th}$ percentile is marked with a vertical line. The median and range of the observed ``dim-state'' flux ratios of our 13 sources with detectable \mgii\ are shown in yellow, with arbitrary vertical scaling. Also shown are the predicted ``bright-state'' flux ratios under the variable obscuration scenario, for different extinction laws (again with various arbitrary vertical scalings): the \citet{Maiolino01} law (green), and MW-like laws from \citet{ODonnell94} with $R_V = 4.4$ (magenta) and $R_V = 3.1$ (red). The \mgii-to-\ha\ ratios observed in the dim states of our CL-AGNs, which are typical of unobscured AGNs, disfavor the obscuration scenario as the main driver of their extreme spectral variability.}
    \label{fig:mgii_ha_hist}
\end{figure}

Table \ref{tab:lines} presents the resulting calculations for several extinction laws. We adopt the relatively shallow AGN-appropriate law of \citet[][see also \citealt{Li07}]{Maiolino01}, and also test steeper MW-like laws with $R_V = 4.4$ and $R_V = 3.1$ \citep{ODonnell94}.
All objects we examined have observed ``dim-state'' \mgii-to-\ha\ flux ratios above the median value in the DR16 quasar catalog (0.35; \citealt{WS22}), with a median ratio of 0.56 for our sample. When using the extinction law of \citet{Maiolino01}, the median predicted bright-state \mgii-to-\ha\ flux ratio is 1.16. For comparison, in the DR16 quasar catalog, ratios exceeding 1.16 are seen in less than 7\% of cases. The discrepancy grows even larger when adopting steeper extinction curves. For MW-like extinction with $R_V=3.1$, the median ratio is 3.0, compared to less than 2\% of the DR16 quasar catalog sources that show such ratios.
These results are further illustrated in Figure \ref{fig:mgii_ha_hist}, which shows the distribution of broad \mgii-to-\ha\ flux ratios for DR16 quasars \citep[][]{WS22}, along with the median and range of the observed dim-state ratios and the predicted bright-state ratios under the extinction scenario.

We therefore conclude that variable obscuration is unlikely to be the primary driver of the state changes in our sources. This conclusion agrees with several previous observational studies that have disfavored the obscuration scenario \cite[e.g.,][see also \citealt{RT23} for a review]{Denney14, LaMassa15, MacLeod16, Duffy25b}.
Instead, the persistence of broad \mgii\ emission in the dim state is best explained by variability in the accretion flow (see, e.g., \citealt{Guo20_mgii} for one model exploring such a scenario).
Accordingly, CL-AGNs that exhibit \mgii\ emission in their dim state should be considered ``changing-state AGNs'' \cite[see detailed discussion of terminology in ][]{RT23}.

\subsection{Scaling Relations of CL-AGNs}
\label{subsec:scaling_relations_clagn}

\subsubsection{\mbh--\sigs\ and implications for \mbh\ estimates}
\label{subsubsec:msig}

There are two different approaches for investigating the position of CL-AGNs on the \Msig\ relation:
\begin{enumerate}[label=(\alph*)]

\item Adopt a virial factor $f$ based on RM AGN studies (Equation~\ref{eq:cho23}) to derive \mbh\ estimates for CL-AGNs, and compare their location on the \Msig\ plane to that of inactive galaxies and other AGNs. It is important to note that $f$ in RM studies is typically calibrated by assuming that RM AGNs, on average, follow the \Msig\ relation of inactive galaxies.

\item Assume that CL-AGNs follow the \Msig\ relation of inactive galaxies, and derive the appropriate virial factor $f$ for this population.

\end{enumerate}
In the following section, we examine both approaches.

\begin{figure*}
    \centering
    \includegraphics[width=1.0\textwidth]{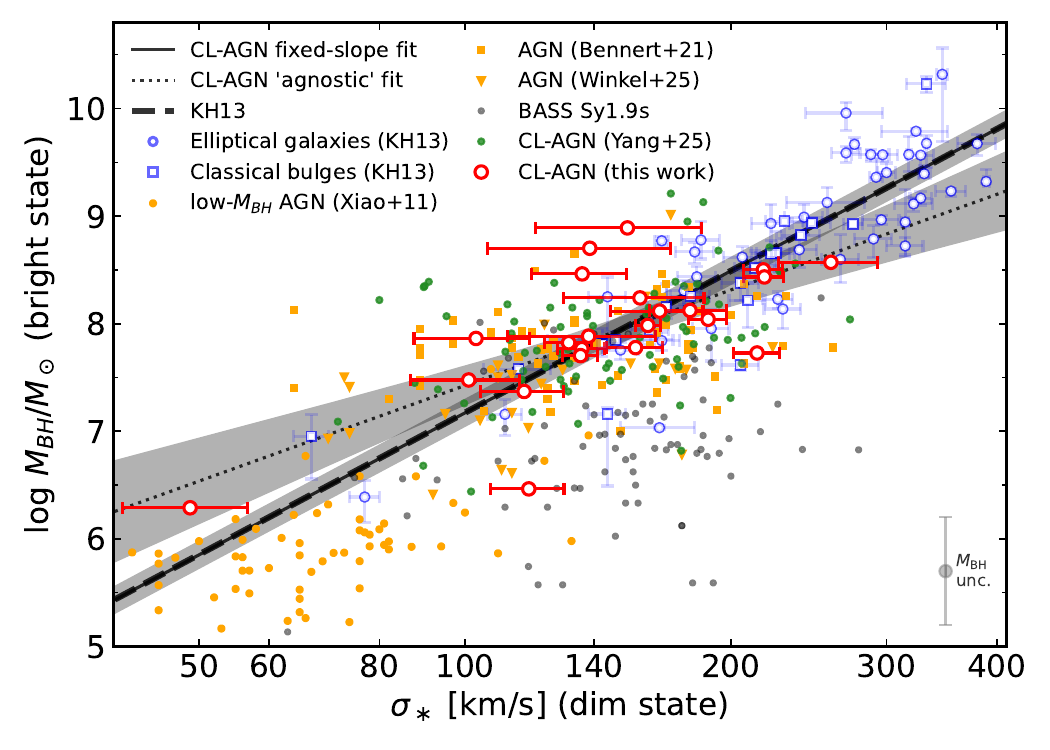}
    \caption{The \Msig\ relation of our CL-AGNs (red symbols). Comparison samples include elliptical galaxies and classical bulges (\KH, blue), BASS type 1.9 AGNs \cite[][gray]{Koss22, MejiaRestrepo22}, low-\mbh\ AGN from \citet{Xiao11}, AGNs from \citet{Bennert21} and \citet[][orange]{Winkel25}, and `turn-on' CL-AGNs from \citet[][green]{Yang25}. The relation from \KH\ (dashed line), our fit where the slope is fixed to the \KH\ value (solid), and our `agnostic' fit where all parameters are free to vary (dotted) are indicated. The uncertainties for our fits are represented by the gray regions. The \mbh\ of our sample, \citet{Xiao11}, and BASS were estimated using the prescription given in Equation \ref{eq:cho23}. For all other samples, \mbh\ values were adopted directly from the respective works. The typical 0.5\,dex uncertainty associated with single-epoch methods is indicated in the bottom right corner. For clarity, uncertainties on the \sigs\ measurements of the comparison AGN samples are not displayed. Our CL-AGNs follow the \Msig\ relation of inactive galaxies and other AGN samples, but are significantly offset from the BASS type 1.9 AGNs.}
    \label{fig:M_sigma}
\end{figure*}

Figure \ref{fig:M_sigma} presents our CL-AGN sample in the \Msig\ plane, where \mbh\ is measured from the bright (legacy SDSS) state, assuming $f=1.12$ (Section \ref{subsub:bh_mass_method}), while \sigs\ is measured from the dimmer (X-shooter/GMOS) state. We also show measurements for elliptical galaxies \cite[see][and references therein; hereafter KH13]{KH13} and various samples of local broad-line AGNs \cite[][]{Xiao11, Bennert21, Koss22, MejiaRestrepo22, Winkel25}.
The relation from \KH, namely $\log(\mbh/10^9\Msun)=-0.51+4.4\times\log(\sigs/200\,\kms)$, with 0.29 dex intrinsic scatter (their equation 3), is also presented.

Most of our CL-AGNs lie within the intrinsic scatter of the \KH\ relation, with no clear systematic offset. To quantify any possible deviation, we define $\Delta \log M=\log M_{\sigma}-\log M_{\rm BLR}$, where $M_{\sigma}$ is the BH mass inferred from a given \Msig\ relation, and $M_{\rm BLR}$ is the mass derived from single-epoch mass prescriptions (see Section \ref{subsub:bh_mass_method}). For the \KH\ relation, the median offset is 0.03\,dex ($16^{\rm th}$--$84^{\rm th}$ percentiles: $-0.58$ to 0.30\,dex), consistent with no deviation from the \KH\ relation.

To assess any possible offset while properly accounting for measurement uncertainties, we fit the data using the Bayesian modeling package \texttt{PyMC} \citep{PyMC}, where we fix the slope and intrinsic scatter to the \KH\ values ($4.384$ and $0.29\,$dex, respectively), and allow only the intercept to vary.\footnote{In all the fits presented below, we take the \mbh\ uncertainty to be 0.5\,dex \cite[][]{Shen13}, assume a uniform prior for the fitted intercept (and slope where applicable), and verify model convergence and the reliability of the fit.}
The resulting `fixed-slope' model fit is
\begin{align}
\label{eq:fixed_slope}
    \log\left(\frac{\mbh}{10^9\,\Msun}\right) &= - \left(0.51 \pm 0.13\right) + 4.4\nonumber \\
    &\quad \times\log\left(\frac{\sigs}{200 \, \kms}\right)\,\, .
\end{align}
The fit is shown as a solid line in Figure \ref{fig:M_sigma}, with the shaded region representing the associated uncertainty on the intercept.
This fit corresponds to an offset of $0.00\pm0.13\,$dex from the \KH\ relation, consistent with the simple median offset reported above.
Thus, we find no evidence that our CL-AGNs deviate from the \KH\ relation of inactive galaxies.

We then perform a second, independent fit where all parameters (intercept, slope, and intrinsic scatter) are left free. In this `agnostic' approach, the resulting model fit is
\begin{align}
\label{eq:agnostic}
    \log\left(\frac{\mbh}{10^9\,\Msun}\right) &= - \left(0.69 \pm 0.16\right) + \left(2.95\pm0.81\right) \nonumber \\
    &\quad  \times\log\left(\frac{\sigs}{200 \, \kms}\right),
\end{align}
with an intrinsic scatter of $0.16\pm0.12$\,dex (dotted line in Figure \ref{fig:M_sigma};
shaded region represents the associated uncertainty).
This approach yields a shallower slope compared to \KH\ at the $1.7\,\sigma$ level, and a lower intercept at the $1.1\,\sigma$ level.
While the statistical significance of the slope difference is modest, the trend is qualitatively similar to that reported in previous studies of RM AGNs, which found that the \Msig\ relation is shallower for RM sources \cite[e.g.,][]{Woo13, Woo15, Batiste17}.
However, given the large uncertainty on the fitted slope, a larger sample spanning a wider range in \sigs\ and \mbh\ is needed to conclusively determine whether CL-AGNs follow a shallower \Msig\ relation than inactive galaxies.

\begin{figure}
    \centering
    \includegraphics[width=1.0\columnwidth]{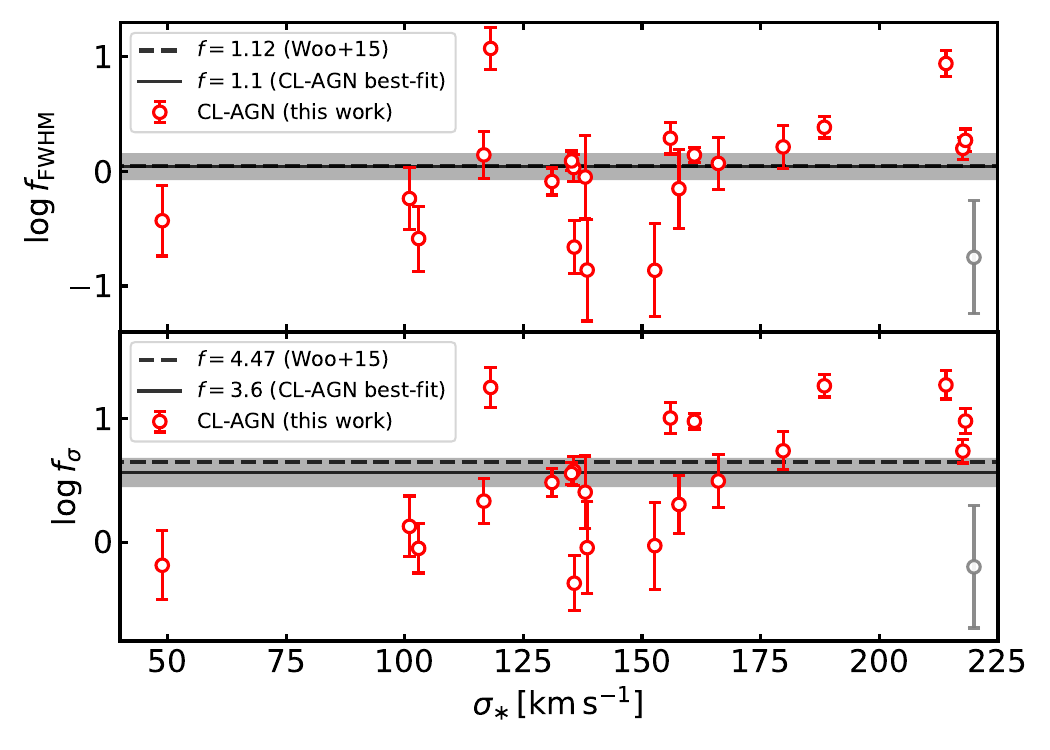}
    \caption{The virial scaling factor $f$ as a function of \sigs\ for our CL-AGN sample, assuming CL-AGNs follow the \KH\ \Msig\ relation for inactive galaxies. {\it Top:} $f$ values derived using the broad-line FWHM as the BLR velocity proxy. {\it Bottom:} $f$ values using the line dispersion ($\sigma$) instead. Horizontal dashed lines mark the values from \citet{Woo15}; solid lines and shaded regions show our inferred $f$ and the accompanying uncertainties. Error bars reflect uncertainties in \sigs; uncertainty from the single-epoch virial product systematics and the \Msig\ relation scatter is indicated in the bottom right ($\sim0.5\,$dex). Our inferred $f$ values are consistent with the RM-based estimates of \citet{Woo15} within the uncertainties.}
    \label{fig:f_vs_sigma}
\end{figure}

Next, we relax the assumption that the virial scaling factor $f$ for CL-AGNs is fixed at $1.12$ \citep[as suggested by][]{Woo15}. Instead, assuming that CL-AGNs, on average, follow the \KH\ \Msig\ relation for inactive galaxies, we infer the value of $f$ implied by our data. Figure \ref{fig:f_vs_sigma} shows the implied $f$ for each CL-AGN in our sample as a function of \sigs. The top panel shows $f$ for a virial product where the broad \ha\ emission-line FWHM is used as a proxy for BLR velocity, while the bottom panel presents $f$ in the case where the proxy is taken to be the line dispersion, $\sigma$. 
Our fitting model yields $f_{\mathrm{FWHM}}=1.1\pm0.3$ when using FWHM, and $f_\sigma=3.6^{+1.1}_{-0.9}$ when using the line dispersion.
For comparison, the values reported by \citet{Woo15} are $f_{\mathrm{FWHM}}=1.12$ and $f_\sigma=4.47$, both of which are consistent with our results within the uncertainties.
All inferred $f$ values for our CL-AGNs are listed in Table \ref{tab:quantities}.

\subsubsection{\mbh--\mstar}
\label{subsubsec:mbh_mstar}

\begin{figure*}
    \centering  \includegraphics[width=1.0\textwidth]{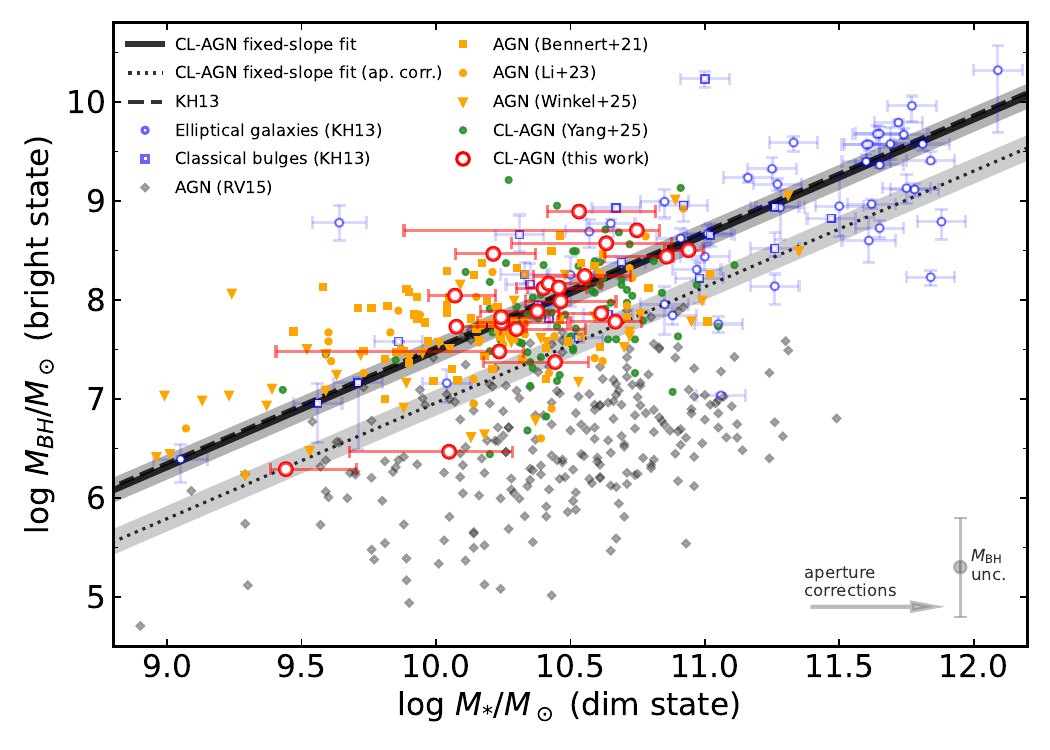}
    \caption{The \MM\ relation for our CL-AGNs (red symbols). Comparison samples include elliptical galaxies and classical bulges (\KH, blue), SDSS low-luminosity AGNs (\RV, gray), AGN samples from \citet{Bennert21}, \citet{Li23}, and \citet[][orange]{Winkel25}, and `turn-on' CL-AGNs from \citet[][green]{Yang25}. Also included are the relation from \KH\ (dashed line), and our fits where the slope is fixed to the KH13 value with and without aperture-corrected stellar masses (solid and dotted, respectively). The uncertainties for our fits are represented by the gray regions. The \mbh\ of our and the \RV\ samples were estimated using the prescription given in Equation \ref{eq:cho23}. For all other samples, \mbh\ values were adopted directly from the respective works. The typical uncertainty associated with single-epoch methods (0.5\,dex) is indicated in the bottom right corner. The arrow marks the median aperture correction estimated for the CL-AGN \mstar\ values (0.46\,dex; see Appendix \ref{app:aperture_corrections}). For clarity, uncertainties on the \mstar\ measurements of the comparison AGN samples are not displayed. Our CL-AGNs follow the \MM\ relation of inactive galaxies and other AGN samples, but are offset from the \RV\ sample.}
    \label{fig:M_vs_Mstar}
\end{figure*}

Figure \ref{fig:M_vs_Mstar} shows the \MM\ relation of inactive elliptical galaxies and classical bulges from KH13 alongside our CL-AGN sample, where \mbh\ is derived from the bright legacy SDSS spectra,
while \mstar\ is derived from the dim SDSS-V spectra.
For comparison, we also include the broad-line, nearby ($z<0.055$) AGN sample from \citet[][hereafter RV15]{RV15}, which reports total stellar masses, as well as the AGN samples of \citet{Bennert21}, \citet{Li23}, and \citet{Winkel25}. These latter works used imaging to decompose their hosts into components, and for our comparison, we use the bulge stellar masses reported in those works.
We also show the relation from \KH, namely, $\log(\mbh/10^9\Msun)=-0.310+1.17\times\log(M_{\rm{\ast}}/10^{11}\,\Msun)$, with 0.28 dex intrinsic scatter (see their equation 10), where \mstar\ refers to the bulge mass.
Our CL-AGN sample appears broadly consistent with the \KH\ \MM\ relation of inactive galaxies, with a median BH-to-stellar mass fraction of 0.38\,\% ($16^{\rm th}$--$84^{\rm th}$ percentiles: 0.15 to 0.88\,\%) for our sample compared to 0.51\,\% ($16^{\rm th}$--$84^{\rm th}$ percentiles: 0.11 to 1.23\,\%) for \KH.

To quantify any potential offset while accounting for measurement uncertainties, we fit the data while only allowing the intercept to vary, keeping the slope and intrinsic scatter fixed to the \KH\ values ($1.17$ and $0.28\,$ dex, respectively).
The resulting `fixed-slope' fit is
\begin{align}
\label{eq:fixed_slope}
    \log\left(\frac{\mbh}{10^9\,\Msun}\right) &= - \left(0.35 \pm 0.13\right) + 1.17\nonumber \\
    &\quad \times\log\left(\frac{M_\ast}{10^{11}\,\Msun}\right)\,\, .
\end{align}
The fit is shown as a solid line in Figure \ref{fig:M_vs_Mstar}, with the shaded region representing the associated uncertainty on the intercept, and is consistent with the \KH\ fit within the uncertainties.

We stress that our \mstar\ measurements may be susceptible to non-negligible light losses, as they are derived from spectra obtained with 2\,\arcsec\ fibers (see Section \ref{subsec:aperture_effects}). For our sample, we estimate that these light losses may lead the total host masses to be underestimated by $\Delta\log M_\ast\simeq 0.19$--0.77\,dex, with a median of 0.46\,dex (see Appendix \ref{app:aperture_corrections} for details). The corresponding median correction on $\log\,M_\ast$ is indicated by a horizontal arrow in Figure \ref{fig:M_vs_Mstar}.
These light losses (and aperture corrections) do not affect the \mbh\ estimates, since these are derived from pointlike emission and the SDSS spectra are PSF calibrated.
The dotted line in Figure \ref{fig:M_vs_Mstar} shows our fixed-slope fit to the \MM\ relation of the CL-AGN sample after scaling up $M_\ast$ based on the estimated light losses. The resulting fit is
\begin{align}
\label{eq:fixed_slope}
    \log\left(\frac{\mbh}{10^9\,\Msun}\right) &= - \left(0.87 \pm 0.13\right) + 1.17\nonumber \\
    &\quad \times\log\left(\frac{M_\ast}{10^{11}\,\Msun}\right)\,\, .
\end{align}

Figure \ref{fig:M_vs_Mstar} compares our CL-AGNs to samples of classical bulges and elliptical galaxies (except for \RV\, who reports total stellar masses). In this context, our noncorrected stellar masses, measured within the 2\arcsec\ SDSS fiber, may serve as a closer approximation to the bulge masses of the CL-AGN hosts, whereas the aperture-corrected values are intended to represent the total stellar mass of the galaxy.
A precise decomposition of the different host components for our CL-AGNs would require high-quality imaging of these systems in their new, dim state, which is beyond the scope of this work.

\subsubsection{Comparison to other samples}
\label{subsubsec:scaling_comparison}

Figures \ref{fig:M_sigma} and \ref{fig:M_vs_Mstar} show our CL-AGNs on the \Msig\ and \MM\ planes, compared with several AGN samples from the literature that have also been analyzed in the context of these scaling relations.

Our CL-AGNs are broadly consistent with other AGN samples occupying similar BH mass regimes \cite[e.g.,][]{Bennert21, Li23, Winkel25}. In particular, both \citet{Li23} and \citet{Winkel25} used RM-based BH mass estimates and found their samples to follow the same scaling relations as inactive galaxies, similar to what we find for our CL-AGNs.
This result suggests that our dimming CL-AGNs in their bright state, despite their strong variability, do not differ from typical AGNs in terms of the largely virialized structure of their BLRs, and in the reliability of their broad lines as probes of BH mass.
These results are consistent with previous studies that examined the \Msig\ relation in CL-AGNs and likewise found them to align with the general AGN population \citep[e.g.,][]{Yu20, Jin22, Yang25}. However, those works relied on lower-resolution SDSS spectroscopy ($R \simeq 2000$) for their \sigs\ measurements, whereas our analysis is based on higher-resolution X-shooter and GMOS data. In addition, the largest studies \citep{Jin22, Yang25} focused exclusively on ``turn-on'' events, while our work explores ``turn-off'' cases.

On the other hand, our CL-AGNs (along with the aforementioned samples) appear to have higher \mbh\ at fixed \sigs\ and \mstar\ compared to the BASS Sy1.9 \cite[][]{Koss22, MejiaRestrepo22} and \RV\ samples, respectively.
The BASS Sy1.9 and \RV\ samples consist of AGNs with host-dominated spectra, selected specifically to enable reliable measurements of \sigs\ and \mstar, respectively. These samples predominantly probe the more abundant population of lower-mass AGNs, with median $\log(\mbh/M_\odot)$ values of $7.0^{+0.7}_{-0.5}$ (BASS Sy1.9s) and $6.6\pm0.6$ (\RV).
In contrast, our CL-AGN subsample was drawn from the more luminous, optically selected SDSS DR16 quasars, which tend to populate the rarer, high-\mbh\ end of the AGN distribution, with median $\log(\mbh/M_\odot)$ values of $8.0^{+0.5}_{-0.4}$ and $7.8\pm0.6$, respectively (see the middle and right panels of Figure \ref{fig:sample_dr16}).\footnote{The trend seen in Figure~\ref{fig:sample_dr16}, whereby CL-AGNs probe slightly lower luminosities than the broader SDSS DR16 quasar population, reflects the tendency of CL-AGNs to have lower Eddington ratios on average; see \Z\ and references therein for a detailed discussion.}

These differences likely reflect several selection biases. In flux-limited surveys such as SDSS, the most luminous quasars---and by extension our CL-AGNs, selected among them---are preferentially detected in galaxies where the BH is unusually massive for the host's stellar mass, rather than in the intrinsically rarer, very massive galaxies that could host such BHs in the absence of scatter in the scaling relations (see \citealt{Lauer07} for discussion).
Conversely, the host-dominated selection of the \RV\ and BASS Sy1.9s samples biases them in the opposite direction, toward systems with comparatively low $\mbh / M_\ast$ ratios, since AGNs with less massive black holes are more likely to have spectra in which the stellar emission of the host galaxy dominates.

In addition to these selection effects, the \RV\ and BASS Sy1.9 samples may also suffer from systematic underestimation of \mbh, due to dust reddening. In such host-dominated systems, significant extinction in the nuclear regions can attenuate the broad-line emission while leaving the stellar continuum largely unaffected, leading to lower observed \Ha\ luminosities, and consequently, underestimated BH masses. \citet{MejiaRestrepo22} showed that in obscured sources, identified via high X-ray column densities, the \Ha\ emission is substantially weaker than in unobscured AGNs, producing \mbh\ underestimations of over 1\,dex for type 1.9 AGNs. Similarly, \citet{Caglar23} found that applying X-ray-based extinction corrections reduces (though not eliminates) the observed mass offset.
Our CL-AGNs, in contrast, appear largely unobscured, as indicated by the persistent detection of broad \mgii\ emission even in their dim states (see Section~\ref{subsec:nature_clagn}). Therefore, while dust-related \mbh\ discrepancies may affect certain samples (i.e., the BASS Sy1.9 and \RV\ ones shown in Figures \ref{fig:M_sigma} and \ref{fig:M_vs_Mstar}), they are not relevant for our CL-AGN sample.

Disentangling the combined impact of these various selection and extinction effects requires a careful statistical treatment, and, given the modest size of our CL-AGN sample, is beyond the scope of this work.
Nevertheless, our results indicate that luminous, optically selected SDSS quasars, including our CL-AGNs, are consistent with the \Msig\ and \MM\ relations of inactive galaxies. More importantly, we find no evidence that CL-AGNs deviate from other SDSS quasars in these scaling relations, supporting the interpretation that CL-AGNs represent a transient phase of otherwise normal AGN activity.


\begin{figure*}
    \centering
    \includegraphics[width=0.48\textwidth]{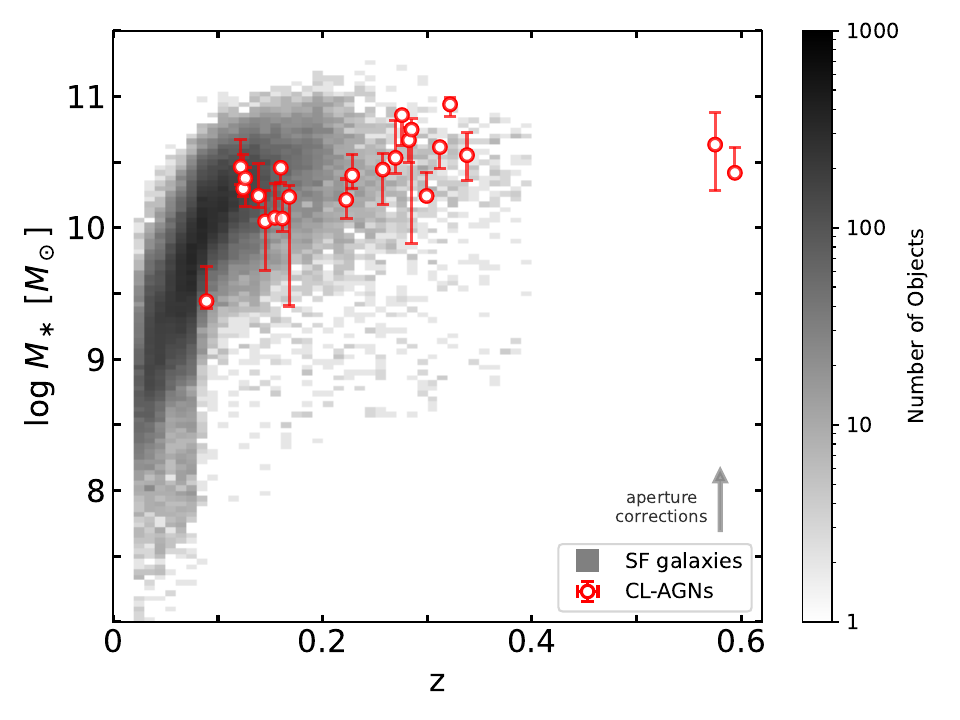} 
    \includegraphics[width=0.48\textwidth]{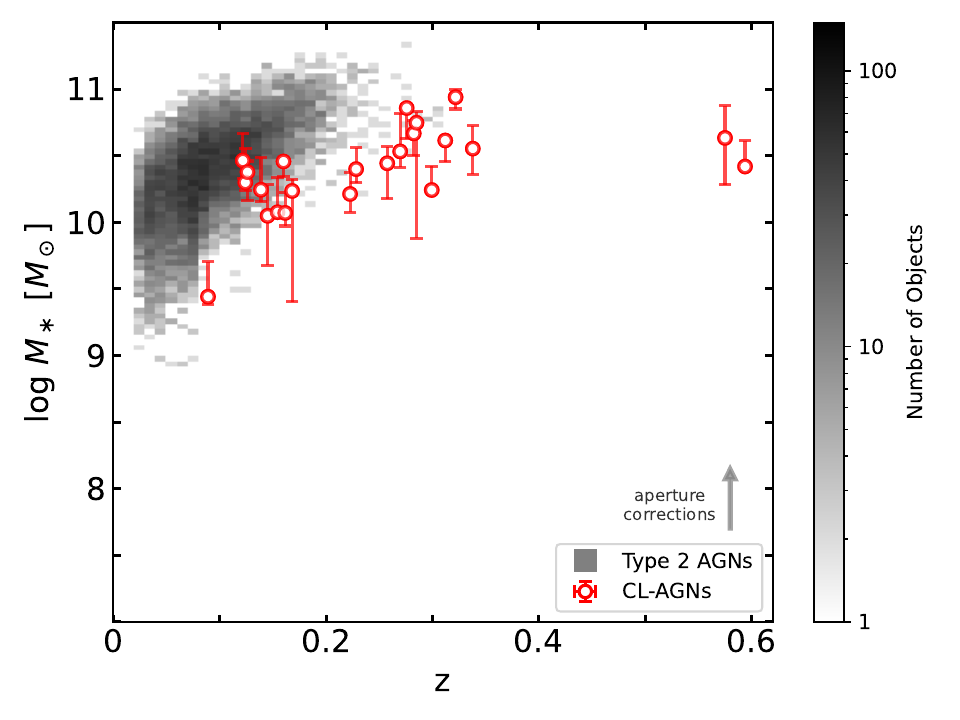}
    \caption{CL-AGNs (red symbols) along with SF galaxies (left panel) and type 2 AGN (right panel; gray regions) in the \mstar--$z$ plane. \mstar\ values for all samples were derived using \texttt{FIREFLY}: for CL-AGNs from their SDSS-V spectra, and for the comparison samples from the DR14 SDSS \texttt{FIREFLY} catalog \citep{Comparat17}. SF galaxy and type 2 AGN classifications were obtained from the SDSS DR7 \cite[][]{Abazajian09_DR7} MPA/JHU catalog \cite[][]{Kauffmann03, Brinchmann04}. The arrow marks the median aperture correction estimated for the CL-AGN \mstar\ values (0.46\,dex; see Appendix \ref{app:aperture_corrections}). The corresponding corrections for the comparison samples are smaller by only 0.08\,dex on average. Both panels show that dimming CL-AGNs, which were originally selected as bright quasars in SDSS and now observed in their faint states, probe lower-\mstar\ regimes at a given redshift than SDSS type 2 AGNs, thereby providing access to previously underrepresented hosts.}
    \label{fig:M_vs_z}
\end{figure*}

\begin{figure*}
    \centering
    \includegraphics[width=0.48\textwidth]{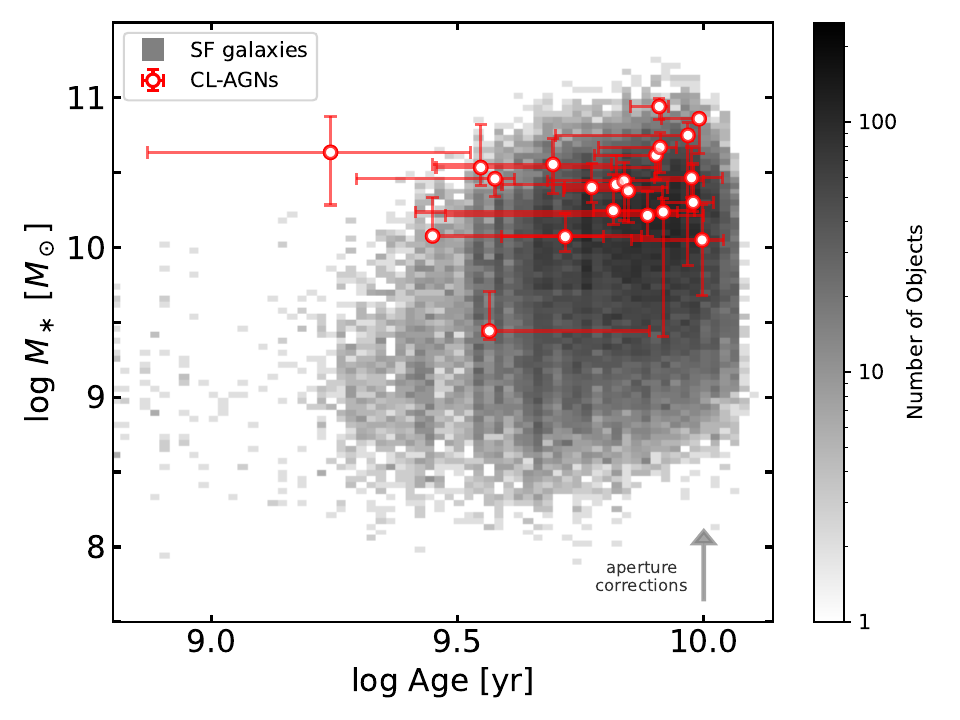}   
    \includegraphics[width=0.48\textwidth]{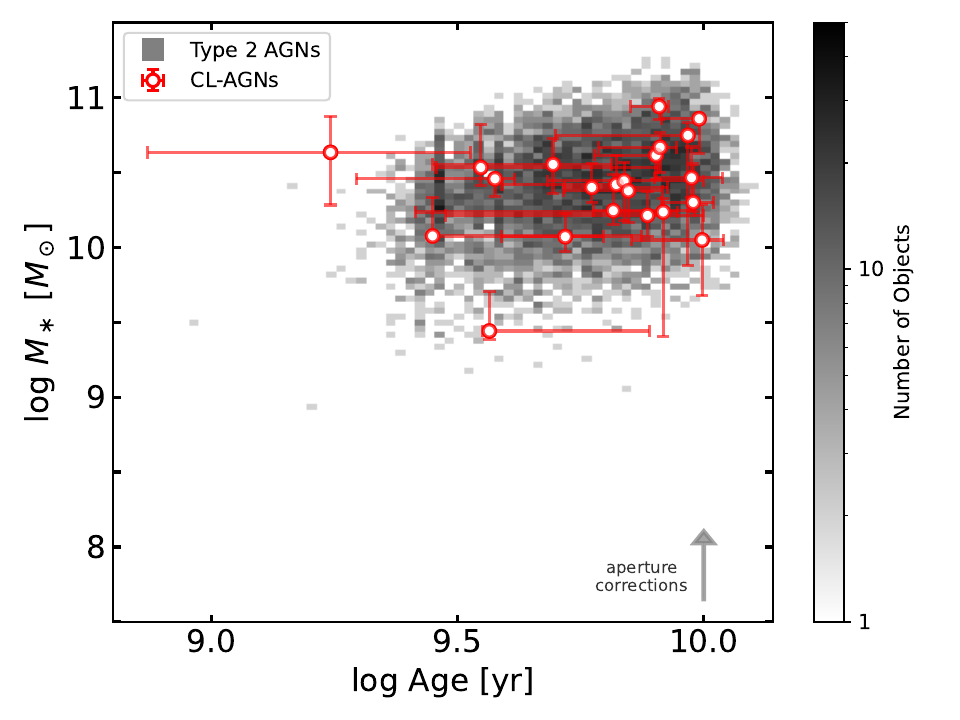}
    \caption{CL-AGNs (red symbols) along with SF galaxies (left panel) and type 2 AGN (right panel; gray regions) in the \mstar--$t_{\rm \ast, M}$ plane. \mstar\ and mass-weighted stellar ages for all samples were derived using \texttt{FIREFLY}: for CL-AGNs from their SDSS-V spectra, and for the comparison samples from the DR14 SDSS \texttt{FIREFLY} catalog \citep{Comparat17}. SF galaxy and type 2 AGN classifications were obtained from the SDSS DR7 \cite[][]{Abazajian09_DR7} MPA/JHU catalog \cite[][]{Kauffmann03, Brinchmann04}. The arrow marks the median aperture correction estimated for the CL-AGN \mstar\ values (0.46\,dex; see Appendix \ref{app:aperture_corrections}). The corresponding corrections for the comparison samples are smaller by only 0.08\,dex on average. Our CL-AGNs align with type 2 AGN hosts in the \mstar--$t_{\rm \ast, M}$ plane.}
    \label{fig:M_vs_age}
\end{figure*}

\subsection{Stellar population properties of CL-AGNs}
\label{subsec:host_results}

This section compares our CL-AGN sample to SDSS DR7 type 2 AGNs and SF galaxies. In the figures presented here, when relevant, aperture corrections are indicated by arrows at the bottom right (see Appendix~\ref{app:aperture_corrections} for details).
Although the SDSS-V spectra for our CL-AGNs were obtained with 2\,\arcsec\ fibers and the SDSS DR7 reference spectra with 3\,\arcsec\ fibers, the DR7 galaxies are at somewhat lower redshifts (median $z\sim0.08$ versus 0.23 for our sample). As a result, as shown in Appendix~\ref{app:aperture_corrections}, the resulting aperture corrections for the CL-AGNs are larger by only $\sim0.08$\,dex on average. Thus, the aperture effects impact both the comparison and CL-AGN samples in a similar way and do not affect the conclusions of this section.

\begin{figure*}
    \centering
    \includegraphics[width=0.48\textwidth]{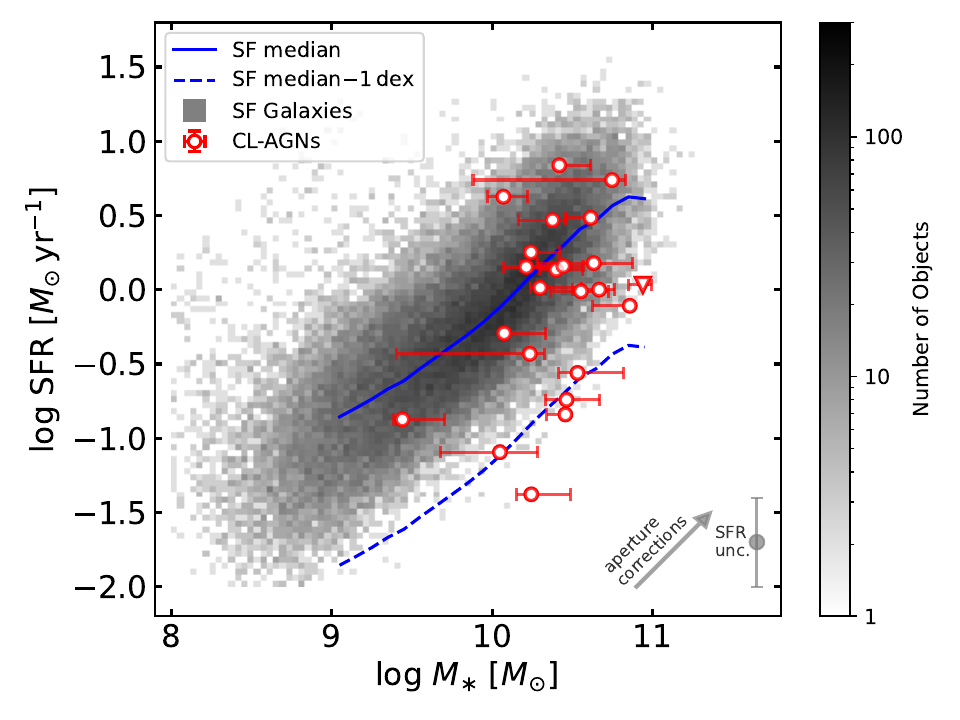}   
    \includegraphics[width=0.48\textwidth]{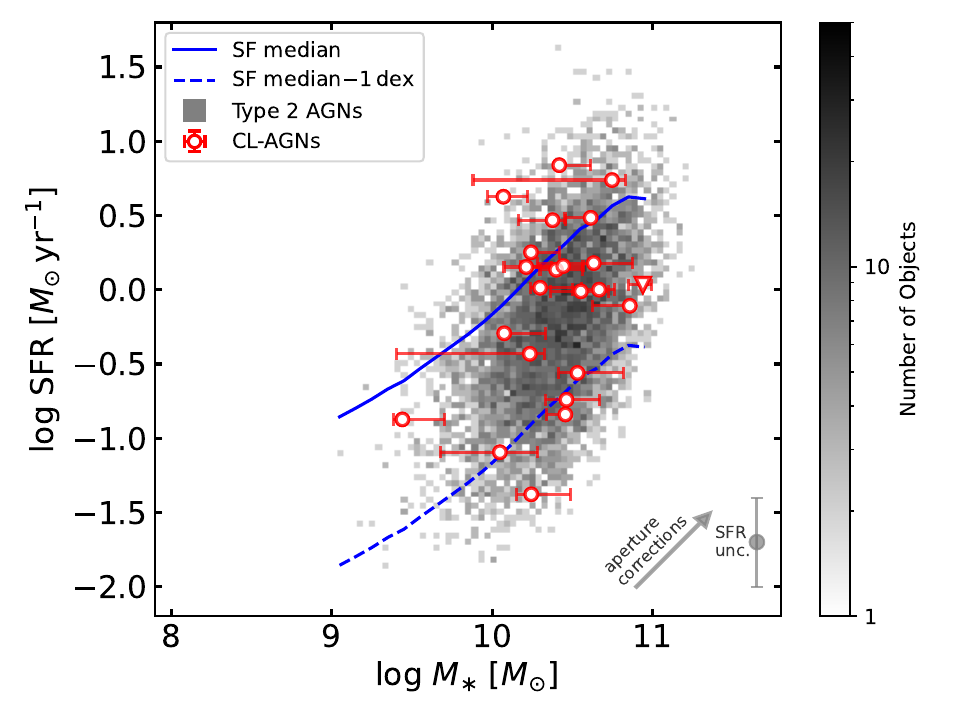}
    \caption{CL-AGNs (red circles; inverted triangles indicate upper limits) compared with SF galaxies (left panel) and type 2 AGNs (right panel; gray regions) in the SFR-\mstar\ plane. For both panels, the solid blue line marks the moving median of the SF population, and the dashed blue line 1\,dex below, roughly marks the boundary of the quenched-galaxy regime. SFRs for all samples were derived using the \oii-based estimator of \ZH\ (Section~\ref{subsubsec:sfr}), while stellar masses were obtained from \texttt{FIREFLY} fits. For CL-AGNs, both quantities were measured directly from our SDSS-V spectra; for the comparison samples, emission-line measurements and classifications were taken from the SDSS DR7 MPA/JHU catalog \citep{Kauffmann03,Brinchmann04}, and \mstar\ values from the SDSS DR14 \texttt{FIREFLY} catalog \citep{Comparat17}. The nominal uncertainty of the \oii-based SFR estimator (0.3\,dex) is shown in the lower right of each panel. The arrow denotes the median aperture correction applied to the CL-AGN \mstar\ and SFR values (0.46 and 0.45\,dex, respectively; see Appendix~\ref{app:aperture_corrections}). The corresponding corrections for the comparison samples are smaller by only 0.08\,dex on average. Our CL-AGNs occupy the same region of the SFR--\mstar\ plane as type 2 AGN hosts.}
    \label{fig:SFR_vs_M}
\end{figure*}

Figure \ref{fig:M_vs_z} shows our CL-AGN sample in the \mstar--$z$ plane, compared with large samples of SF galaxies and type 2 AGNs, both drawn from SDSS DR7 (left and right panels, respectively). 
Our CL-AGNs span $\log(\,\mstar/M_\odot)=\,$9.4--10.9, with a median of 10.4.
In both panels, CL-AGNs occupy systematically lower \mstar\ at a given $z$ relative to the SDSS DR7 SF galaxy and type 2 AGN populations. 
This offset may reflect differences in the galaxy- and quasar-selection criteria in SDSS rather than an intrinsic distinction between the populations: dimming CL-AGNs, which were originally selected as bright quasars in SDSS and now observed in their faint states, probe lower-\mstar\ regimes at a given redshift than SDSS type 2 AGNs, allowing access to previously underrepresented hosts.
 
\begin{figure*}
    \centering
    \includegraphics[width=0.48\textwidth]{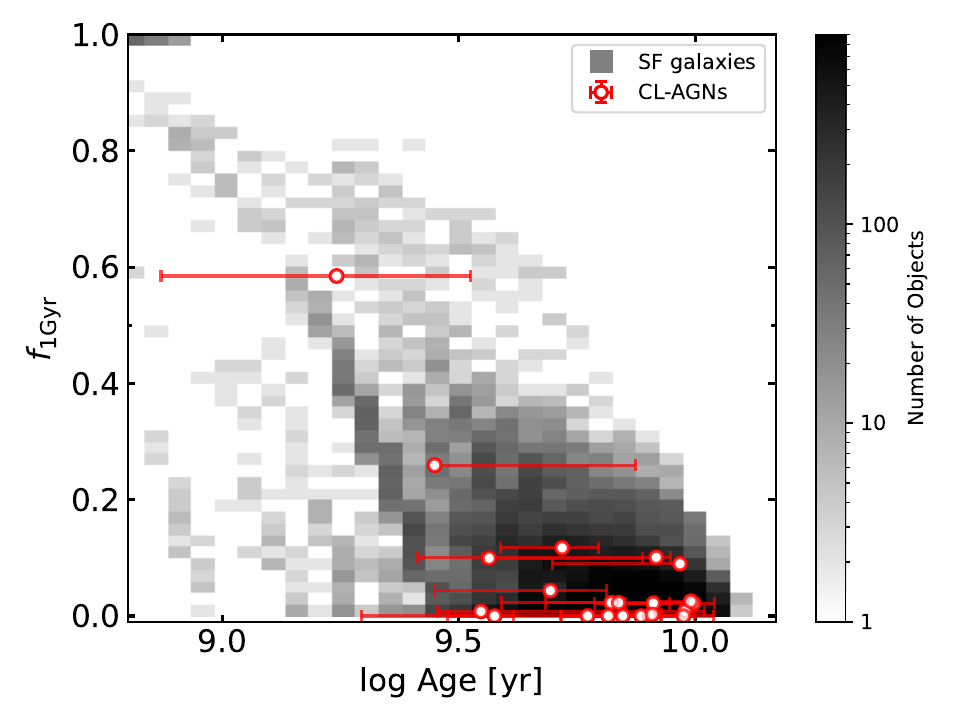}   
    \includegraphics[width=0.48\textwidth]{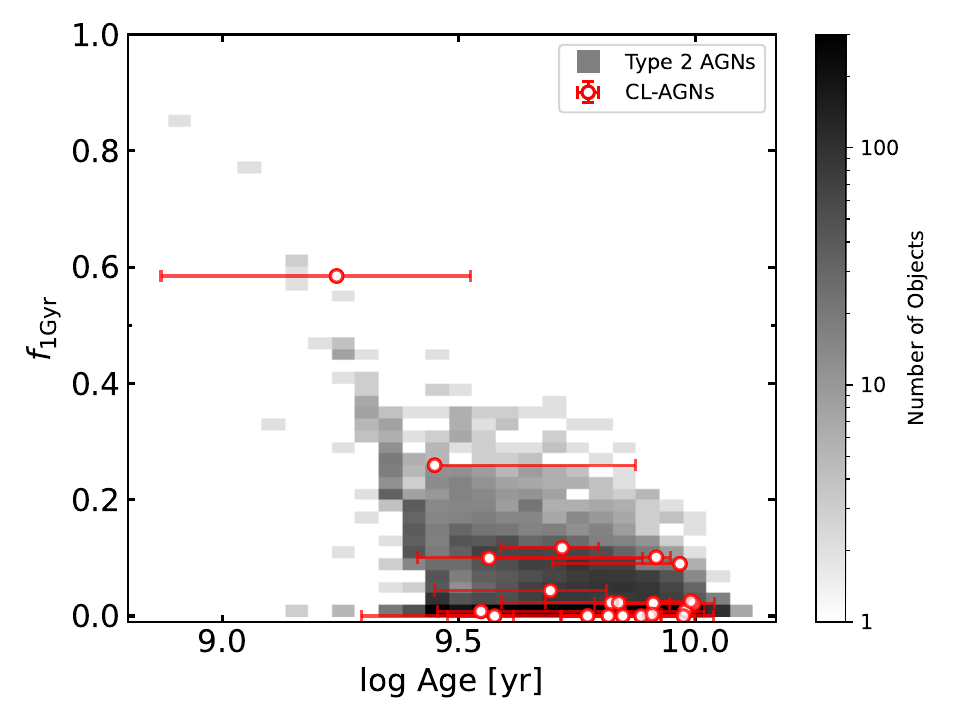}
    \caption{CL-AGNs (red symbols) along with SF galaxies (left panel) and type 2 AGN (right panel; gray regions) in the $f_{\rm 1Gyr}$--age plane. $f_{\rm 1Gyr}$ and mass-weighted stellar ages for all samples were derived using \texttt{FIREFLY}: for CL-AGNs from their SDSS-V spectra, and for the comparison samples from the DR14 SDSS \texttt{FIREFLY} catalog \citep{Comparat17}. SF galaxy and type 2 AGN classifications were obtained from the SDSS DR7 \cite[][]{Abazajian09_DR7} MPA/JHU catalog \cite[][]{Kauffmann03, Brinchmann04}. Our CL-AGNs exhibit similar young-star fractions to those of type 2 AGN hosts, with only a few sources showing evidence of recent starburst (i.e., $f_{\rm 1Gyr}>0.1$). One source lies above the $f_{\rm 1Gyr}$ distribution of the type 2 AGN sample, but it is also one of our higher-redshift objects ($z=0.575$), outside the comparison redshift range, making it difficult to draw firm conclusions from this case.}
    \label{fig:fractions}
\end{figure*}

The left panel of Figure \ref{fig:M_vs_age} shows the CL-AGN sample in the \mstar--$t_{\rm \ast, M}$ plane, compared with SDSS DR7 SF galaxies.
The mass-weighted stellar ages span 1.7--9.9\,Gyr, with a median of 6.9\,Gyr.
Our CL-AGNs generally lie within the SF distribution, but preferentially sample its higher-mass and older-age end. 
The right panel of Figure \ref{fig:M_vs_age}, which compares the CL-AGNs with typical type 2 AGNs on the same plane, reveals that both samples occupy a similar range in both \mstar\ and $t_{\rm \ast, M}$. 
This behavior suggests that the trend seen in the left panel reflects the general tendency of AGNs in the local Universe to reside in galaxies with somewhat higher stellar masses and older stellar ages, rather than indicating that CL-AGNs are a distinct population (among AGNs) in this regard. 
This is further supported by KS tests comparing the CL-AGN and type 2 AGN samples, finding no statistically significant difference in either \mstar\ or $t_{\rm \ast, M}$ ($p_{\rm KS}=0.75$ and 0.21, respectively).

Figure \ref{fig:SFR_vs_M} shows our CL-AGN sample in the SFR--\mstar\ plane.
For our sample, the SFRs range from 0.04 to $7\,M_\odot\,\rm{yr}^{-1}$ with a median value of $1\,M_\odot\,\rm{yr}^{-1}$.
For context, we indicate the moving median of the SF galaxy population based on our adopted SFR \oii\ prescription, as well as a line offset by $-1\,$ dex, below which systems are roughly considered quiescent \cite[e.g.,][]{Donnari19,Morselli19}.
The left panel, comparing CL-AGNs with SF galaxies, shows that CL-AGNs preferentially occupy the high-mass end of the distribution, with some sources lying below the distribution of SF galaxies.
The right panel, comparing CL-AGNs with type 2 AGNs,\footnote{Our type 2 AGN panel does not show the most massive quenched galaxies. This reflects a disagreement between \oii-based SFRs and the D4000-based SFRs from the MPA/JHU catalog as noted by \ZH, where the former tends to assign higher SFRs to objects classified as quenched in the MPA/JHU catalog.} shows that CL-AGNs are consistent with the type 2 AGN population, and that the difference seen in the left panel simply reflects the broader distinction between AGNs and SF galaxies. This is further supported by a KS test comparing the SFR distributions of the CL-AGN and type 2 AGN samples, yielding no statistically significant difference between the two populations ($p_\mathrm{KS}=0.17$).

Finally, Figure \ref{fig:fractions} shows the mass-weighted fraction of young ($<1$\,Gyr) stars, $f_{\rm 1Gyr}$, as a function of $t_{\ast, M}$ for our CL-AGNs, compared with SF galaxies and type 2 AGNs. We find four systems with evidence of recent star formation episodes (i.e., $f_{\rm 1Gyr}>0.1$). Three of these lie within the distribution of type 2 AGNs, with $f_{\rm 1Gyr}=0.1$--0.25.
One object (J1104+0118), with $f_{\rm 1Gyr}=0.6$, lies outside the AGN distribution; however, this source is at $z=0.575$, outside the redshift range of the comparison samples (SDSS DR7 AGNs at $z\lesssim0.3$), making it hard to draw conclusions from this case.

Overall, we find no evidence of our CL-AGN sample deviating from the general AGN population in their stellar population properties.

\bigskip\bigskip

\section{Conclusions}
\label{sec:conclusions}

We obtained and analyzed medium-resolution spectroscopy of 23 dimming CL-AGNs identified within SDSS-V data. The spectroscopy, obtained using VLT/X-shooter and Gemini-N/GMOS, was aimed to further study the nature of the CL transitions and the host galaxies of these CL-AGNs. 
These data, together with the SDSS-V spectra, allowed a robust measurement of the stellar velocity dispersions (\sigs) of the hosts, as well as their stellar masses (\mstar), SFRs, and stellar population ages ($t_{\rm \ast, M}$).
Our main findings are:

\begin{enumerate}
    \item The detection of broad \mgii\ emission in the dim states of 13 out of 16 sources with clean and reliable spectral coverage of both \ha\ and \mgii\ indicates that these sources are not obscured, but are instead observed in a low-accretion state (Section \ref{subsec:nature_clagn}, Figure \ref{fig:mgii_ha_hist}).

    \item By adopting a virial scaling factor based on RM AGN studies, we find that our CL-AGNs align with the \Msig\ relation of inactive galaxies and other AGN populations in a similar mass regime, in terms of normalization. This result suggests that, despite their extreme variability, our dimming CL-AGNs in their bright state resemble typical AGNs in the largely virialized structure of their BLRs and in the reliability of their broad lines as tracers of BH mass. Alternatively, assuming that CL-AGNs follow the \Msig\ relation of inactive galaxies, we infer the corresponding virial scaling factors: $f_{\mathrm{FWHM}}=1.1\pm0.3$ when using FWHM, and $f_\sigma=3.6^{+1.1}_{-0.9}$ when using the line dispersion (Section \ref{subsubsec:msig}, Figures \ref{fig:M_sigma} and \ref{fig:f_vs_sigma}).

    \item If the slope of the \Msig\ relation of CL-AGNs is allowed to vary, the best-fitting relation favors a shallower slope ($2.95 \pm 0.81$) than the canonical value derived for inactive galaxies. While the statistical significance of this difference is modest ($\sim\,1.7\sigma$), the inferred trend is qualitatively similar to earlier findings for samples of RM AGNs (Section \ref{subsubsec:msig}, Figure \ref{fig:M_sigma}).
    
    \item Our CL-AGNs are consistent with the \MM\ relation of inactive galaxies, with a median BH-to-stellar mass fraction of 0.38\,\% ($16^{\rm th}$--$84^{\rm th}$ percentiles: 0.15 to 0.88\,\%; Section \ref{subsubsec:mbh_mstar}, Figure \ref{fig:M_vs_Mstar}).
    
    \item The hosts of our CL-AGNs are consistent with those of type 2 AGNs drawn from SDSS DR7, in terms of SFR, stellar age, stellar mass, and the fraction of young stars, with only a few sources showing evidence of recent starburst (Section \ref{subsec:host_results}, Figures \ref{fig:M_vs_age}, \ref{fig:SFR_vs_M}, and \ref{fig:fractions}).
    
\end{enumerate}

Our observations and analysis reinforce the growing consensus that CL-AGN variability is driven by intrinsic changes in the accretion flow rather than by variable obscuration. 
More importantly, the host galaxies of CL-AGNs show no significant differences from those of the general AGN population across a range of physical properties. This behavior suggests that CL-AGNs do not reside in distinct environments, but instead represent typical AGNs caught during episodes of unstable accretion. This result offers a valuable observational opportunity, which our study pursued for this modest sample: CL-AGNs provide access to both AGN-dominated and host-dominated states within the same systems, making them well-suited for studying the connection between SMBHs and their host galaxies.

To further test this interpretation, future work should examine CL-AGNs within the broader context of AGN variability, assessing whether they represent the extreme tail of ordinary stochastic variations or a distinct physical process of accretion-state change. Such studies will clarify whether CL-AGNs truly trace the same underlying population as typical AGNs, or instead form a distinct subclass within the AGN population.

In parallel, a more definitive assessment of the reliability of CL-AGN mass estimates and the degree of BLR virialization in these systems will require a well-defined sample including both dimming and brightening CL-AGNs, with BH mass estimates derived from both their bright and dim spectra. Such a sample will enable a systematic search for trends between accretion state and deviations from standard virial expectations, offering a crucial test of the robustness of current mass-scaling relations.


\begin{acknowledgments}

We thank the anonymous referee for their insightful and constructive comments, which helped us improve the paper.
We acknowledge support from:
the European Research Council (ERC) under the European Union's Horizon 2020 research and innovation program (grant agreement No. 950533; G.Z., B.T.); 
the Israel Science Foundation (grant No. 1849/19; G.Z., B.T.);
FONDECYT Regular \#1230345 (C.R.), \#1231718 (R.J.A.), and \#1200495 (F.E.B.); 
FONDECYT Iniciaci\'on \#11241477 (L.H.G.);
ANID BASAL project FB210003 (R.J.A., F.E.B., C.R.);
ANID Millennium Science Initiative ICN12\_009 (F.E.B., L.H.G.);
H.M.H.T. acknowledges financial support from DGAPA-PAPIIT project AG101725 and CONAHCYT project CF-2023-G-1052;
SNSF Consolidator grant F01$-$13252 (C.R.);
W.N.B. acknowledges funding from NSF grant AST-2407089.
This research was supported by the Excellence Cluster ORIGINS and by the Munich Institute for Astro-, Particle and BioPhysics (MIAPbP), which are funded by the Deutsche Forschungsgemeinschaft (DFG, German Research Foundation) under Germany's Excellence Strategy - EXC 2094 - 390783311.
B.T. acknowledges the hospitality of the Instituto de Estudios Astrof\'isicos at Universidad Diego Portales, the Instituto de Astrof\'isica at Pontificia Universidad Cat\'olica de Chile, and the Institut d'Astrophysique de Paris, where parts of this study have been carried out.

Funding for the Sloan Digital Sky Survey V has been provided by the Alfred P. Sloan Foundation, the Heising-Simons Foundation, the National Science Foundation, and the Participating Institutions. SDSS acknowledges support and resources from the Center for High-Performance Computing at the University of Utah. SDSS telescopes are located at Apache Point Observatory, funded by the Astrophysical Research Consortium and operated by New Mexico State University, and at Las Campanas Observatory, operated by the Carnegie Institution for Science. The SDSS web site is \url{www.sdss.org}.
SDSS is managed by the Astrophysical Research Consortium for the Participating Institutions of the SDSS Collaboration, including Caltech, The Carnegie Institution for Science, Chilean National Time Allocation Committee (CNTAC) ratified researchers, The Flatiron Institute, the Gotham Participation Group, Harvard University, Heidelberg University, The Johns Hopkins University, L'Ecole polytechnique f\'{e}d\'{e}rale de Lausanne (EPFL), Leibniz-Institut f\"{u}r Astrophysik Potsdam (AIP), Max-Planck-Institut f\"{u}r Astronomie (MPIA Heidelberg), Max-Planck-Institut f\"{u}r Extraterrestrische Physik (MPE), Nanjing University, National Astronomical Observatories of China (NAOC), New Mexico State University, The Ohio State University, Pennsylvania State University, Smithsonian Astrophysical Observatory, Space Telescope Science Institute (STScI), the Stellar Astrophysics Participation Group, Universidad Nacional Aut\'{o}noma de M\'{e}xico, University of Arizona, University of Colorado Boulder, University of Illinois at Urbana-Champaign, University of Toronto, University of Utah, University of Virginia, Yale University, and Yunnan University.

\end{acknowledgments}

\facilities{Sloan (SDSS and BOSS), VLT:Kueyen (X-shooter), Gemini:Gillett (GMOS).}

\software{{\tt astropy} \citep{Astropy13,Astropy18,Astropy22},
\texttt{DRAGONS} \citep{DRAGONS},
{\tt FIREFLY} \citep{Firefly},
{\tt Matplotlib} \citep{Matplotlib07}, {\tt Molecfit} \citep{Molecfit1, Molecfit2},
{\tt NumPy} \citep{NumPy20},
\texttt{pandas} \citep{Mckinney10, pandas},
\texttt{pPXF} \citep{Cappellari23}, \texttt{PyMC} \citep{PyMC},
{\tt PyQSOFit} \citep{QSOFit, Shen19}, 
{\tt SciPy} \citep{SciPy20}. 
}

\clearpage
\bibliography{CLAGN_hosts_bib}{}
\bibliographystyle{aasjournal}

\begin{appendix}
\section{Estimating the Effective Spectral Resolution}
\label{app:resolution}

To accurately measure \sigs\ (Section \ref{subsubsec:measuring_sigma}), one must correctly estimate the effective spectral resolution ($R$) of each spectrum.
When the observed object is slit-filling (i.e., the extent of the observed source is larger than the slit width), we take the instrumental resolution to be the nominal one at each slit (Sections \ref{subsec:vlt}, \ref{subsec:gemini}).
For X-shooter, this choice is supported by the work of \citet{Gonneau20}, who demonstrated that the actual spectral resolution is relatively constant with wavelength and closely matches the nominal values.
If, however, the source is not slit-filling, the spectral resolution improves over the nominal value. This is relevant only for our X-shooter data, where the larger slit (1.5\arcsec/1.6\arcsec\ vs. 0.5\arcsec\ for GMOS) is not always filled.

Thus, it was important to correctly determine the source's extent on the detector for each observation.
To accomplish this task, for each of our observations, we choose a wavelength range in the UVB X-shooter arm that is free of strong absorption or emission features, and perform a Gaussian fit to the spatial profile in the corresponding 2D frame across this range. 
The FWHM of the fitted Gaussian provides an estimate of the source's extent on the detector at the mean wavelength of the relevant range, $\lambda_0$ (resulting in $\simeq0.85$--2.2\,\arcsec).
To account for the dependence of seeing ($\Theta$) on wavelength, we extrapolate the source's extent across all wavelengths using the relation \cite[e.g.,][]{SR90}

\begin{equation}
    \Theta\left(\lambda\right) = \Theta\left(\lambda_0\right)\left(\frac{\lambda}{\lambda_0}\right)^{-0.2},
\end{equation}
while limiting the source's extent on the detector to be no larger than the slit width for each arm.
Finally, we use the nominal sampling rate of each X-shooter arm and slit width, along with the pixel scale ($0.2\,\text{\AA} \,\rm{pix}^{-1}$ for X-shooter), and assume that the spectral resolution improves linearly with decreasing slit-filling fraction, to convert the source's extent on the detector in each observation at each wavelength, to the corresponding instrumental resolution at each pixel.

Since many of our objects were observed across two or three visits, with potentially different seeing conditions, we matched the resolution across epochs for each object.
Specifically, for each object, we identified the epoch with the worst seeing, and convolved the spectra from the other epochs with a variable-width Gaussian in order to match this baseline resolution.
We coadded all the epochs for each object, weighting them by the inverse variance.
The resulting spectral resolution at 3950\,\AA\ (i.e., around the Ca~\textsc{ii} H+K absorption feature) for each object is listed in Table \ref{tab:obs}. For X-shooter, we obtained $R$ in the range of 3200--5000, while for GMOS, in the range of 3610--4270.

\section{Aperture Corrections}
\label{app:aperture_corrections}

\begin{figure*}
    \centering
    \includegraphics[width=\textwidth]{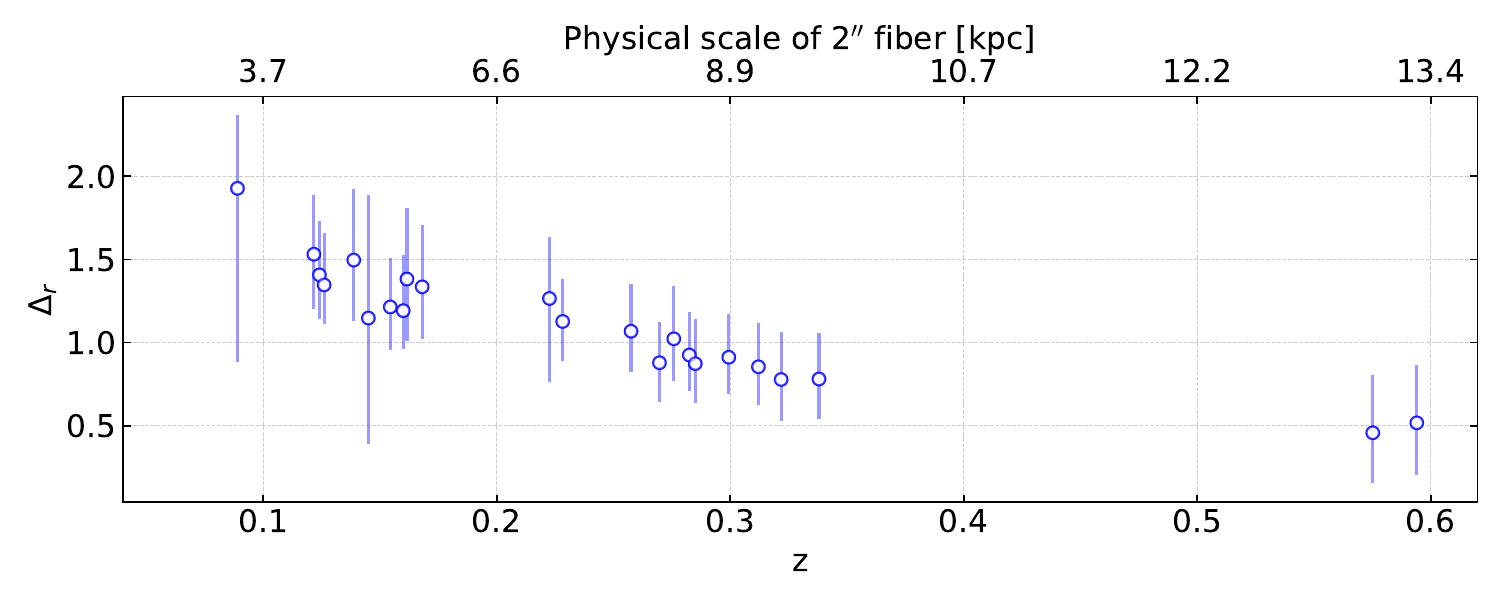}   
    \caption{Median offsets between the synthetic and \texttt{cmodel} $r$-band magnitudes ($\Delta_r$) for our CL-AGN sample, shown as a function of redshift. The points mark the median offsets for each source, and the error bars indicate the $16^{\rm th}$--$84^{\rm th}$ percentile range of the comparison sample. The upper axis shows the physical scale of a 2\arcsec\ SDSS fiber corresponding to each redshift.}
    \label{fig:aperture}
\end{figure*}

In this work, some of the quantities derived from the spectra depend on the total light of the host galaxy (i.e., stellar mass and SFR; see Sections \ref{subsec:ssp} and \ref{subsubsec:sfr}).
Since the dim-state spectra are obtained with an SDSS 2\arcsec\ fiber, and the flux calibration of these SDSS-V spectra is anchored to PSF magnitudes using standard stars in each field, the spectra can underestimate the total light emitted by the host galaxy. We must therefore account for the missing light outside the fiber. This is challenging because dim-state imaging is not available.

To estimate the aperture losses, we computed synthetic $r$-band photometry from the dim-state SDSS-V spectra of our CL-AGNs. For each source, we then randomly drew 1000 comparison galaxies (spectroscopic CLASS = `GALAXY') from the BOSS DR16 catalog. The comparison set was selected to lie within a redshift bin corresponding to $\pm10\%$ in the physical scale of a 2\arcsec\ fiber, and within $\pm0.1$ in synthetic $r$-band magnitude.
For each source, we measured the median offset between the synthetic $r$-band magnitude (derived from their SDSS DR16 spectra) and the \texttt{cmodel} $r$-band magnitude (i.e., the total galaxy light, as reported in the SDSS DR16 photometric database). 
This offset provides a reproducible estimate of the host light missed by the fiber, and does not require re-imaging our CL-AGNs in their dim state.

The resulting median offsets, along with their $16^{\rm th}$ and $84^{\rm th}$ percentiles, are shown as a function of redshift in Figure \ref{fig:aperture}. This method produced aperture corrections in the range of $\Delta_r = 0.46$--1.9\,mag, with a median of 1.2\,mag.

For our comparison sample of SF galaxies and type 2 AGNs, based on the cross-match between the \citet{Comparat17} \texttt{FIREFLY} catalog and the MPA/JHU catalog \citep{Kauffmann03,Brinchmann04}, we also calculated aperture corrections using the same method. The MPA/JHU spectra were obtained with 3\arcsec\ fibers, and the resulting median correction is $\Delta_r = 1$\,mag, corresponding to a difference of only $\sim0.08$\,dex relative to our sample, which does not significantly affect our comparative analysis.

\section{CL-AGN sample spectra}
\label{app:spectra_plots}

Figure \ref{fig:all_spectra} presents the legacy SDSS bright-state spectra, SDSS-V dim-state spectra, and X-shooter/GMOS spectra for our CL-AGN sample.

\begin{figure}
\centering
\begin{tabular}{ccc}
\includegraphics[width=0.3\textwidth]{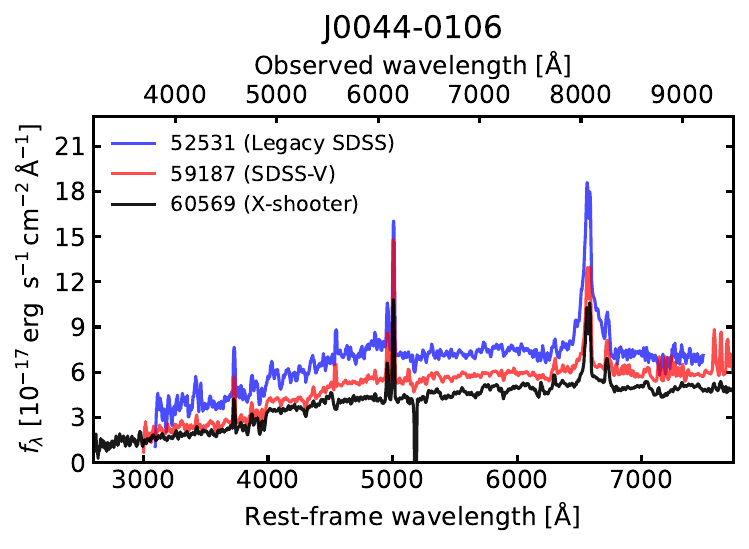} &
\includegraphics[width=0.3\textwidth]{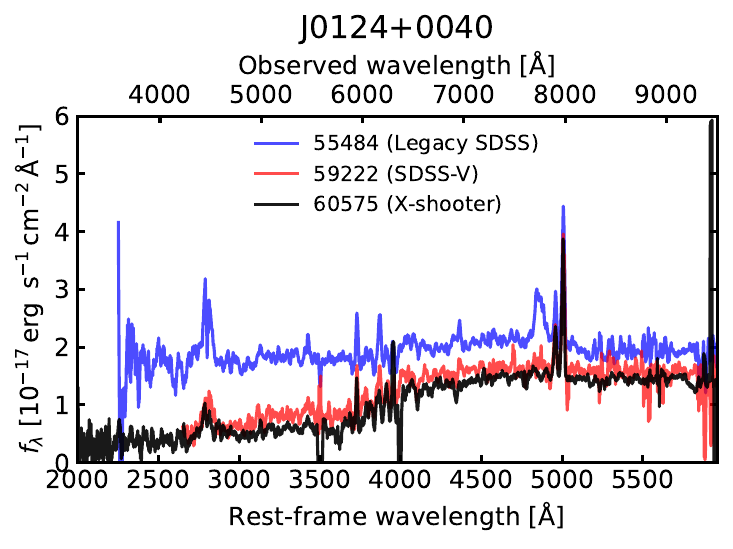} &
\includegraphics[width=0.3\textwidth]{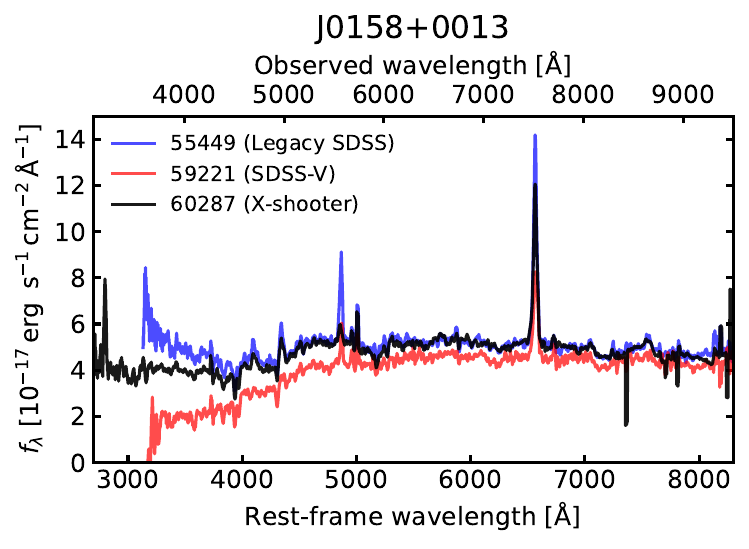} \\

\includegraphics[width=0.3\textwidth]{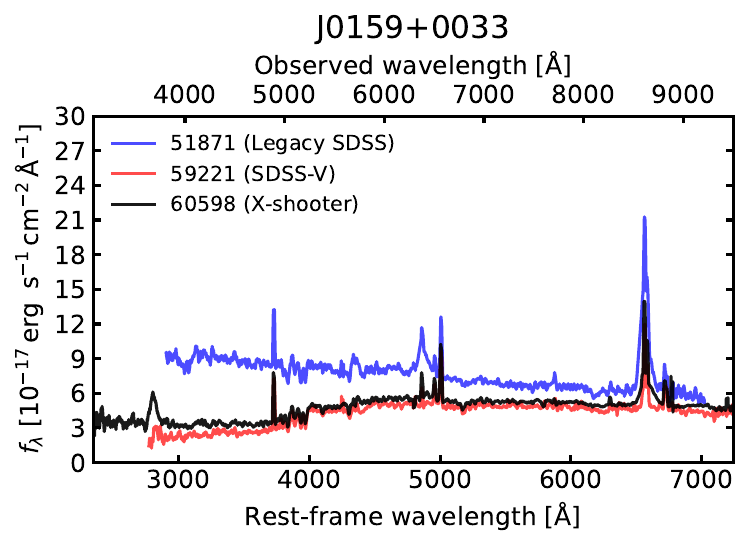} &
\includegraphics[width=0.3\textwidth]{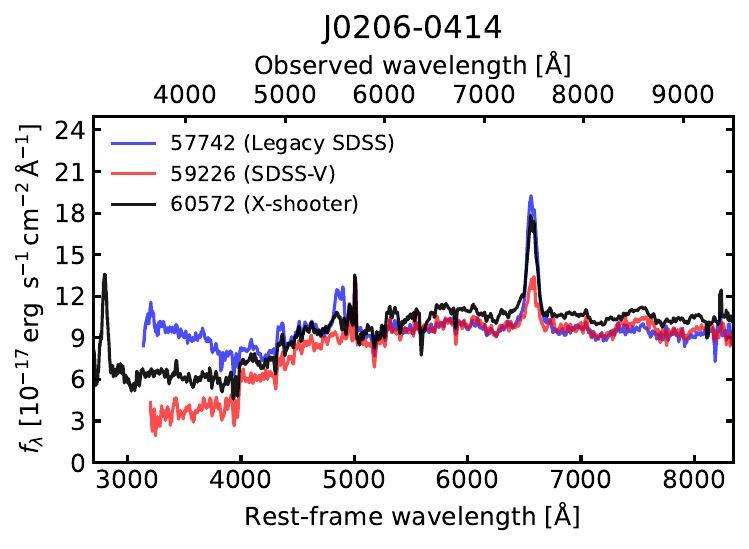} &
\includegraphics[width=0.3\textwidth]{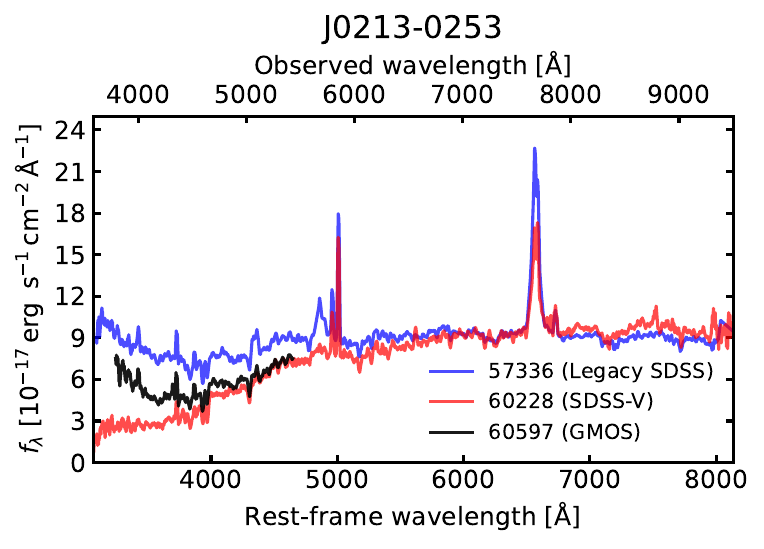} \\

\includegraphics[width=0.3\textwidth]{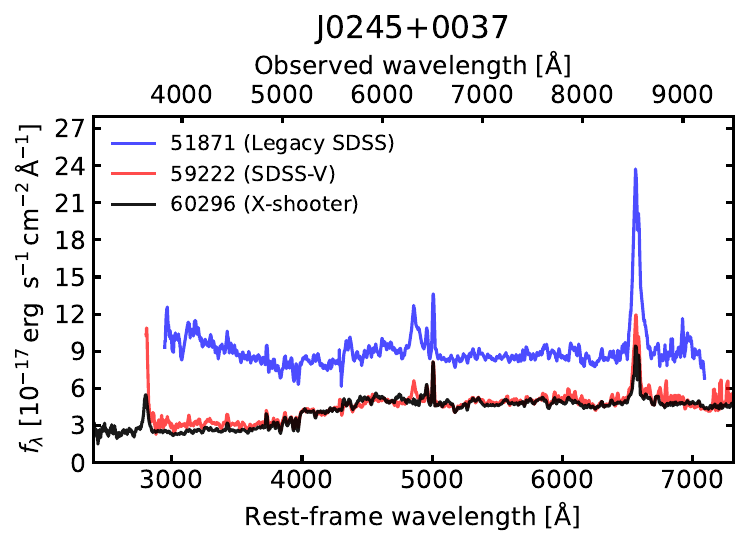} &
\includegraphics[width=0.3\textwidth]{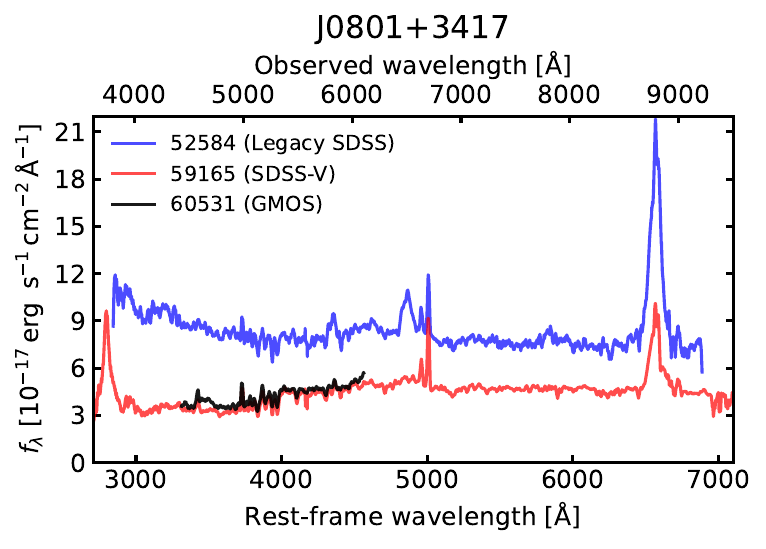} &
\includegraphics[width=0.3\textwidth]{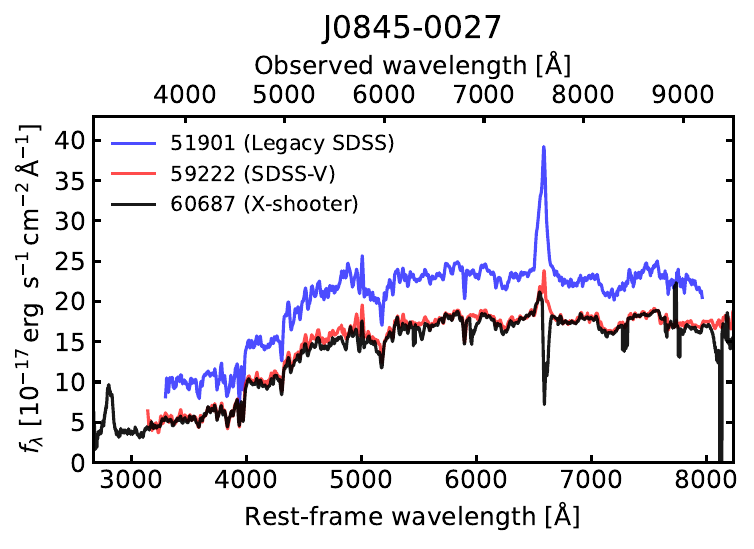} \\

\includegraphics[width=0.3\textwidth]{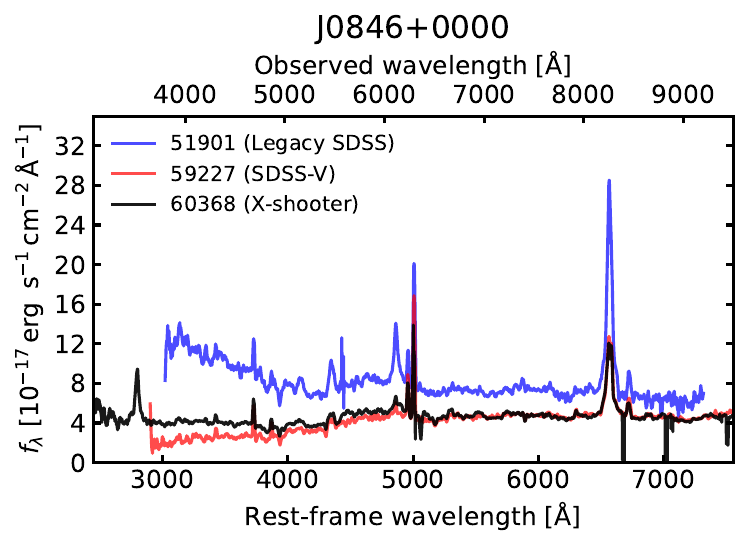} &
\includegraphics[width=0.3\textwidth]{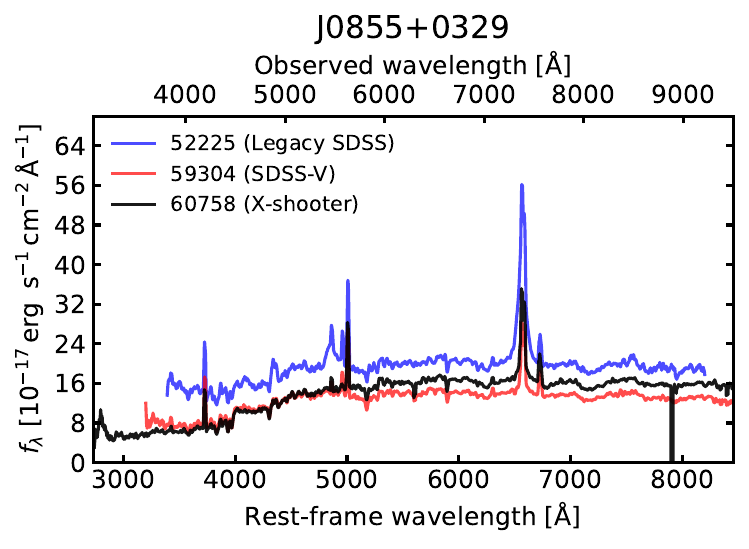} &
\includegraphics[width=0.3\textwidth]{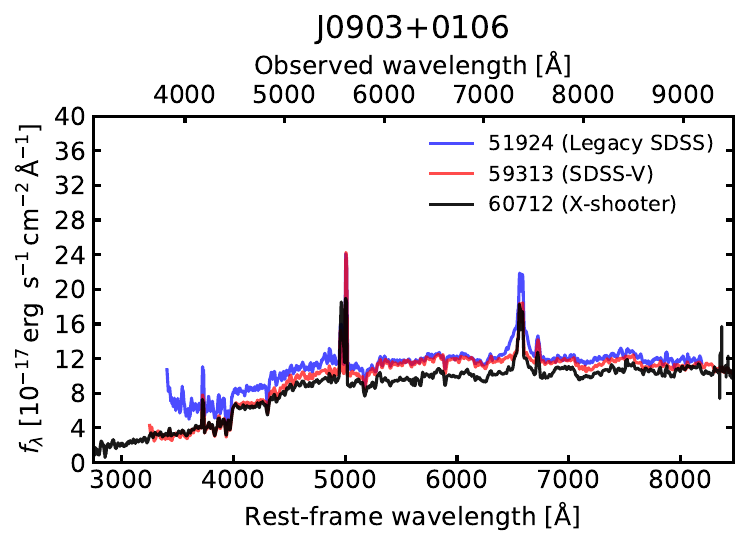} \\

\includegraphics[width=0.3\textwidth]{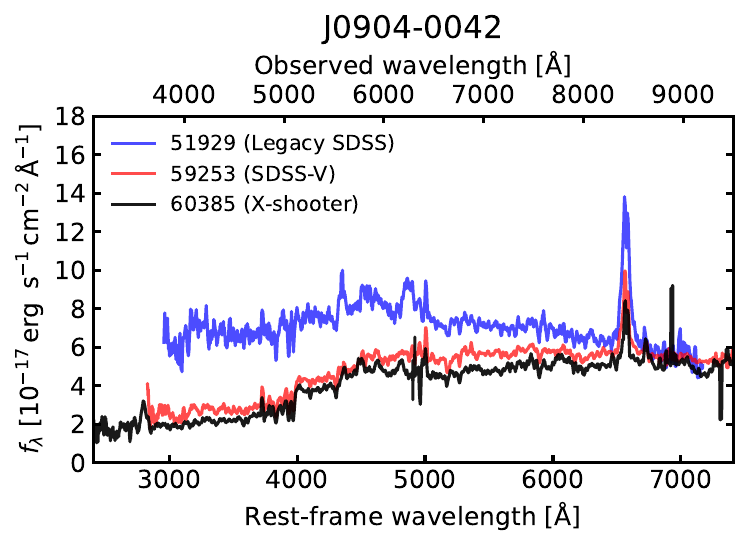} &
\includegraphics[width=0.3\textwidth]{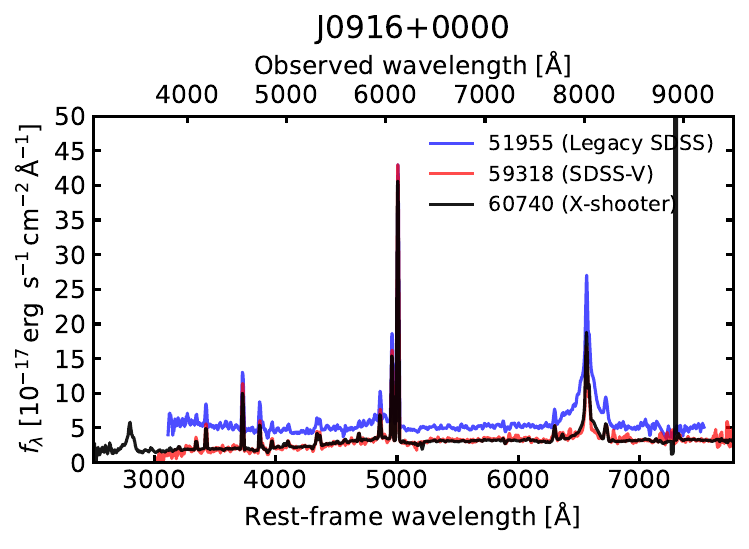} &
\includegraphics[width=0.3\textwidth]{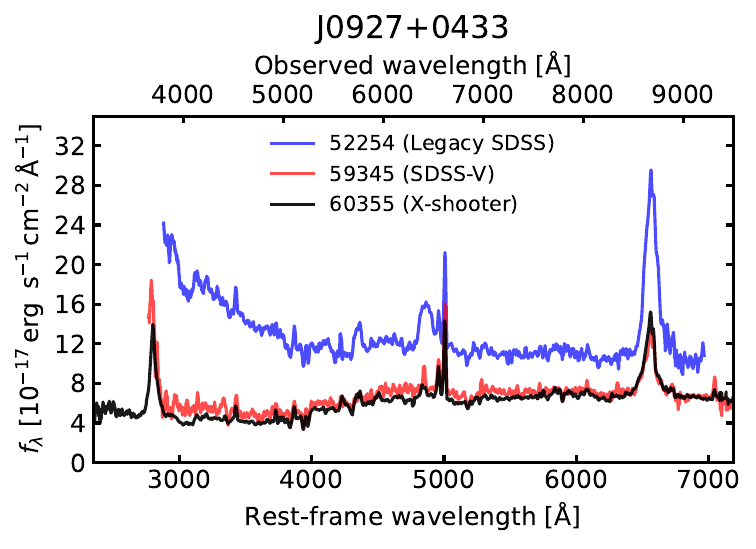} \\

\end{tabular}
\caption{Spectra of our CL-AGN sample (continued on next page). Each panel displays the legacy SDSS bright-state spectrum (blue), SDSS-V dim-state spectrum (red), and X-shooter/GMOS spectrum (black). All spectra were smoothed over $20\,$\AA.}
\label{fig:all_spectra}
\end{figure}

\begin{figure}
\centering
\begin{tabular}{ccc}
\includegraphics[width=0.3\textwidth]{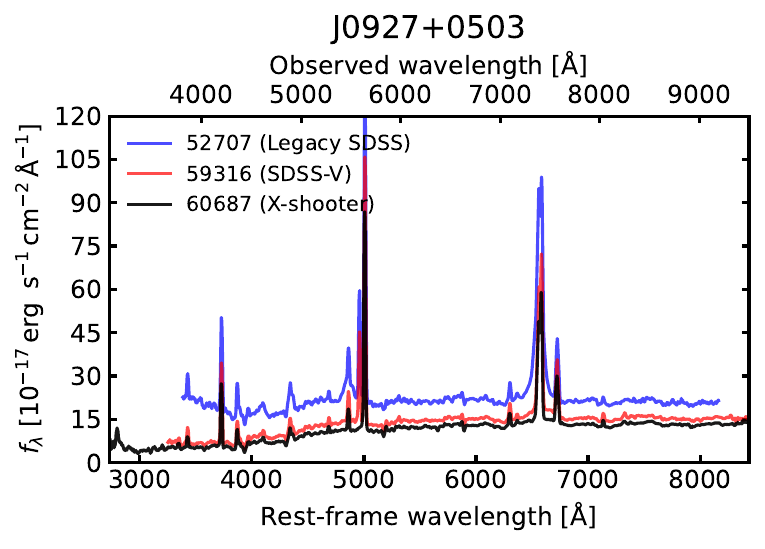} &
\includegraphics[width=0.3\textwidth]{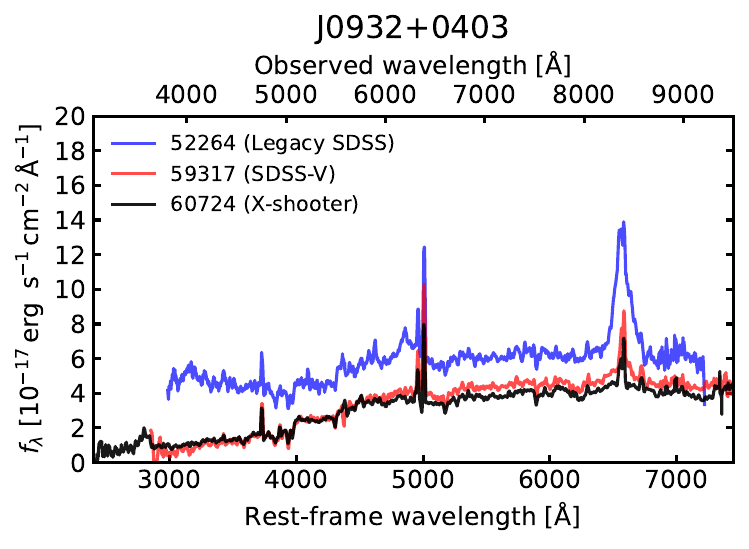} &
\includegraphics[width=0.3\textwidth]{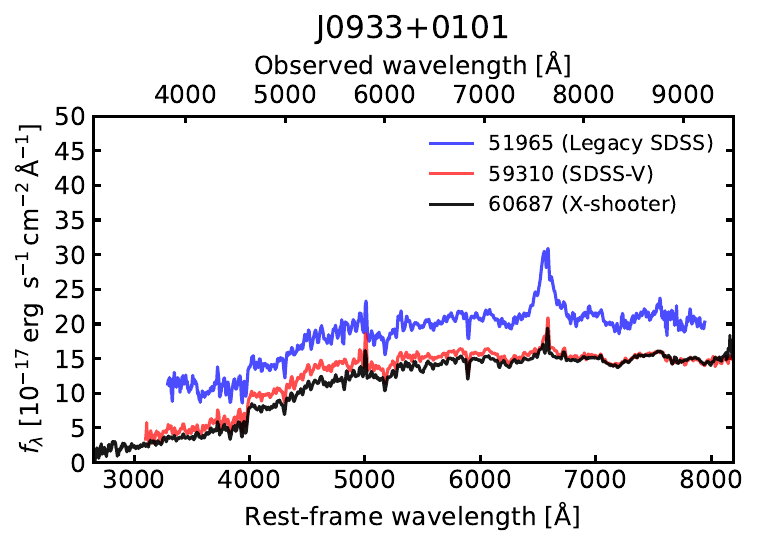}
 \\
 
\includegraphics[width=0.3\textwidth]{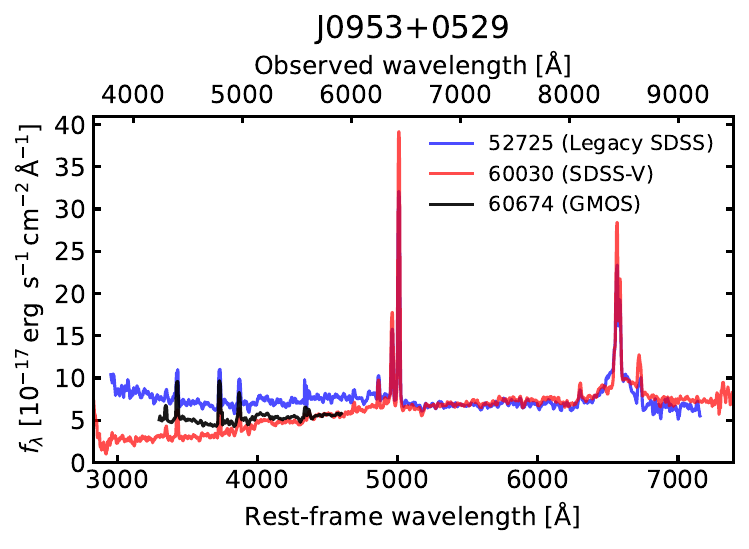} &
\includegraphics[width=0.3\textwidth]{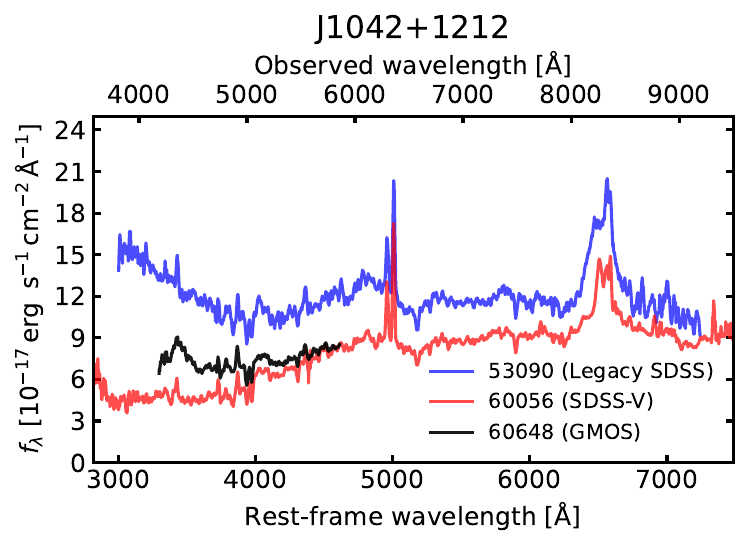} &
\includegraphics[width=0.3\textwidth]{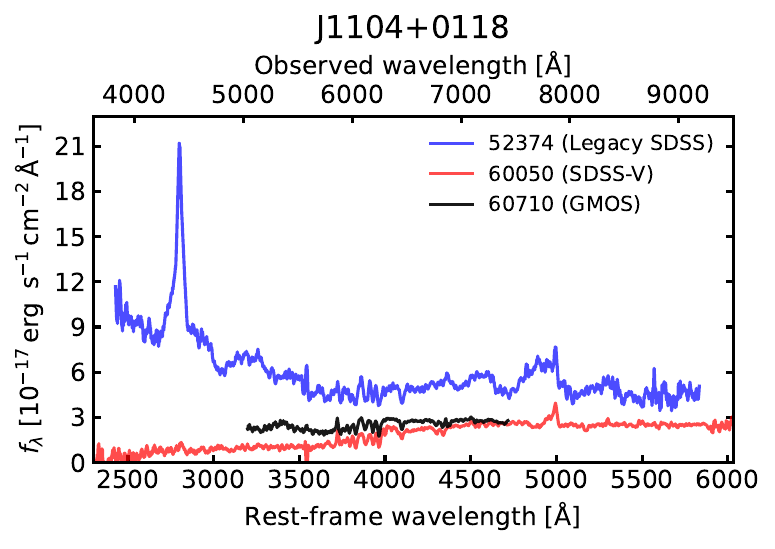} \\

\includegraphics[width=0.3\textwidth]{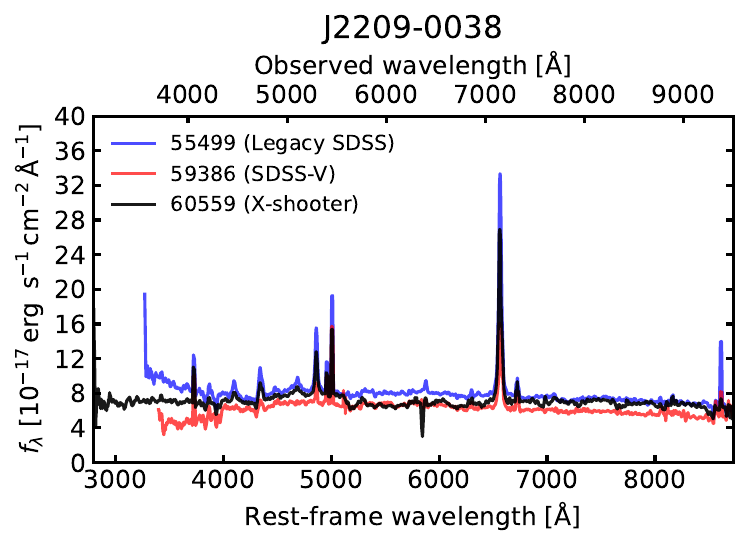} &
\includegraphics[width=0.3\textwidth]{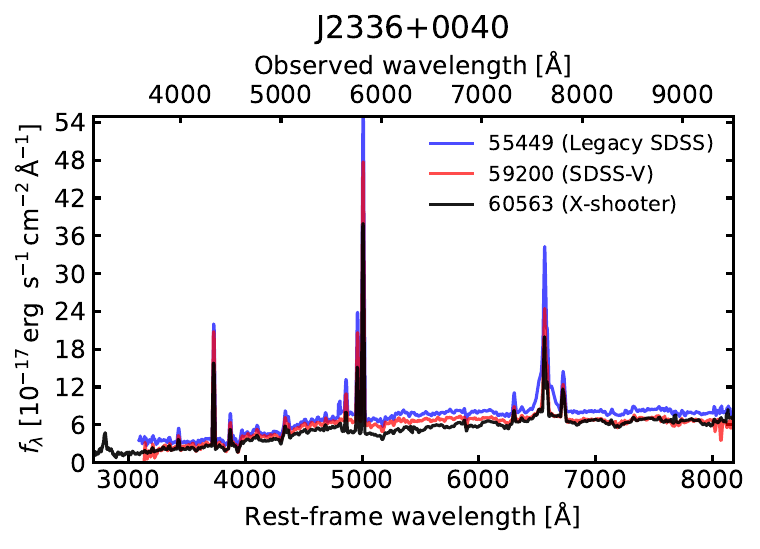}
\\
\end{tabular}
\addtocounter{figure}{-1}
\caption{Spectra of our CL-AGN sample (continued). Each panel displays the legacy SDSS bright-state spectrum (blue), SDSS-V dim-state spectrum (red), and X-shooter/GMOS spectrum (black). All spectra were smoothed over $20\,$\AA.}
\end{figure}

\section{pPXF fits}
\label{app:ppxf_plots}

Figure \ref{fig:ppxf_fits} shows the X-shooter/GMOS dim-state spectra for the CL-AGN sample, along with the corresponding \texttt{pPXF} spectral fits (see Section \ref{subsubsec:measuring_sigma} for details).

\begin{figure}
\centering
\begin{tabular}{ccc}
\includegraphics[width=0.3\textwidth]{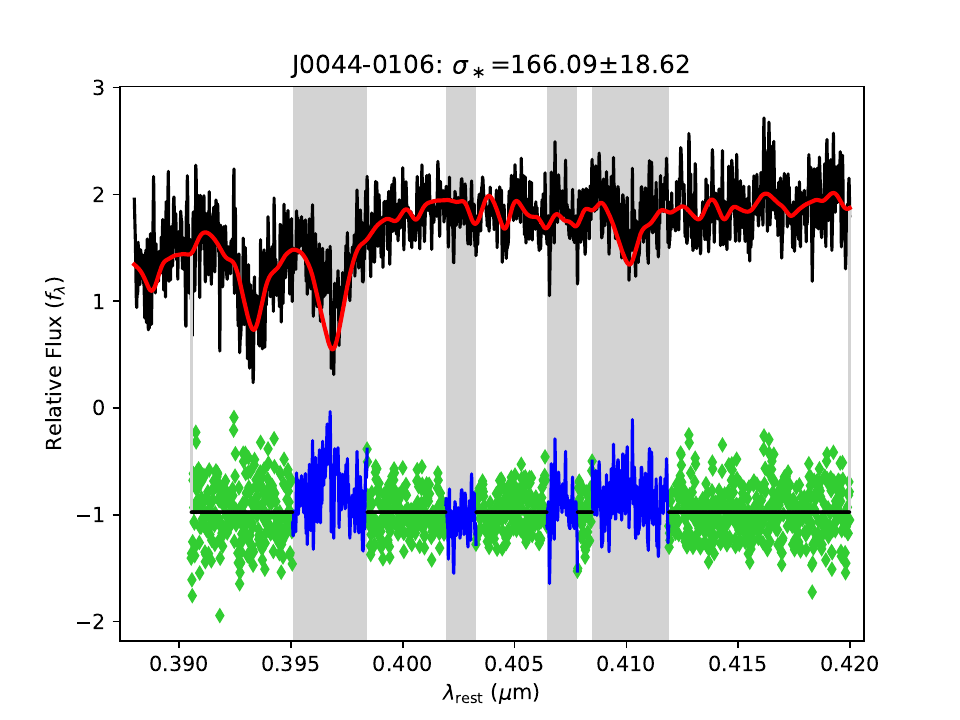} &
\includegraphics[width=0.3\textwidth]{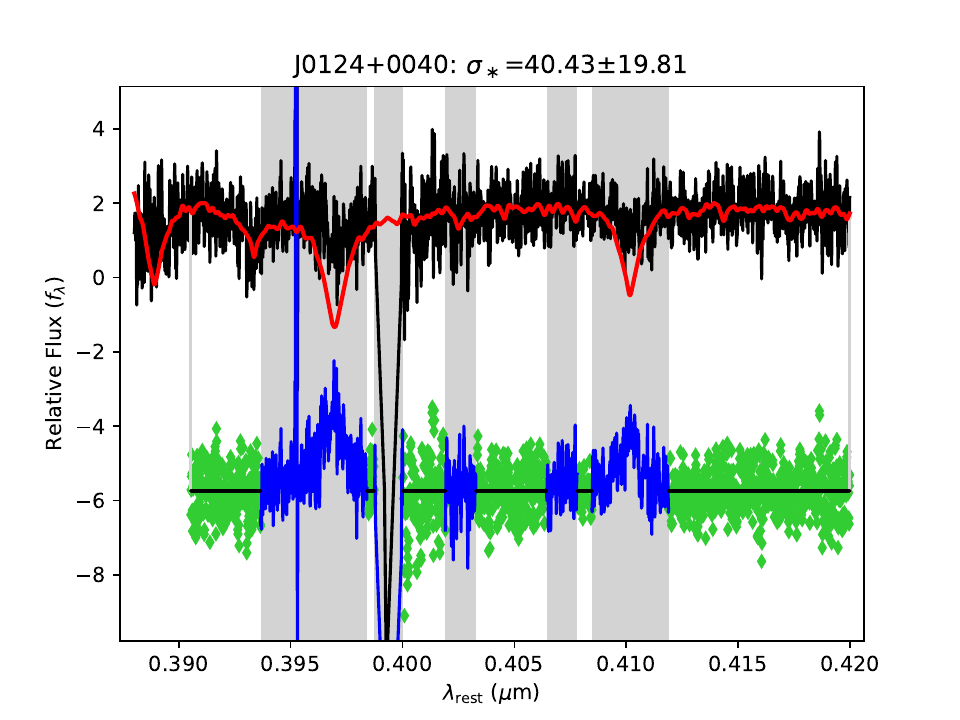} &
\includegraphics[width=0.3\textwidth]{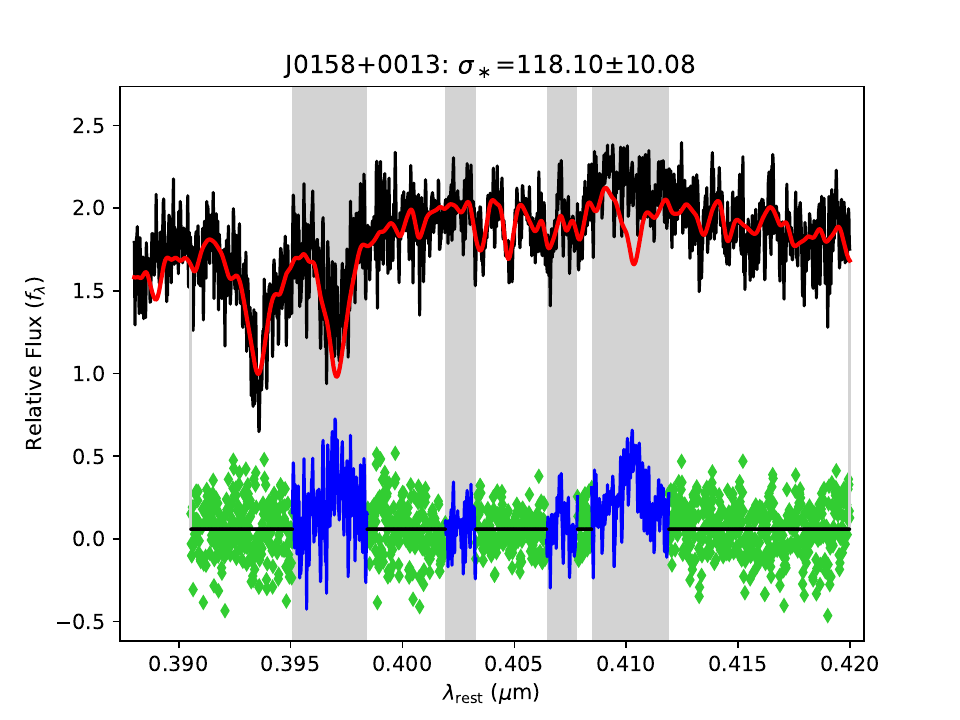} \\

\includegraphics[width=0.3\textwidth]{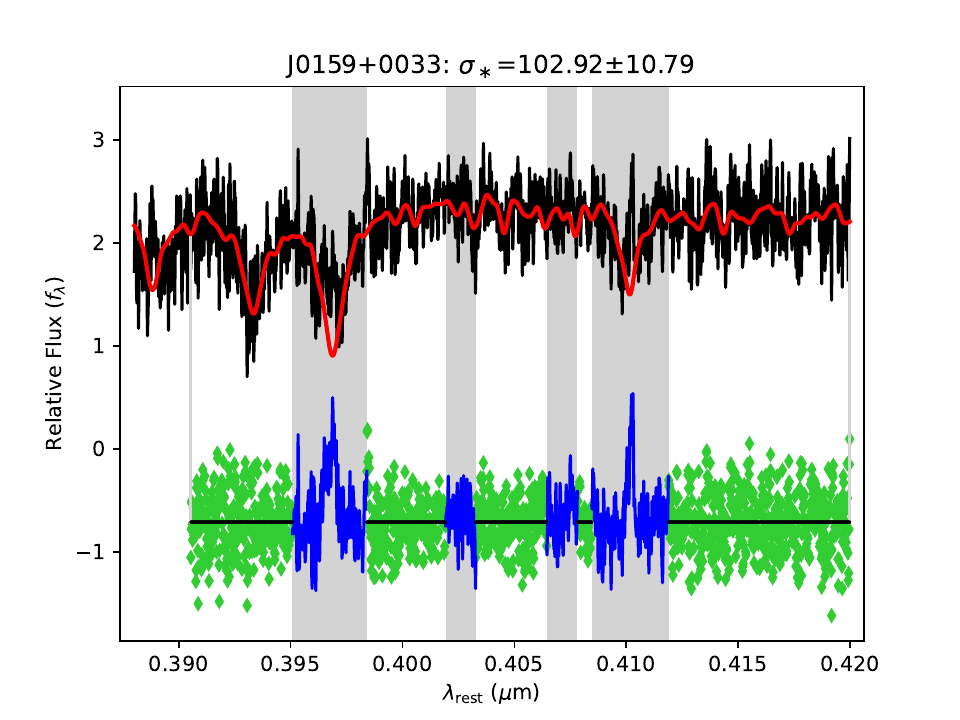} &
\includegraphics[width=0.3\textwidth]{figs/ppxf/J0206-0414_xshooter_ppxf_final_run.pdf} &
\includegraphics[width=0.3\textwidth]{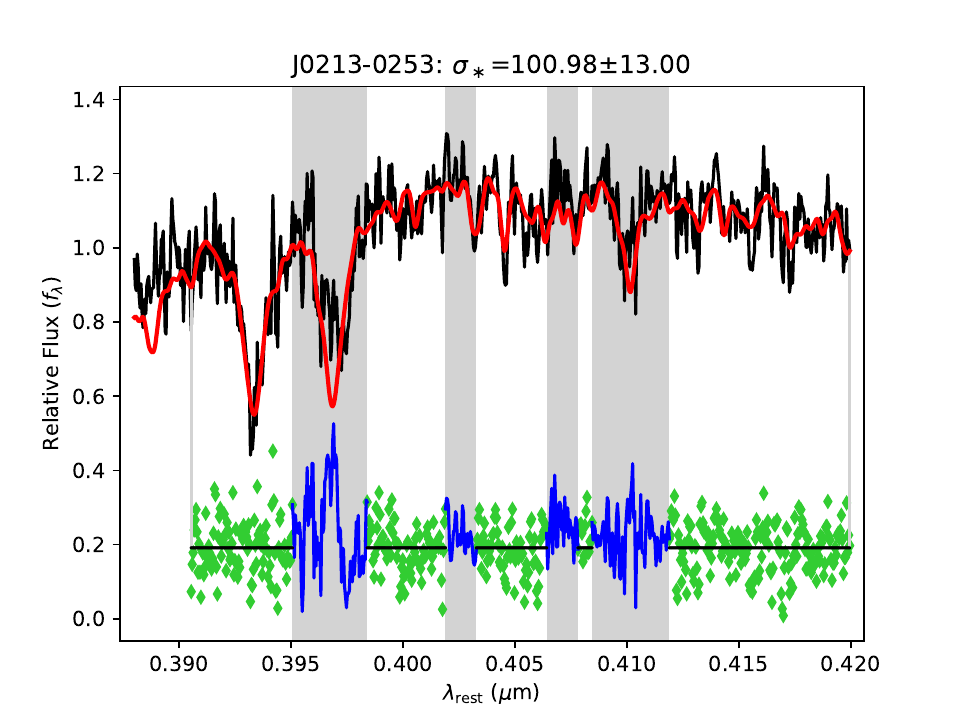} \\

\includegraphics[width=0.3\textwidth]{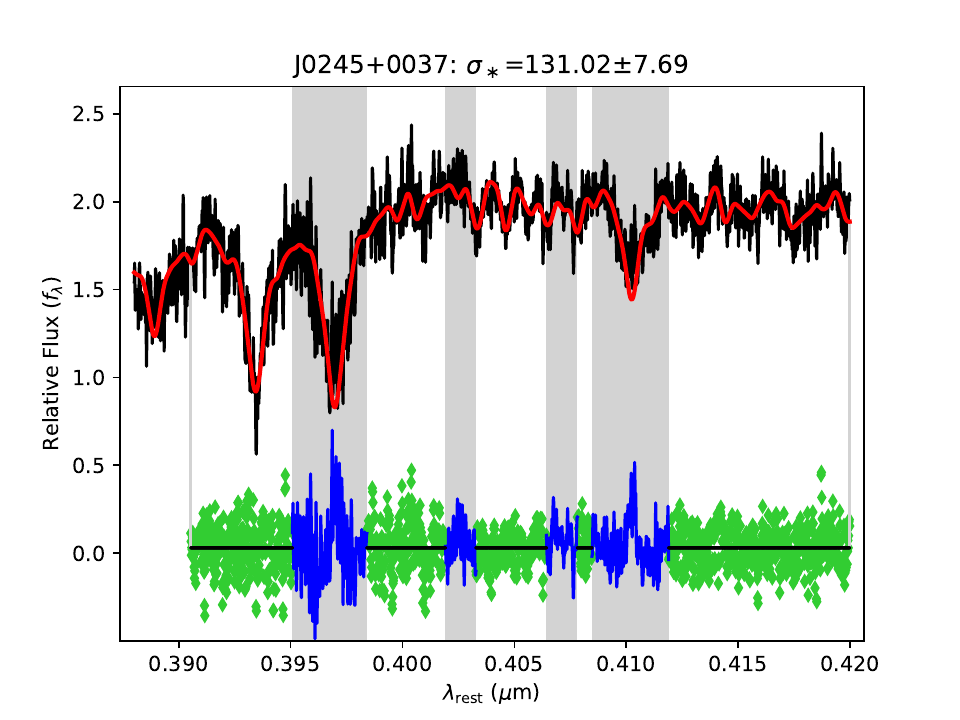} &
\includegraphics[width=0.3\textwidth]{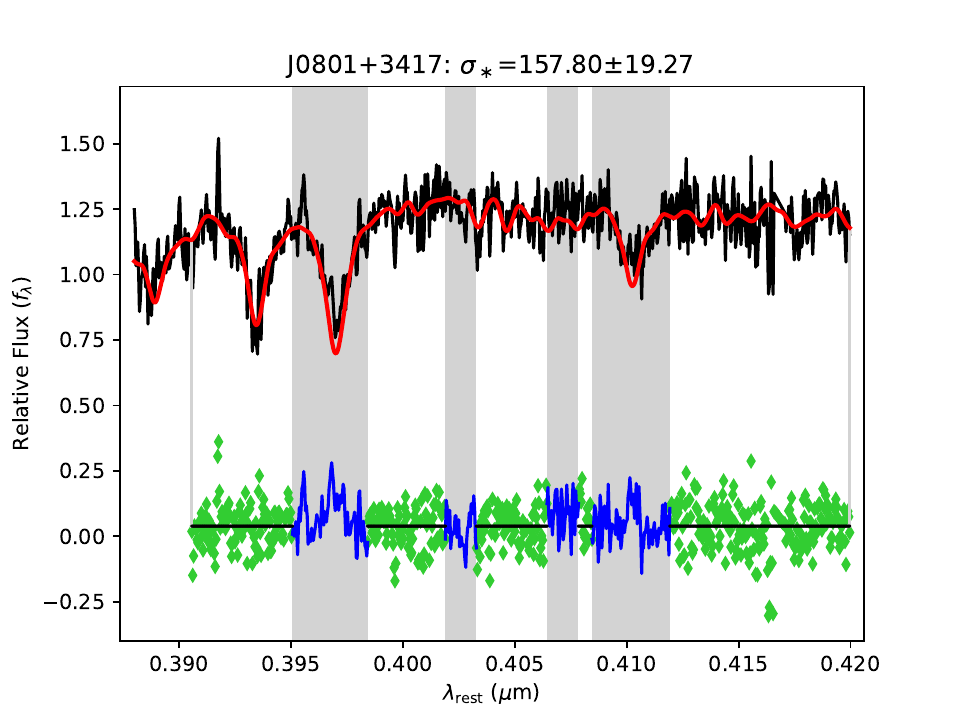} &
\includegraphics[width=0.3\textwidth]{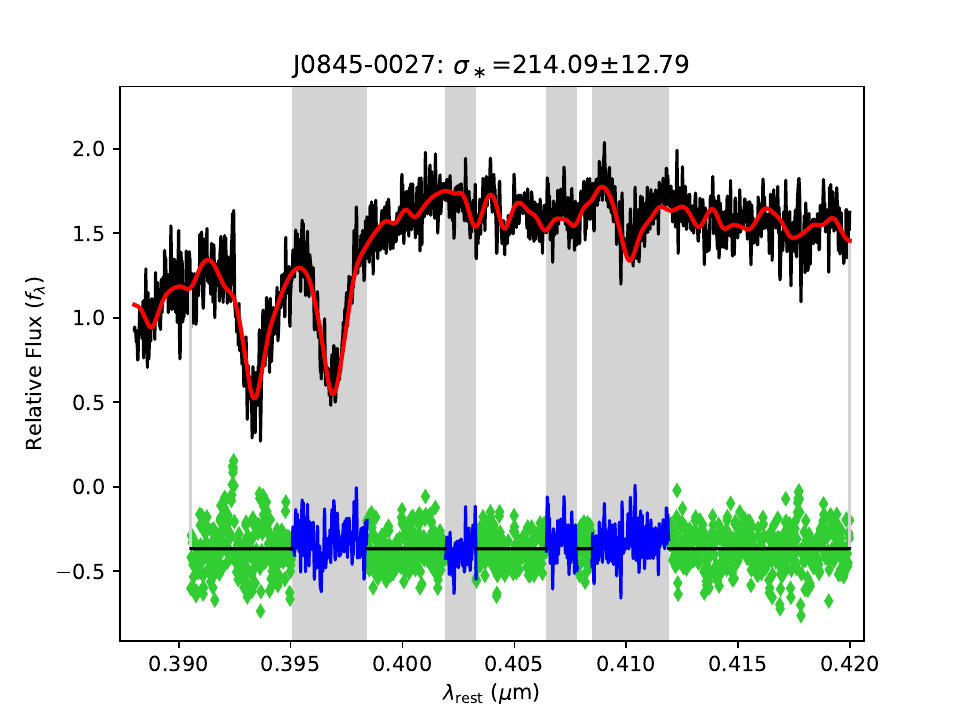} \\

\includegraphics[width=0.3\textwidth]{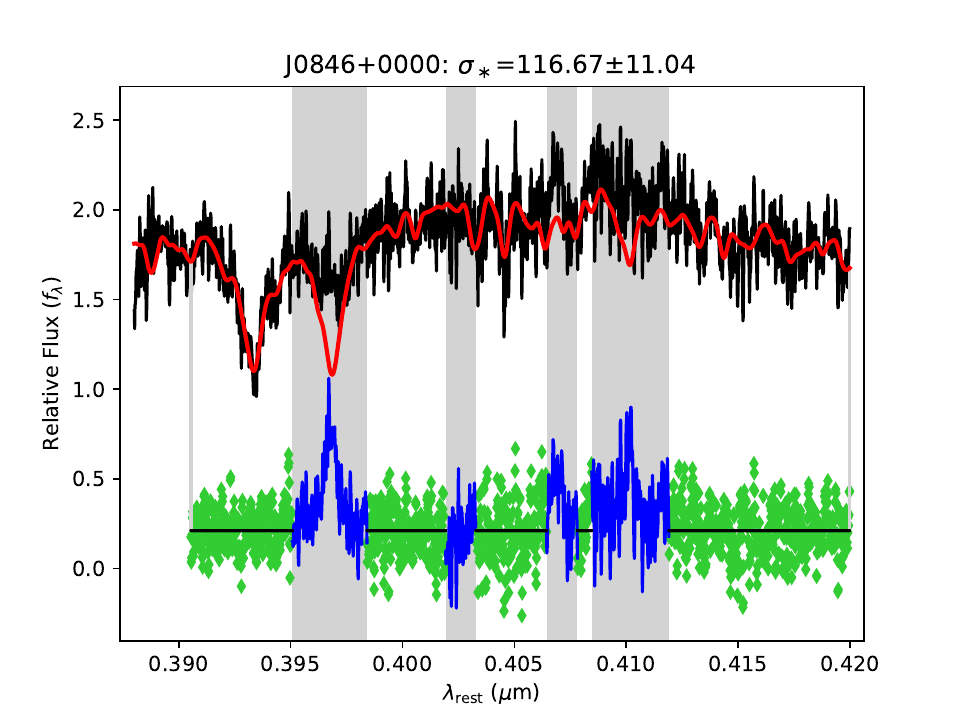} &
\includegraphics[width=0.3\textwidth]{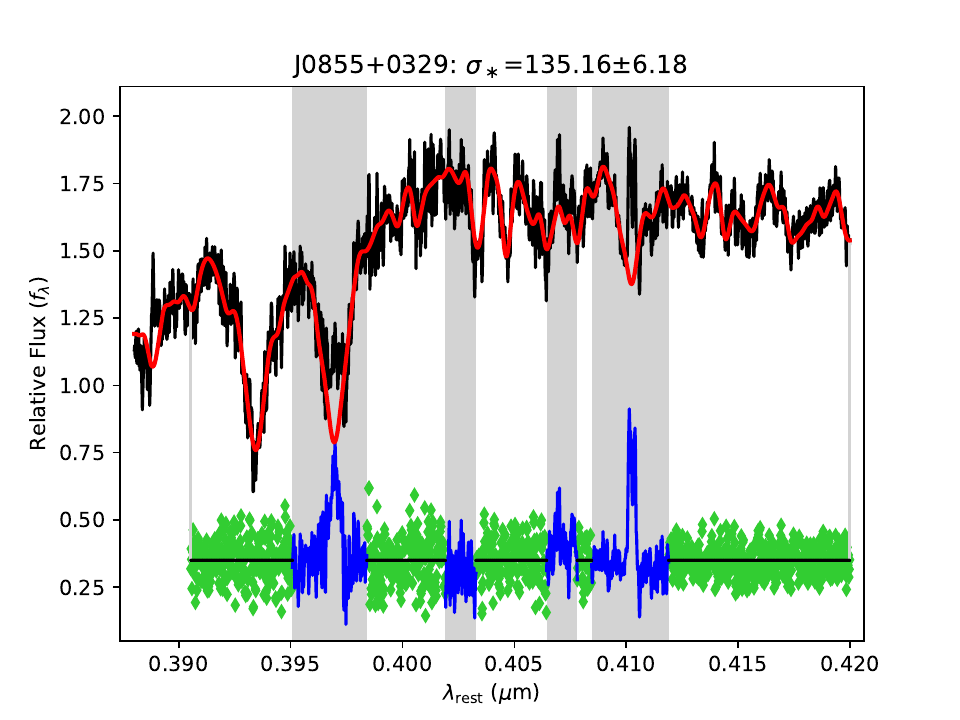} &
\includegraphics[width=0.3\textwidth]{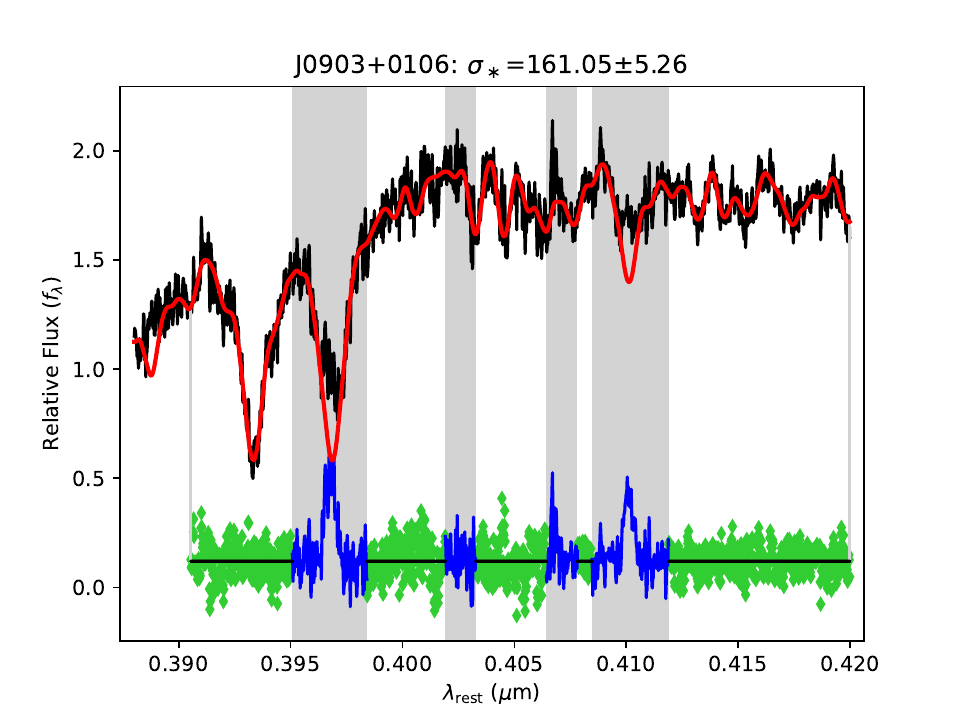} \\

\includegraphics[width=0.3\textwidth]{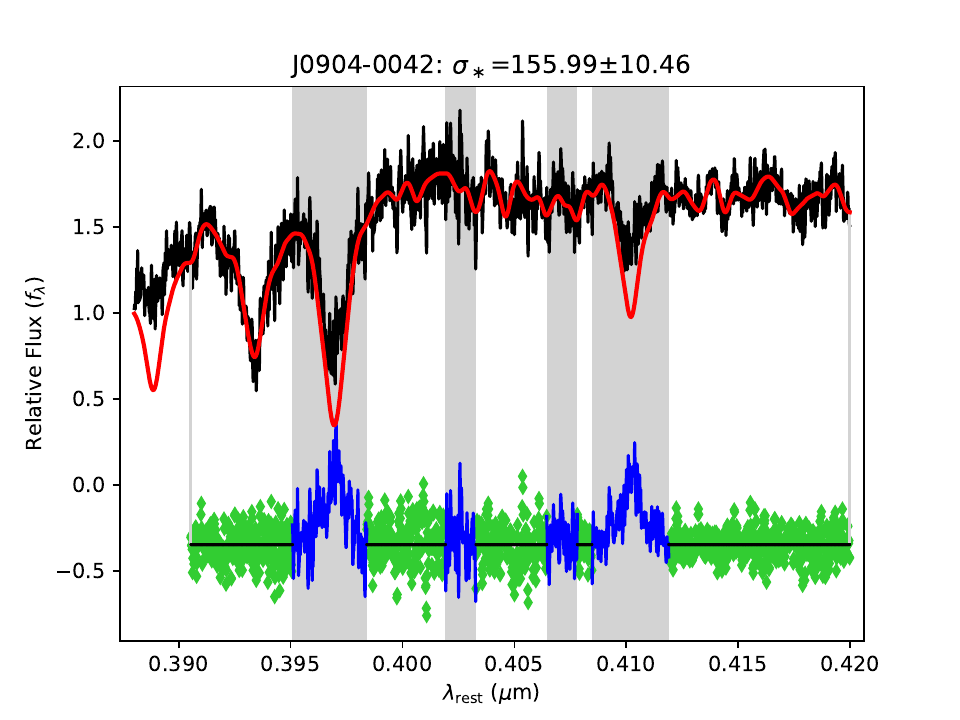} &
\includegraphics[width=0.3\textwidth]{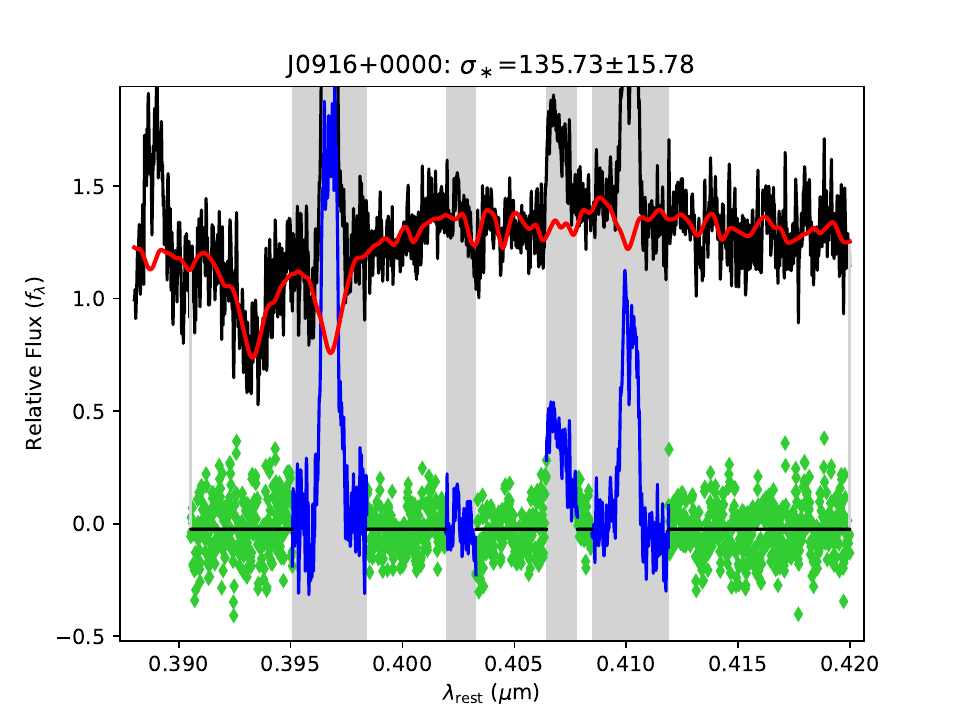} &
\includegraphics[width=0.3\textwidth]{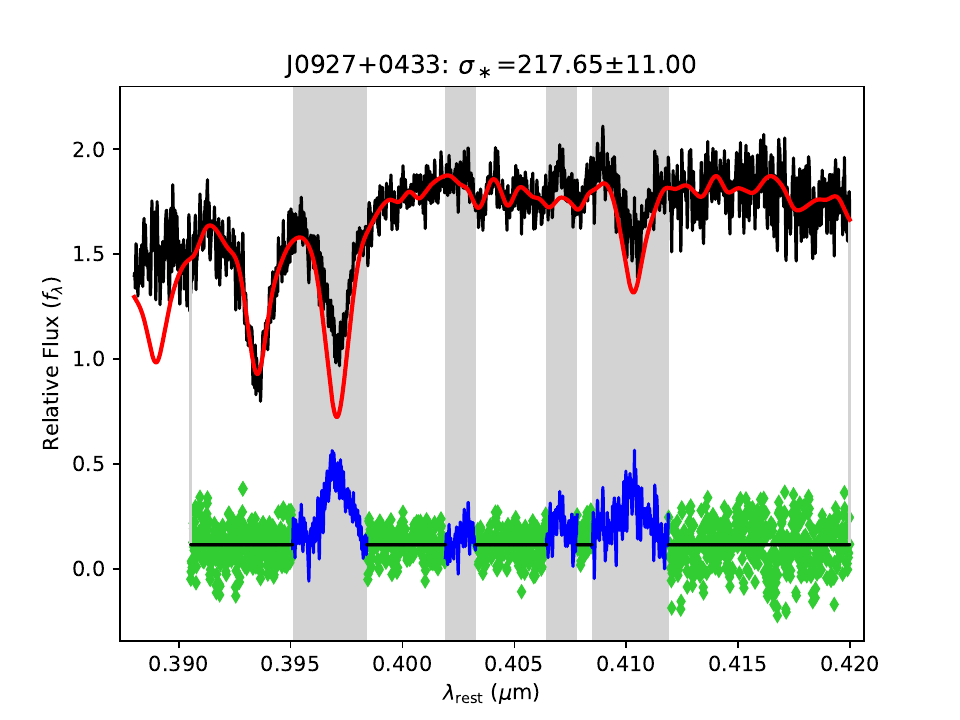} \\

\end{tabular}
\caption{\texttt{pPXF} fitting and \sigs\ measurement for our CL-AGN sample (continued on next page). The X-shooter/GMOS data are shown in black, the best-fitting \texttt{pPXF} model in red, and the residuals are shown near the bottom, where green points were considered for the fit while blue ones were masked out (masked spectral regions are indicated in gray and listed in Table \ref{tab:mask_lines}).}
\label{fig:ppxf_fits}
\end{figure}

\begin{figure}
\centering
\begin{tabular}{ccc}
\includegraphics[width=0.3\textwidth]{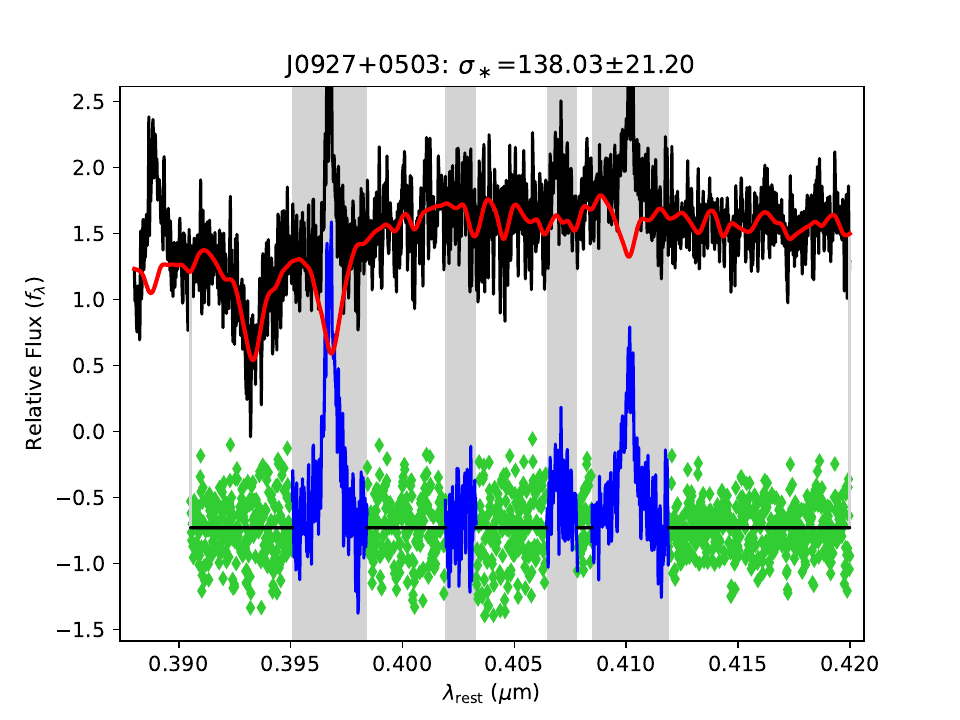} &
\includegraphics[width=0.3\textwidth]{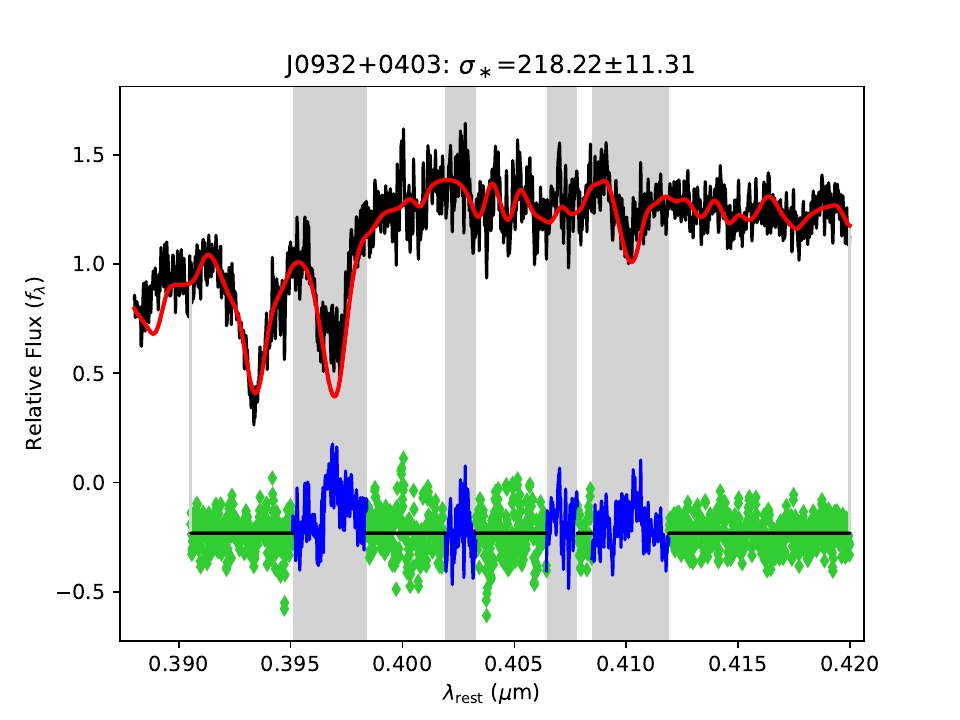} &
\includegraphics[width=0.3\textwidth]{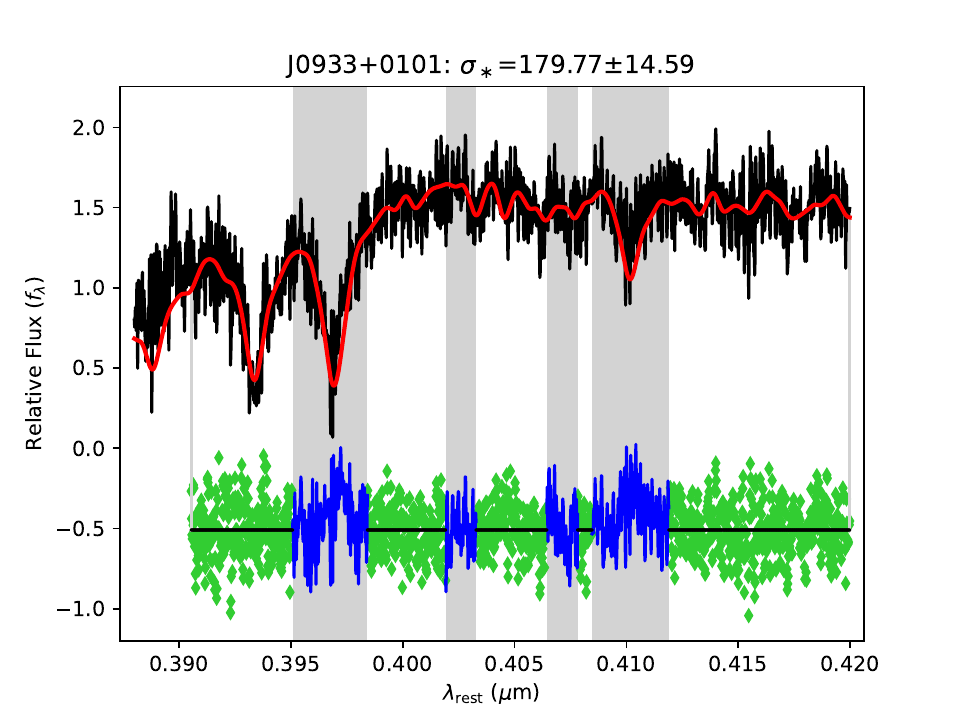}
 \\
 
\includegraphics[width=0.3\textwidth]{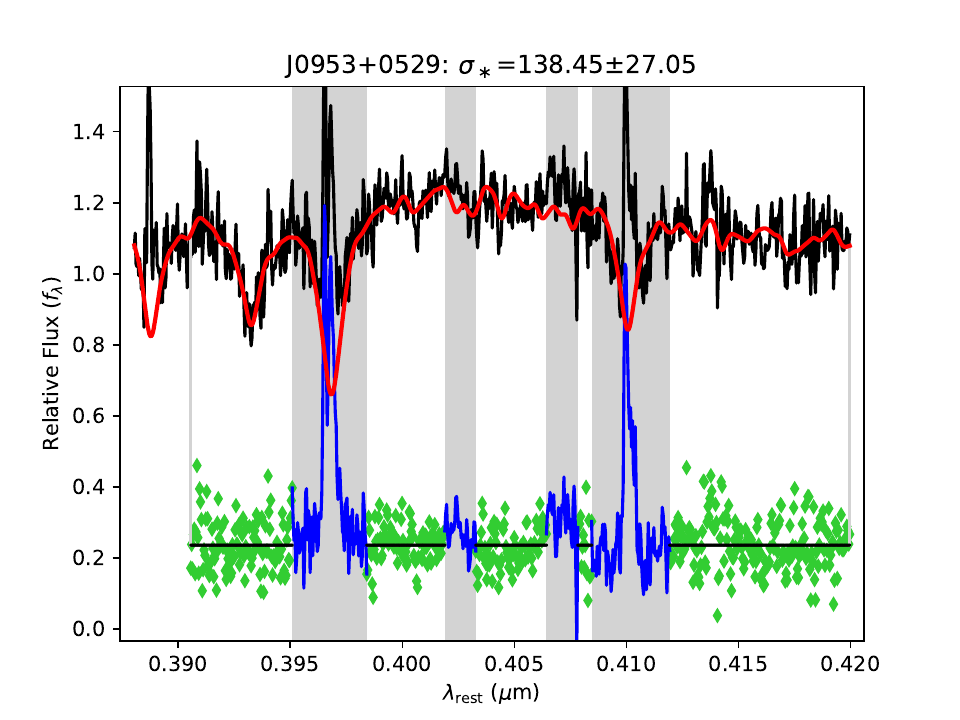} &
\includegraphics[width=0.3\textwidth]{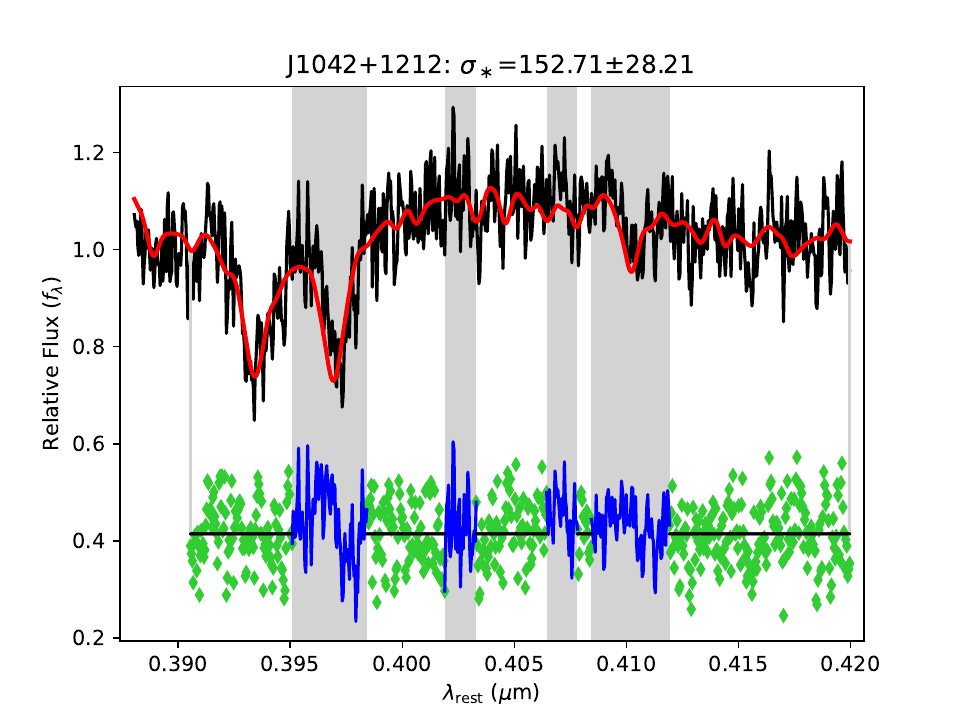} &
\includegraphics[width=0.3\textwidth]{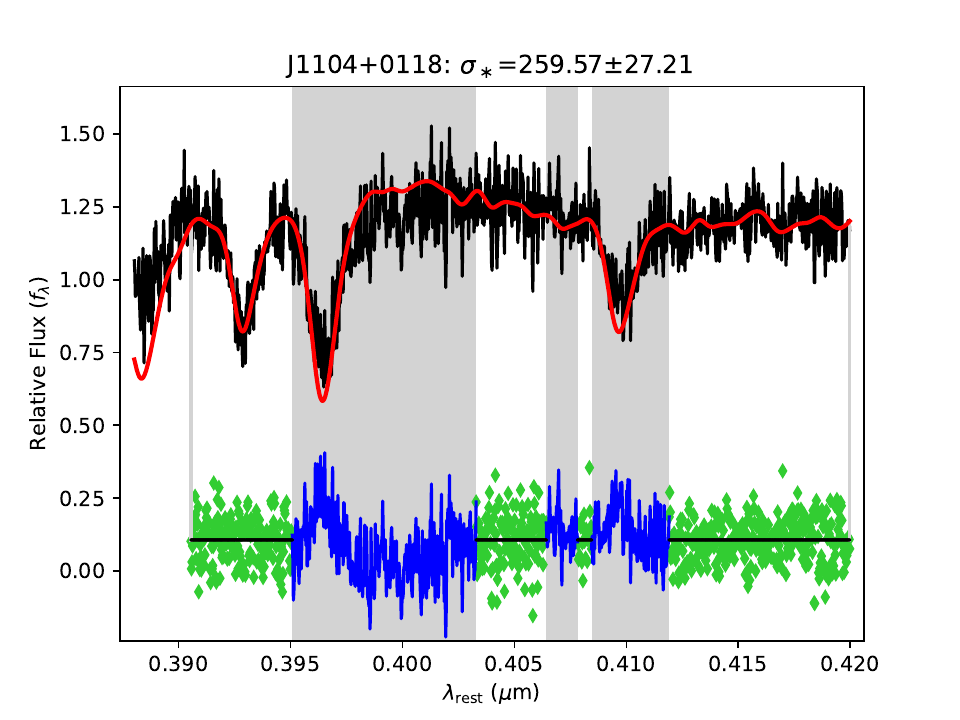} \\

\includegraphics[width=0.3\textwidth]{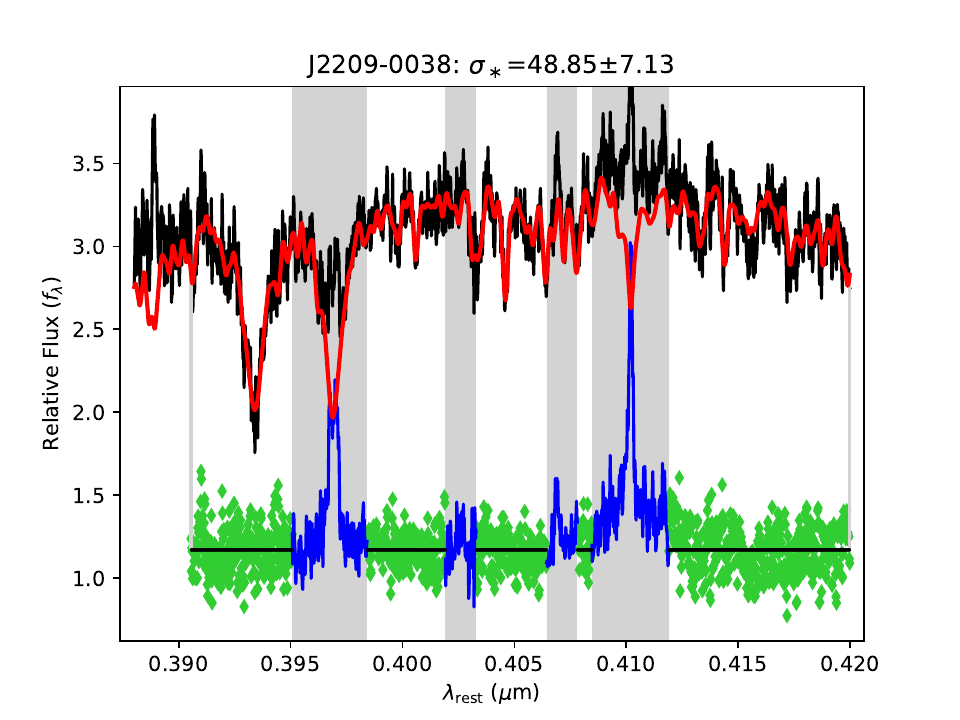} &
\includegraphics[width=0.3\textwidth]{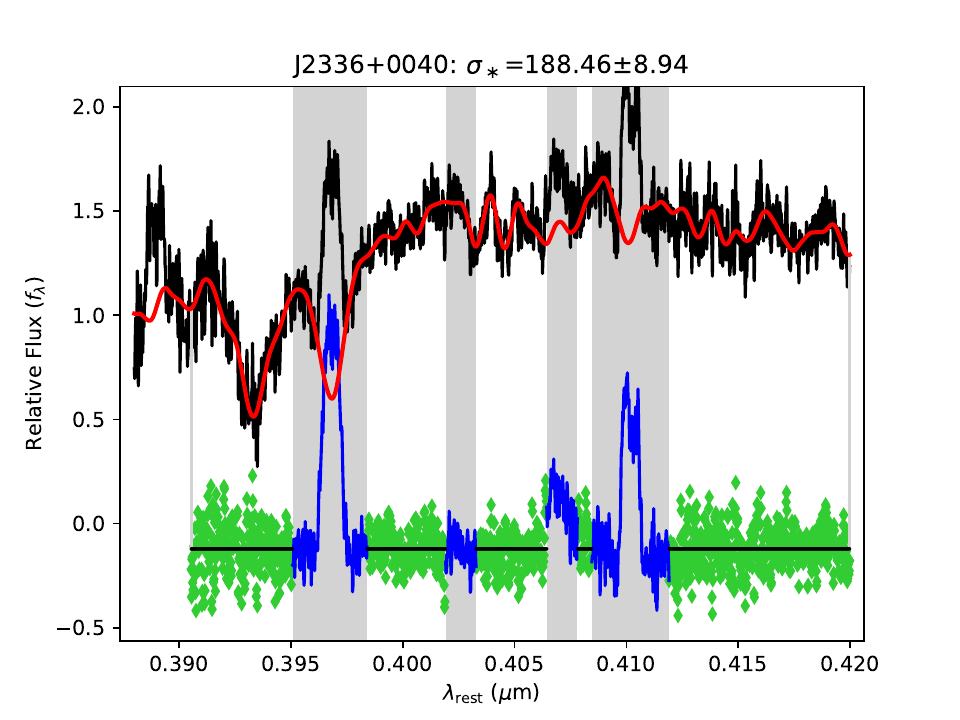}
\\
\end{tabular}
\addtocounter{figure}{-1}
\caption{\texttt{pPXF} fitting and \sigs\ measurement for our CL-AGN sample (continued). The X-shooter/GMOS data are shown in black, the best-fitting \texttt{pPXF} model in red, and the residuals are shown near the bottom, where green points were considered for the fit while blue ones were masked out (masked spectral regions are indicated in gray and listed in Table \ref{tab:mask_lines}).}
\end{figure}
\section{FIREFLY fits}
\label{app:firefly_plots}

Figure \ref{fig:firefly_fits} displays the SDSS-V dim-state spectra for the CL-AGN sample, along with the corresponding \texttt{FIREFLY} spectral fits of the host galaxy (see Section \ref{subsec:ssp} for details).

\begin{figure}
\centering
\begin{tabular}{ccc}
\includegraphics[width=0.3\textwidth]{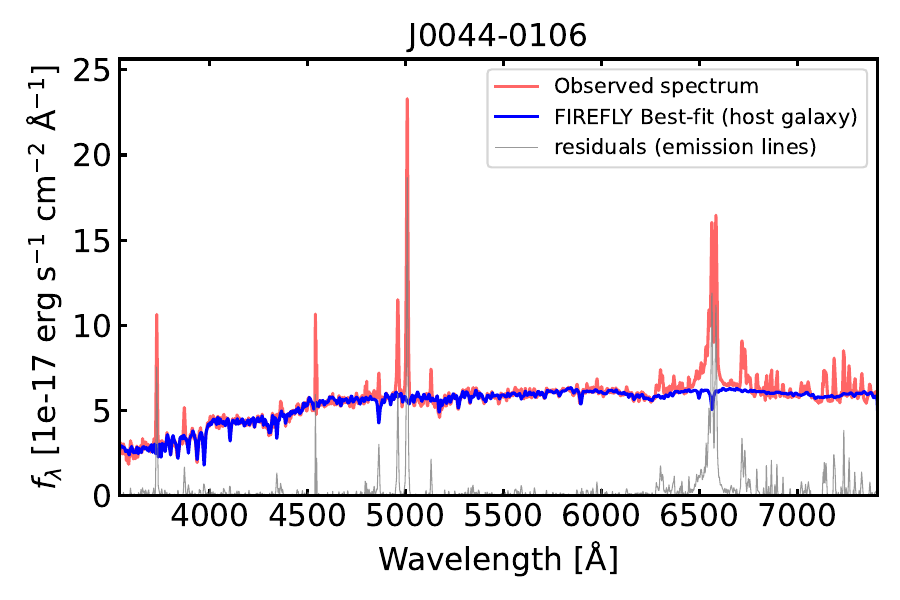} &
\includegraphics[width=0.3\textwidth]{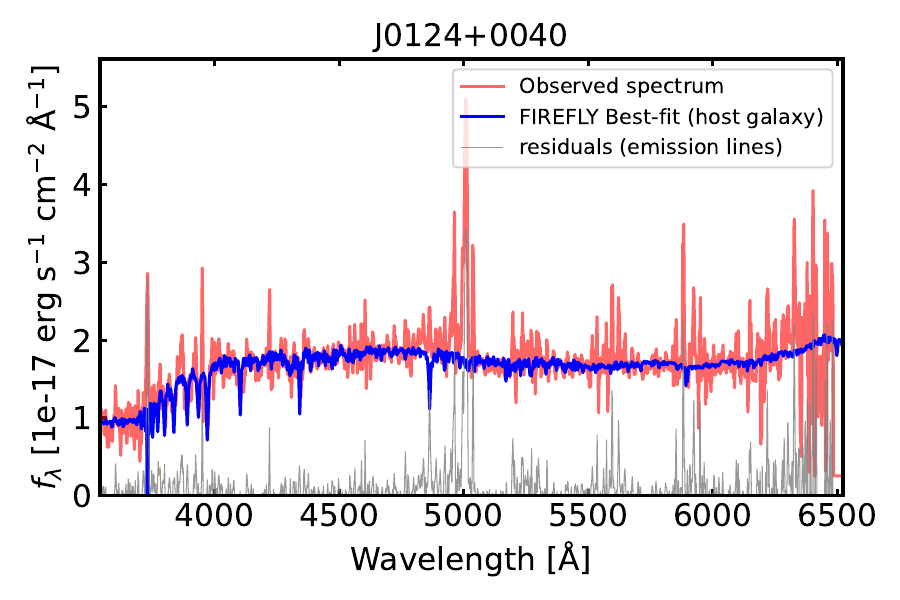} &
\includegraphics[width=0.3\textwidth]{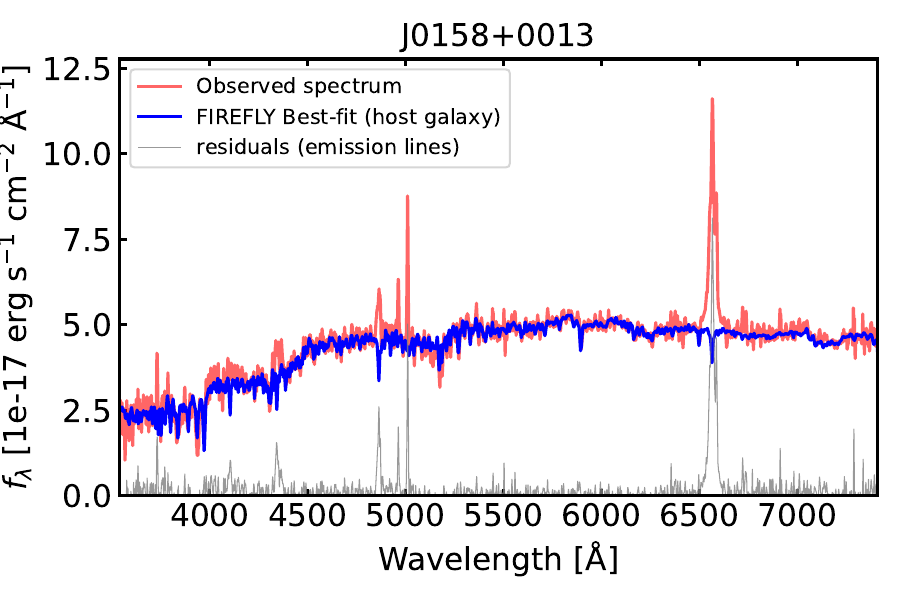} \\

\includegraphics[width=0.3\textwidth]{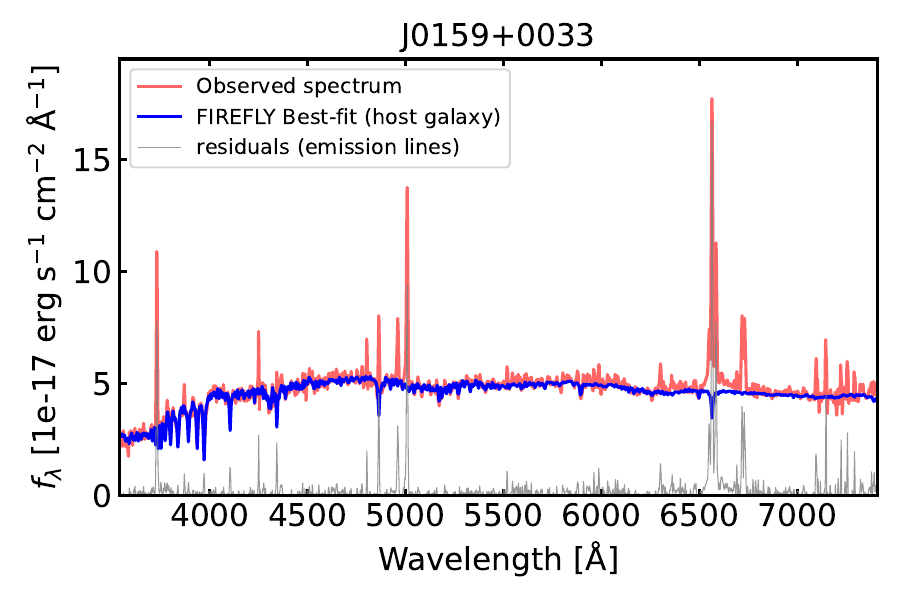} &
\includegraphics[width=0.3\textwidth]{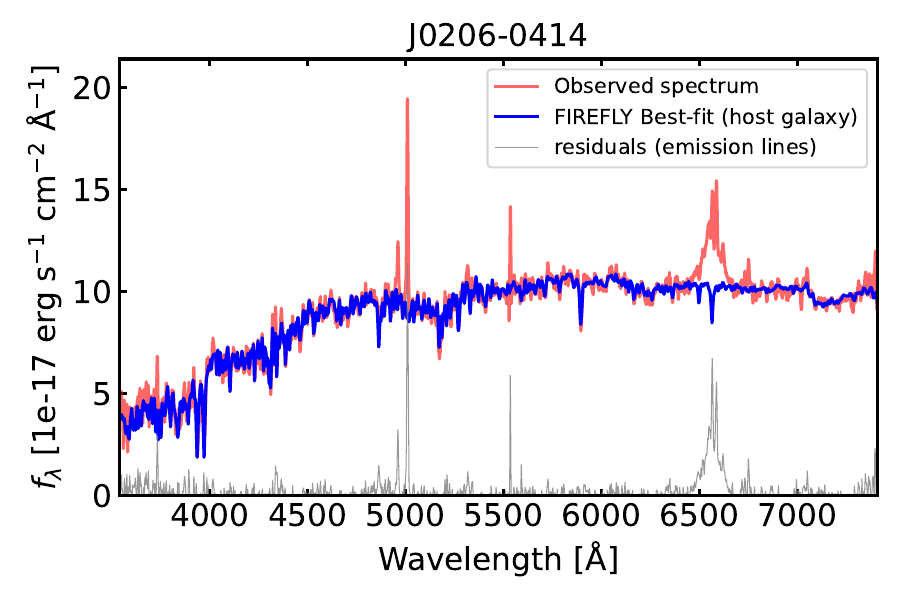} &
\includegraphics[width=0.3\textwidth]{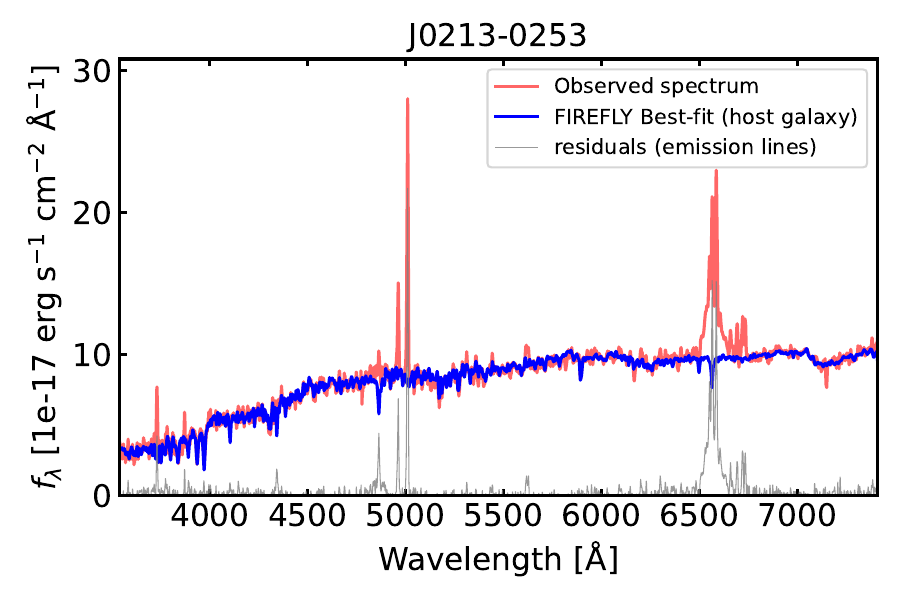} \\

\includegraphics[width=0.3\textwidth]{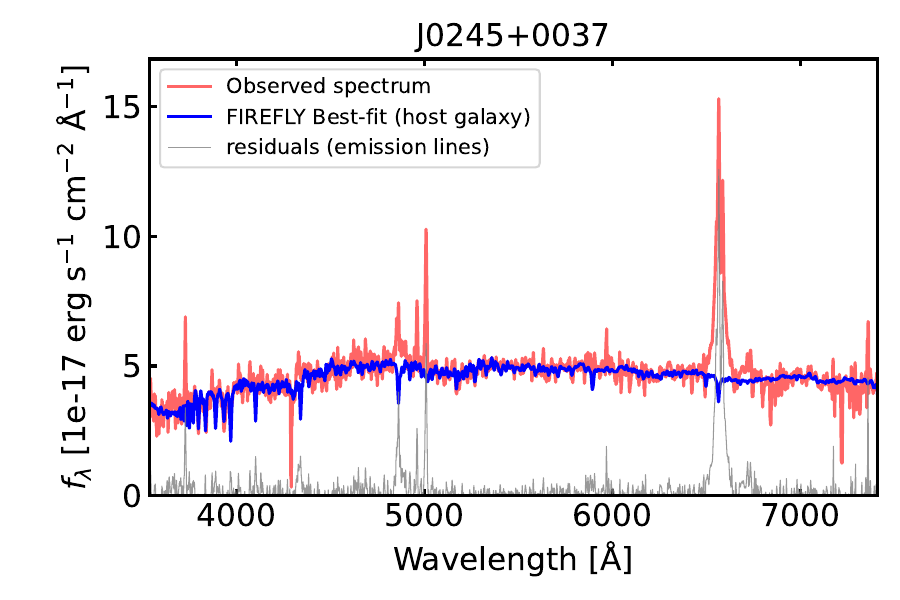} &
\includegraphics[width=0.3\textwidth]{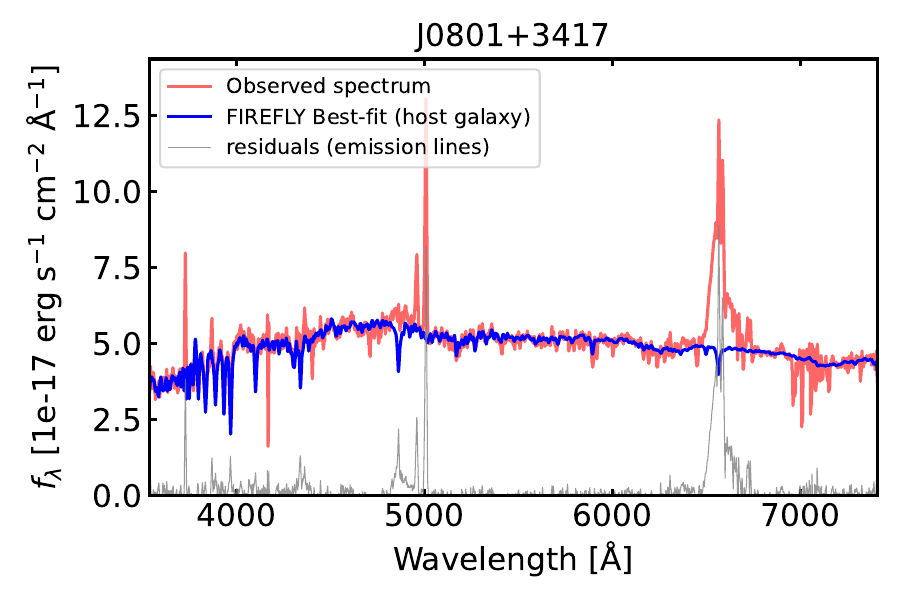} &
\includegraphics[width=0.3\textwidth]{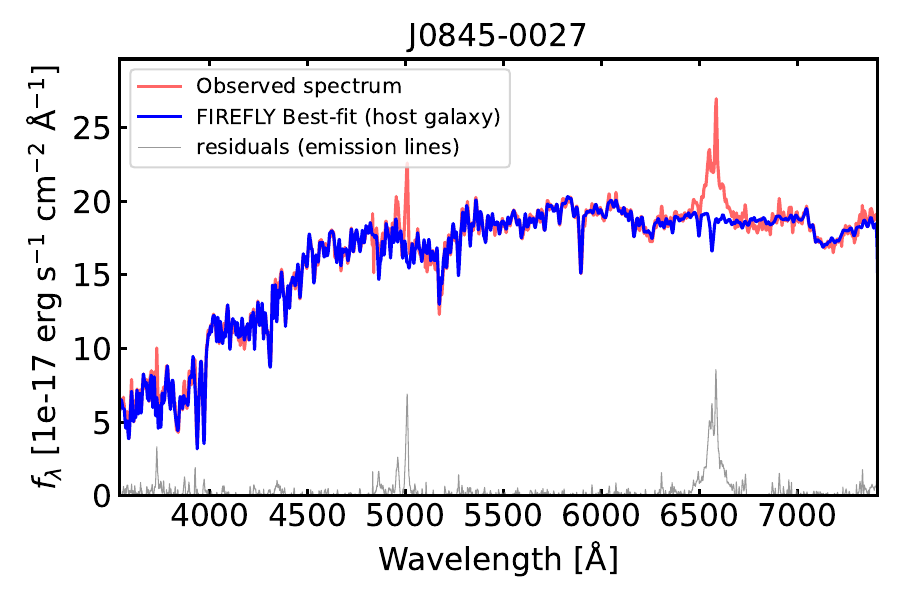} \\

\includegraphics[width=0.3\textwidth]{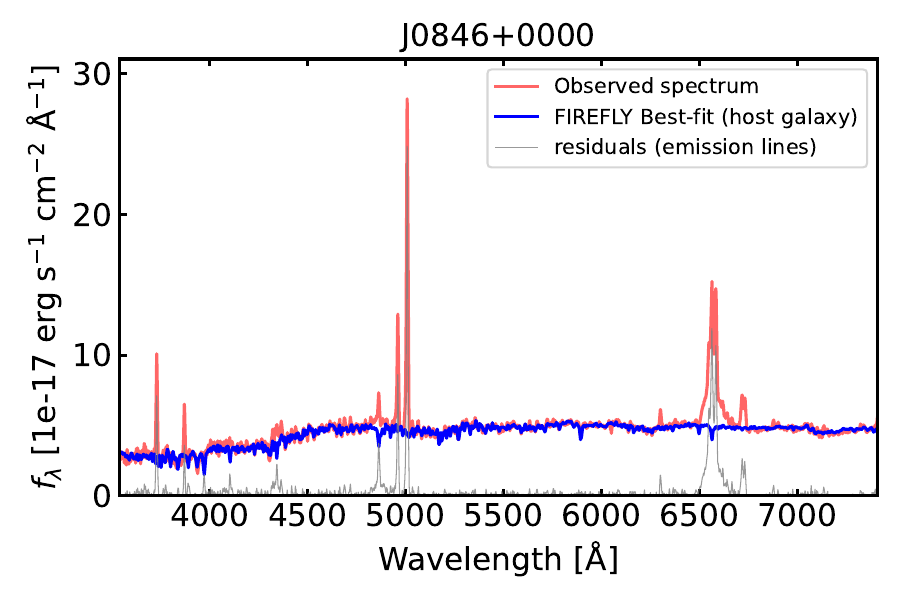} &
\includegraphics[width=0.3\textwidth]{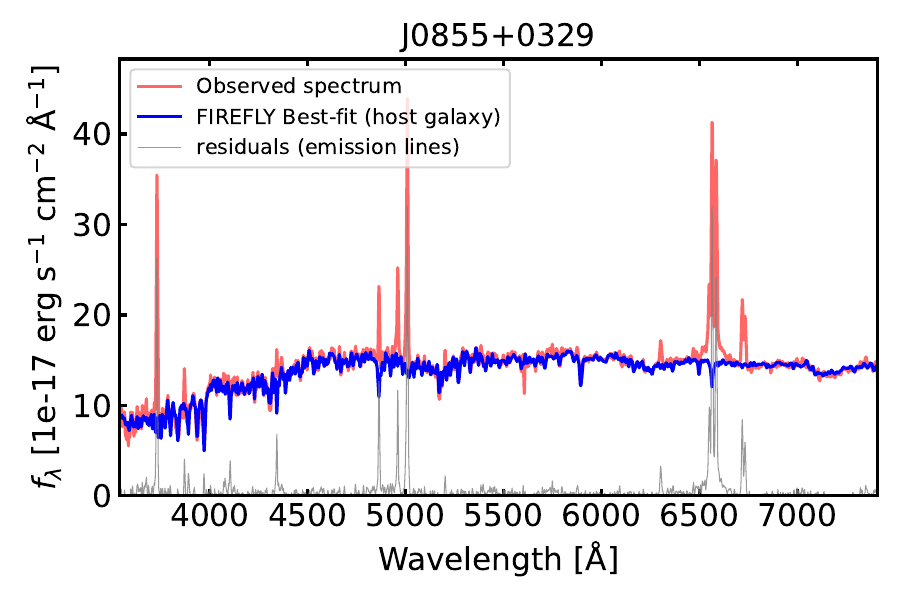} &
\includegraphics[width=0.3\textwidth]{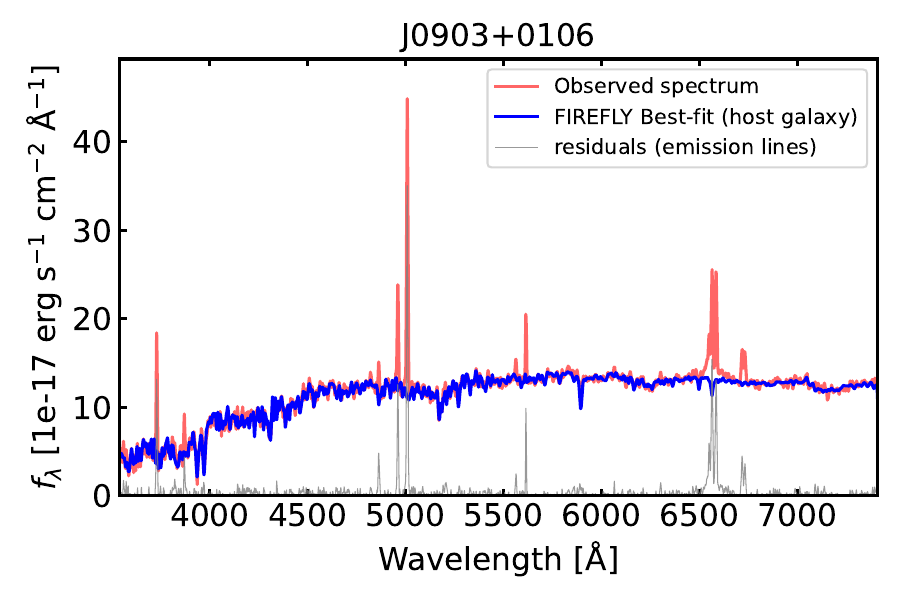} \\

\includegraphics[width=0.3\textwidth]{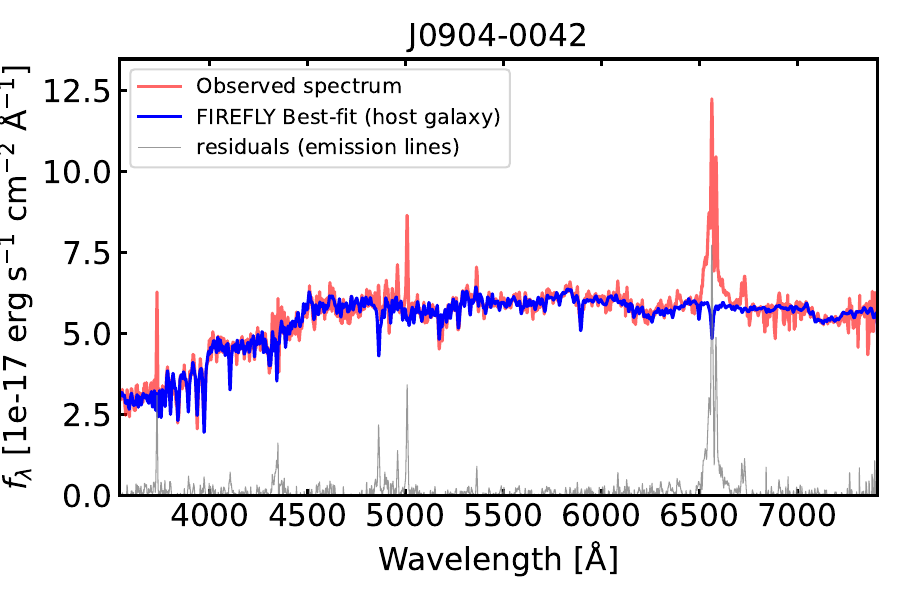} &
\includegraphics[width=0.3\textwidth]{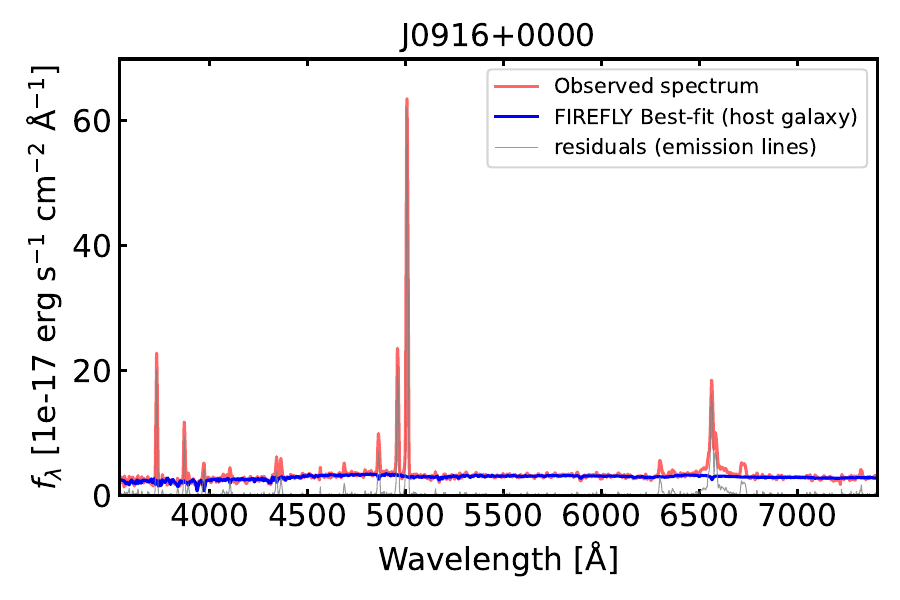} &
\includegraphics[width=0.3\textwidth]{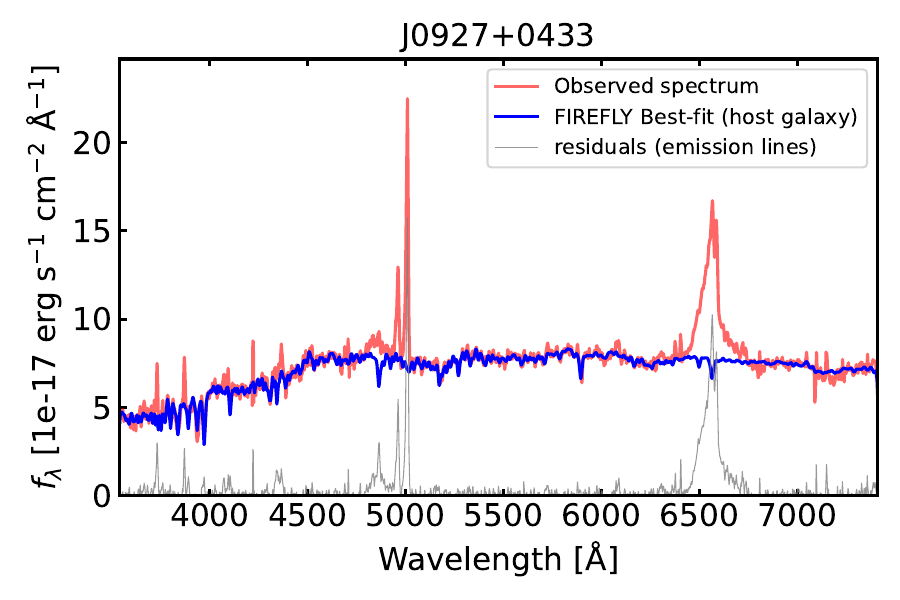} \\

\end{tabular}
\caption{\texttt{FIREFLY} host galaxy spectral fits for the CL-AGN sample (continued on next page). Each panel shows one object: the SDSS-V dim-state spectrum (red; smoothed over 5 pixels), the corresponding \texttt{FIREFLY} best-fitting host galaxy model (blue), and the residuals (gray; also smoothed over 5 pixels). The prominent features visible in the residuals correspond to AGN emission lines that were masked during the \texttt{FIREFLY} fitting.}
\label{fig:firefly_fits}
\end{figure}

\begin{figure}
\centering
\begin{tabular}{ccc}
\includegraphics[width=0.3\textwidth]{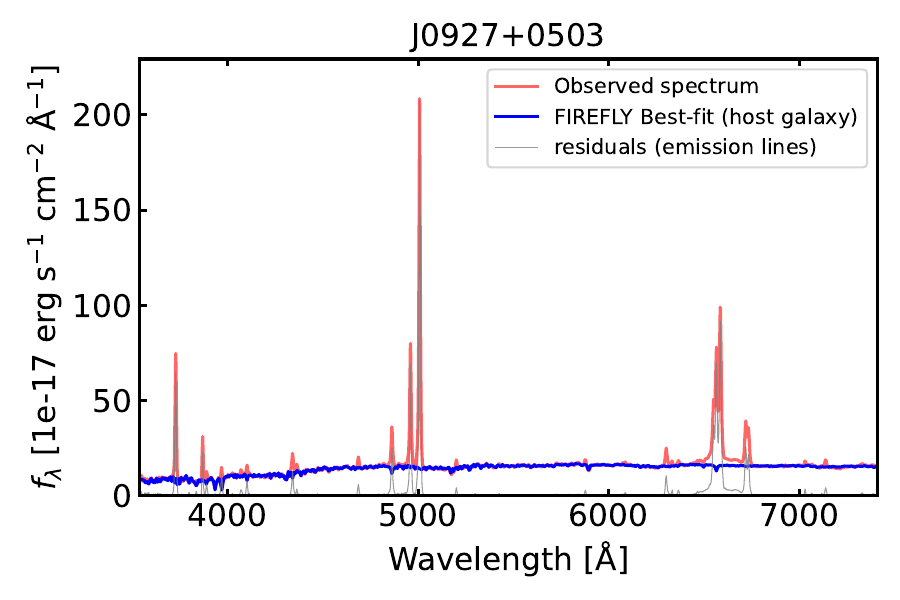} &
\includegraphics[width=0.3\textwidth]{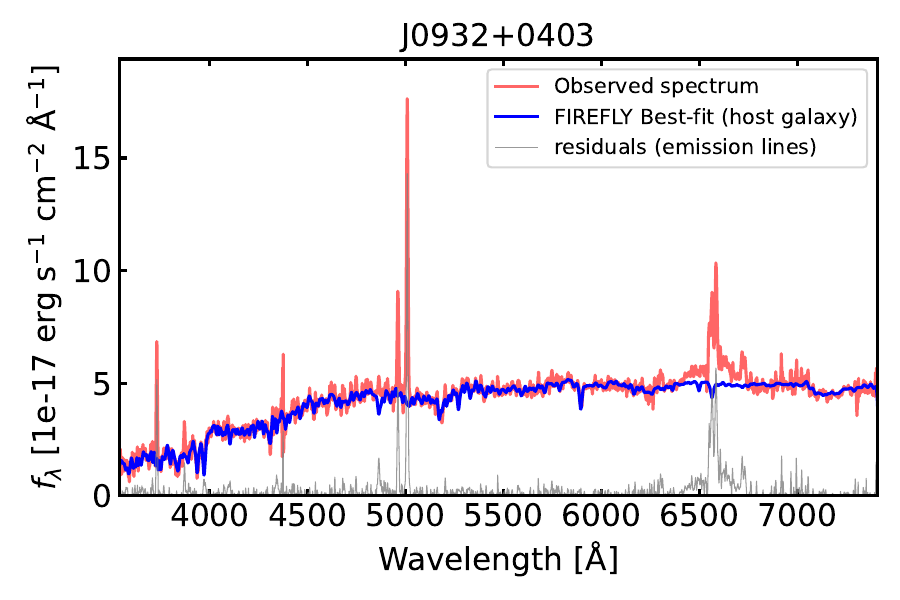} &
\includegraphics[width=0.3\textwidth]{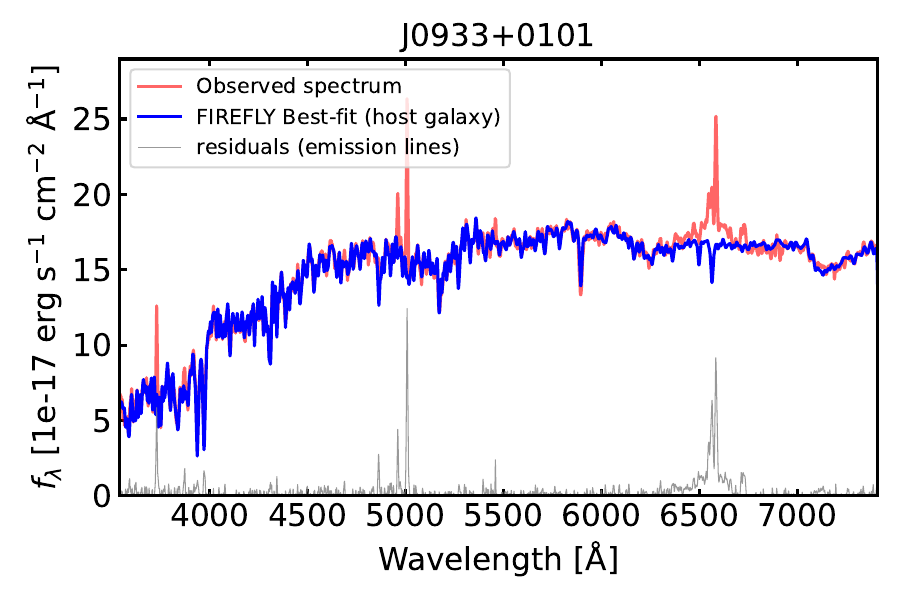}
 \\
 
\includegraphics[width=0.3\textwidth]{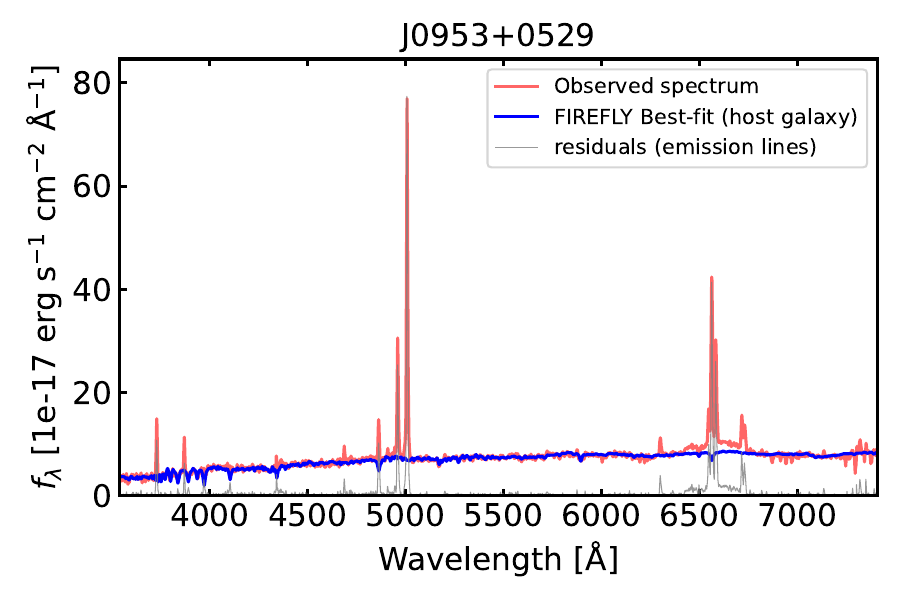} &
\includegraphics[width=0.3\textwidth]{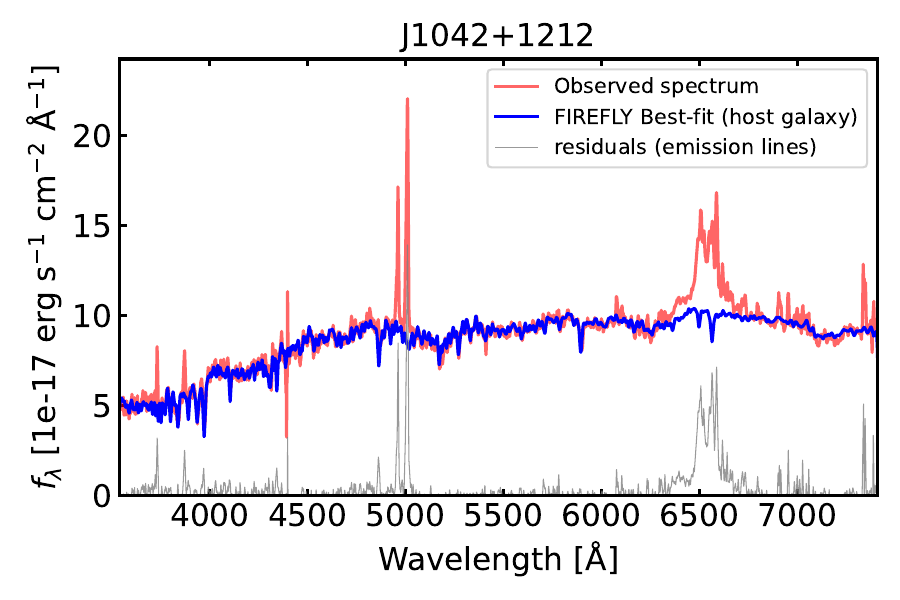} &
\includegraphics[width=0.3\textwidth]{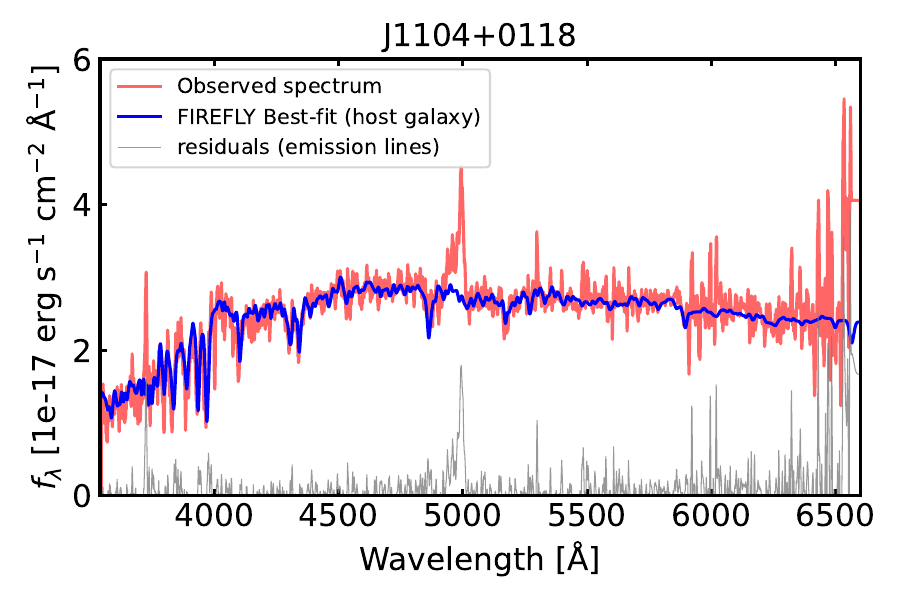} \\

\includegraphics[width=0.3\textwidth]{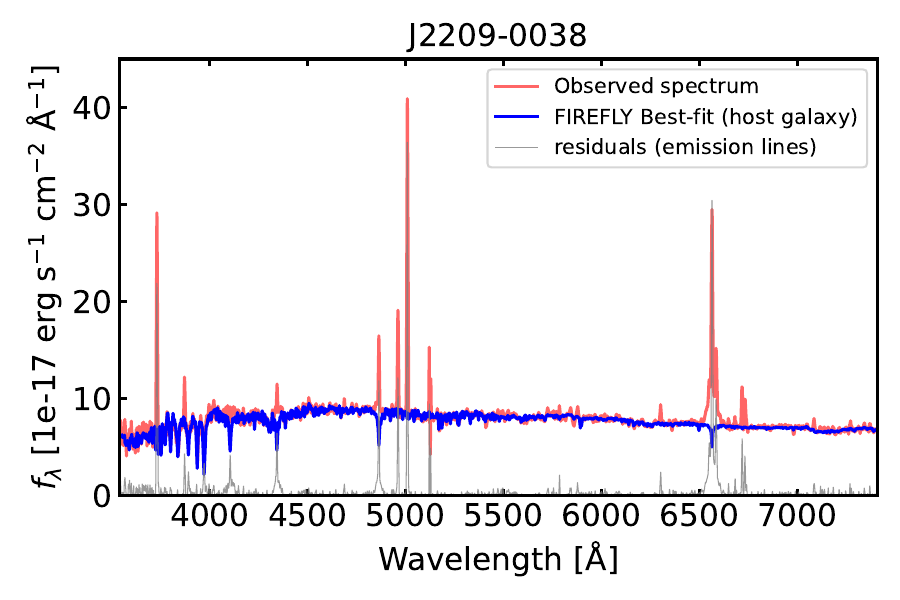} &
\includegraphics[width=0.3\textwidth]{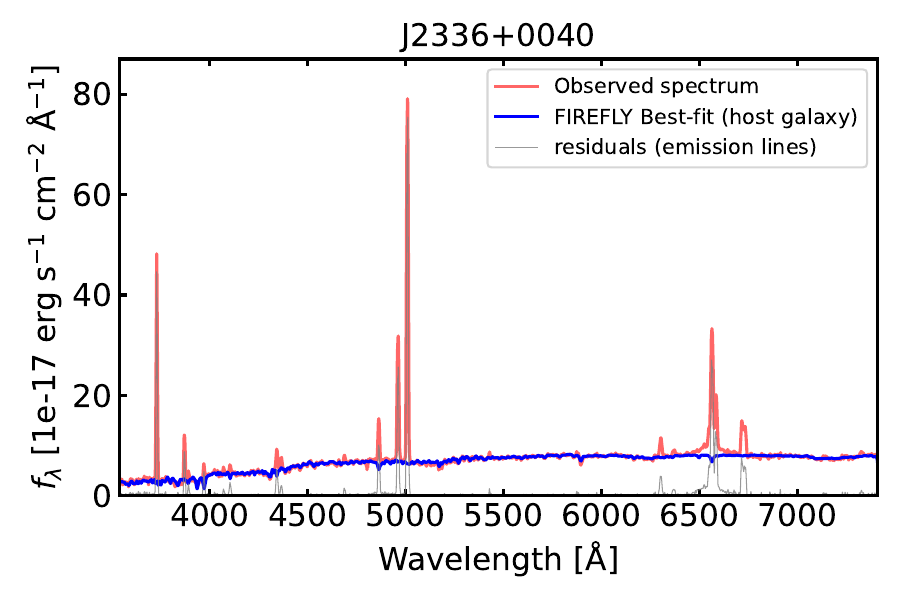}
\\
\end{tabular}
\addtocounter{figure}{-1}
\caption{\texttt{FIREFLY} host-galaxy spectral fits for the CL-AGN sample (continued). Each panel shows one object: the SDSS-V dim-state spectrum (red; smoothed over 5 pixels), the corresponding \texttt{FIREFLY} best-fitting host-galaxy model (blue), and the residuals (gray; also smoothed over 5 pixels). The prominent features visible in the residuals correspond to AGN emission lines that were masked during the \texttt{FIREFLY} fitting.}
\end{figure}

\end{appendix}

\end{document}